\documentclass[aps,nofootinbib,prr,reprint,superscriptaddress,longbibliography]{revtex4-1}
\usepackage{amsmath}
\usepackage{graphicx}
\usepackage{amsfonts}
\usepackage{color}
\usepackage{upgreek}
\usepackage{bm}
\usepackage{lipsum}
\usepackage{comment}
\usepackage{enumerate}
\usepackage{dsfont}

\usepackage[table]{xcolor}
\usepackage{textcomp}
\usepackage{amsmath}
\usepackage{amssymb}
\usepackage{graphicx} 
\usepackage{pgf}
\usepackage{epstopdf}
\usepackage{braket}
\usepackage{mathtools}
\usepackage{times}

\usepackage[retainorgcmds]{IEEEtrantools}
\newcommand{\bea}{\begin{eqnarray}}
\newcommand{\eea}{\end{eqnarray}}

\newcommand{\Tr}{\mathrm{Tr}}

\usepackage[colorlinks,linkcolor=blue,urlcolor=blue,citecolor=blue]{hyperref}

\begin{document}

\title{Separation of relaxation timescales via strong system-bath coupling: Dissipative three-level system as a case study  }

\author{Brett Min}
\email{brett.min@mail.utoronto.ca}
\affiliation{Department of Physics and Centre for Quantum Information and Quantum Control, University of Toronto, 60 Saint George St., Toronto, Ontario, M5S 1A7, Canada}

\author{Matthew Gerry}
\affiliation{Department of Physics and Centre for Quantum Information and Quantum Control, University of Toronto, 60 Saint George St., Toronto, Ontario, M5S 1A7, Canada}

\author{Dvira Segal}
\email{dvira.segal@utoronto.ca}
\affiliation{Department of Chemistry
University of Toronto, 80 Saint George St., Toronto, Ontario, M5S 3H6, Canada}
\affiliation{Department of Physics and Centre for Quantum Information and Quantum Control, University of Toronto, 60 Saint George St., Toronto, Ontario, M5S 1A7, Canada}

\begin{abstract}

We analytically demonstrate that strong system-bath coupling separates the relaxation dynamics of a dissipative quantum system into two distinct regimes: a short-time dynamics that, as expected, accelerates with increasing coupling to the environment, and a slow dynamics that, counterintuitively, becomes increasingly prolonged at sufficiently strong coupling. Using the reaction-coordinate polaron-transform mapping, we uncover the general mechanism behind this effect and derive accurate expressions for both relaxation timescales. Numerical simulations confirm our analytical predictions. From a practical perspective, our results suggest that strong coupling to a dissipative bath can autonomously generate and sustain long-lived quantum coherences, offering a promising strategy for bath-engineered quantum state preparation.

\end{abstract}

\maketitle

\date{\today}

\section{Introduction}

Dissipation, arising from the interaction of quantum systems with an external environment, is typically viewed as detrimental to maintaining delicate quantum superpositions and entanglement. To maintain the coherence of a quantum system, many quantum devices--such as those used in quantum information processing--are designed to be as isolated as possible from their surroundings until measurements are made. Contrary to this intuition, \textit{quantum dissipative state preparation} suggests that carefully-designed system-bath interactions and/or structured environments are a {\it resource} that can drive a quantum system toward states with desired properties \cite{Julio_2011,Barontini_2013,Tomita_2017}.

One aspect of dissipative state preparation involves engineering a steady state, specifically, allowing the quantum system to interact with its environment in a controlled manner such that, in the long-time limit, the system evolves toward an equilibrium state with desired properties. This includes engineering of pure states useful for quantum metrology~\cite{Kouzelis_2020}, rapid generation of many-body entanglement between multiple qubits~\cite{Lim_2024}, stabilization of long-range fermionic order~\cite{Neri_2025}, cooling of a quantum simulator via a dissipative driving of auxiliary spin degrees of freedom~\cite{Raghunandan_2020}, generation of steady-state coherence via a non-Markovian bath~\cite{Wu_2025}, engineering  magnetic orders in quantum spin chains \cite{Min_PRL}, and expanding the parameter regimes for a symmetry-protected topological order immersed in thermal environment \cite{Min_PRB}.

Beyond the steady state, the transient dynamics of dissipative systems also offers a wealth of opportunities for control. A prominent example is the phenomenon of \textit{metastability}, in which quantum coherence, correlations, or ordering exists for long but finite times before eventual thermalization \cite{Macieszczak_2016,Rose_2016,Kouzelis_2020,Hwang_2017,Maciezczak_2021,Nick_PRA,Mori_2021}. For instance, in many-body settings, dissipation can induce a hierarchy of relaxation timescales, giving rise to rich nonequilibrium phases and dynamical crossovers \cite{Cabot_2021,Cabot_2022,Macieszczak_2016,Kouzelis_2020,Rose_2016,Lim_2024,Hwang_2017,Maciezczak_2021,Wolff_2020,Giorgi_2023,Gerry_2024,Brumer_2014}. Such long-lived prethermal states and metastable plateaus can counterintuitively be {\it extended} by dissipation rather than destroyed by it~\cite{Cabot_2022,Hwang_2017,Wolff_2020}. This has opened up new perspectives on time-crystalline behavior, dissipation-assisted magnetic ordering, and other dynamic phenomena, suggesting that quantum coherence can survive far longer than expected. 

Understanding dynamical behaviors such as relaxation times hinges on the spectral structure of the \textit{Liouvillian}, the generator of open-system dynamics. The eigenspectrum of the Liouvillian in the complex plane governs dissipative timescales and reveals how multiple relaxation channels interplay \cite{Zhou_2021,Luitz_2020,Luitz_2022,Luitz_2024,Presilla_2021,Marko_2015,Sommer_2021,Rose_2022,Zheng_2023,Shu_2023,Yang_2024,Martina_2025,Zhou_2023,Rossatto_2016,Shao_2023,Nazir_2025}. For instance, short- and long-time relaxations are associated with the large and small real parts of the Liouvillian spectrum, respectively. The gap between the zero eigenvalue and the eigenvalue with the next smallest real part is known as the \textit{Liouvillian gap} \cite{Mori_2020,Mori_2023,Mori_2024,Yuan_2021}, which plays a key role in determining the asymptotic relaxation timescale and signals dissipative phase transitions when it closes~\cite{Kessler_2012,Moos_2012,Horstmann_2013,Minganti_2018,Haga_2021}. In addition, by exploiting the \textit{Liouvillian exceptional point}, where the slowest decay mode coalesces with a faster one, relaxation toward steady state can be accelerated, allowing the observation of the quantum \textit{Mpemba effect} \cite{Weibin_2023,Zhang2025}.

On a separate note, there has been a growing experimental and theoretical effort to explore and harness the effects of strong system–bath coupling (SBC) \cite{Pino_2015,Damanet_2019,Qu_2025s,Poletti_2012,Rosso_2021,Liu_2025,Blais_2011,Raghunandan_2020,Nick_PRXQ,Nick_PRB,Min_PRB,Min_PRL,Jitian_2025,Garwola_PRB,Brenes_2025,Brenes_JCP,Nazir18,Cao_2016,Napoli_2025,Mochida_2024,Masuki_2024,Ashida_2023,Masuki_2023,Masuki_2022,Ashida_2022,Ashida_2021}. For instance, in cavity quantum electrodynamics (QED) setups, an atom or material placed inside a cavity can interact strongly with the vacuum field, offering an alternative route to engineering and controlling useful quantum properties of the embedded system.

In this paper, our aim is to shed light on the role of strong SBC in determining the relaxation times of a dissipative system. For dissipative systems under the Markovian approximation, the Liouvillian is typically constructed from the Lindblad quantum master equation (QME) \cite{Manzano_2020}. However, one of the crucial assumptions in deriving the Lindblad QME is that the system–bath coupling must be weak. Furthermore, the system's energy eigenvalues must be well separated to justify the so-called \textit{secular approximation}. As a result, both numerical and analytical techniques have been developed to extend the standard Lindblad formalism to regimes where these approximations no longer hold \cite{Ankerhold_2013,Becker_2021,Poletti_2019,Rudner_2020,Betzholz_2022,Yang_2024,Neri_2025,Poletti_2012,Trushechkin_2022}.

To circumvent this issue, we employ the recently developed \textit{reaction-coordinate polaron-transform} (RCPT) approach, a Markovian embedding technique that allows strong system–bath coupling effects to be incorporated into an effective system Hamiltonian, which is then weakly coupled to the residual bath. Specifically, we apply this mapping technique to a three-level system coupled to a structured bosonic environment at arbitrary coupling strength and study how strong coupling influences the relaxation timescales. We emphasize that our setup includes a fully microscopic description of the system, the bath, and their interaction, and that we analytically derive the relaxation timescales as functions of the SBC strength. 

Our main findings are: (i) strong SBC leads to the emergence of two distinct timescales. The shorter time scale becomes even faster as the coupling strength increases, while the longer timescale slows down exponentially with increasing coupling. (ii) The branching of the two timescales is attributed to the anisotropy induced in the two dissipative channels by strong coupling. (iii) Between the two timescales, a metastable state emerges, characterized by long-lived coherence. At  sufficiently strong coupling, steady-state coherences persist. (iv) We provide a rigorous microscopic description that captures the development of this long-lived metastable state both in the weak-coupling regime---particularly for nearly degenerate energy levels \cite{Brumer_2014, Gerry_2024}---and in the strong-coupling regime for a general spectrum of the three-level system, under appropriate limiting conditions.

The rest of the paper is organized as follows: In Sec.~\ref{sec: Model}, we introduce a minimal impurity model consisting of a three-level system coupled to its environment at arbitrary coupling. In Sec.~\ref{sec: RCPT mapping}, we transform the model into a frame where the strong SBC becomes weak, restoring the validity of a perturbative treatment of the effective interaction. In Sec.~\ref{sec: time scales}, we extract the Liouvillian spectrum, thus the system's lifetimes, from a Lindblad QME and analyze their dependence on system and bath parameters. In Sec.~\ref{sec: Numerical simulation}, we numerically simulate dissipative dynamics to confirm that the timescales obtained from the Liouvillian spectrum accurately describe the behavior of the system. We conclude in Sec.~\ref{sec: conclusion}.

\begin{figure}[htbp]
\fontsize{6}{10}\selectfont 
\centering
\includegraphics[width=1.0\columnwidth]{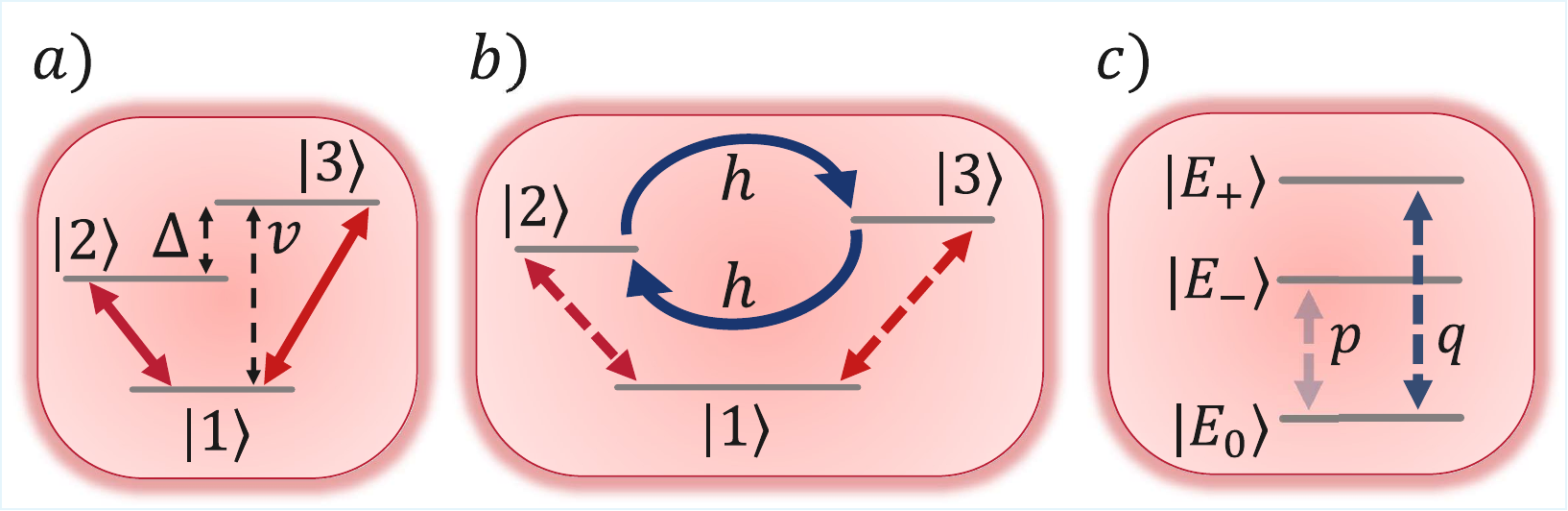}
\caption{(a) A schematic diagram of the three-level system in its initial representation. It consists of a single ground state $\ket{1}$ at zero energy and two excited states, $\ket{2}$ and $\ket{3}$. The splittings $v$ and $v - \Delta$ separate the levels $\ket{1} \leftrightarrow \ket{3}$ and $\ket{1} \leftrightarrow \ket{2}$, respectively. Thick red arrows indicate transitions induced by interaction with a thermal bath at temperature $T$. (b) The model after the RCPT mapping, which reorganizes the energy levels and introduces a non-diagonal term in the effective system Hamiltonian. This latter term connects $\ket{2}$ and $\ket{3}$ with amplitude $h$ (blue curly arrows). The three-level system still interacts with the bosonic environment via the same coupling operator as in the original representation. This operator induces the same transition in the system, but with a modified amplitude, indicated by dashed red arrows. (c) Effective system in a diagonal basis. In this representation, the bath induces anisotropy between transitions $\ket{E_0}\leftrightarrow\ket{E_-}$ and $\ket{E_0}\leftrightarrow\ket{E_+}$, with transition matrix elements denoted by $p$ (fading dashed arrow) and $q$ (thick dashed arrow), respectively.}
\label{fig:figure 1}
\end{figure}

\section{Model}
\label{sec: Model}

In this section, we introduce a minimal model of choice for a quantum system strongly coupled to its environment. It consists of a three-level system (a ground state at zero energy with two non-degenerate excited states) interacting with a bosonic environment. The total Hamiltonian is given as
\begin{equation}
\label{eq: total Hamiltonian original}
    \hat{H} = \hat{H}_S+\hat{H}_I+\hat{H}_B,
\end{equation}
where 
\begin{equation}
\begin{aligned}
    \hat{H}_S =& (v-\Delta)\ket{2}\bra{2}+v\ket{3}\bra{3},\\
    \hat{H}_I = & \hat{S}\sum_kt_k(\hat{c}^\dagger_k+\hat{c}_k),\quad
    \hat{H}_B =& \sum_k\nu_k\hat{c}^\dagger_k\hat{c}_k.
    \end{aligned}
\end{equation}
Here, $v$ is the level splitting between the ground state $\ket{1}$ and the second excited state $\ket{3}$, while $v-\Delta$ separates $\ket{1}$ and the first excited state $\ket{2}$ with $\Delta < v$. The system-bath coupling operator, 
\bea
\hat{S}= \frac{\ket{1}\bra{2}+\ket{1}\bra{3}}{\sqrt{2}}+\text{H.c.},
\label{eq:S}
\eea
induces transitions between 
the ground state and the excited manifold. 
The bath is described by a collection of harmonic oscillators with $\{\hat{c}_k\}$ as the canonical bosonic operators, where $\nu_k$ and $t_k$ are the frequencies and the SBC strengths between the system's operator $\hat{S}$ and the  displacements of the bath's modes, respectively. The effect of bath on the system is typically described by a spectral density function defined as $J(\omega)=\sum_kt^2_k\delta(\omega-\nu_k)$. Fig.~\ref{fig:figure 1}(a) provides a schematic diagram of the model. 

In the case where $\Delta\ll v$, amounting to a near-degeneracy between the two excited states, this system corresponds to the so-called ``V'' model. This model has been shown to exhibit long-lived metastable states, often involving coherences in the energy eigenbasis, during the relaxation towards equilibrium \cite{Ivander_2023, Gerry_2024, Brumer_2014}, as well as coherences that persist at steady state in nonequilibrium setups \cite{Ivander_2022, Gerry_2023, Koyu_2021}.
Our study does not restrict $\Delta$ only to this regime of near degeneracy. 

\section{RCPT mapping and the Effective Hamiltonian}
\label{sec: RCPT mapping}

Our objective is to study the dynamics of the three-level system without restricting the interaction energy with the bath to be weak. 
Although numerous numerically exact techniques can tackle this problem, such as hierarchical equations of motion \cite{Tanimura_2020,Lambert_2023} and  influence functional path integral and Monte Carlo approaches \cite{Kundu23,Segal_2010,Theta1,Theta2,Gribben_2020}, our goal here is to gain analytical insight into the problem. The polaron-transformed quantum master equation, based on the polaron mapping for redefining the system-bath interaction energy, is one of the leading approaches in this respect. However, this method is typically cumbersome for analytical work even for minimal models \cite{Segal02,Jang08,Ahsan13,Cao15,Ahsan24}.
The recently developed
\textit{reaction-coordinate polaron-transform} (RCPT) approach
\cite{Nick_PRXQ} offers a powerful, elegant and transparent alternative. The method was shown to correctly capture both the weak and ultra-strong coupling limits of the equilibrium spin-boson model \cite{Nick_PRXQ}.
The RCPT technique has been previously applied to study steady-state properties of impurity systems \cite{Nick_PRXQ,Nick_PRB}, bath engineering of magnetic order in quantum spin chains \cite{Min_PRL}, dissipative fermionic chain models with symmetry-protected topological order \cite{Min_PRB}, bath-generated couplings and synchronization of non-interacting spins \cite{Brenes_JCP}, extensions to multiple baths with non-commuting system-bath couplings \cite{Garwola_PRB}, optimal heat transport \cite{Brenes_2025}, and suppression of decoherence via a frustration of dissipation \cite{Jitian_2025}. Here, for the first time, the RCPT method is applied to analyze the role of strong coupling to the lifetime of dissipative systems.

In order to extract the characteristic timescales of relaxation dynamics under strong system-bath coupling, we must first construct the Liouvillian superoperator $\hat{\mathcal{L}}$ for our dissipative three-level system model. To do so, we begin by formulating the Lindblad QME based on the model described in Eq.~\eqref{eq: total Hamiltonian original}. However, because the Lindblad QME is valid only in the weak SBC regime, we cannot directly construct $\hat{\mathcal{L}}$ from Eq.~\eqref{eq: total Hamiltonian original} where the coupling terms $t_k$ are assumed to lead to an arbitrarily strong coupling energy. Hence, we must move to a frame in which the SBC appears weak, allowing the dissipation to be treated perturbatively.  

To this end, we apply the RCPT approach on Eq.~\eqref{eq: total Hamiltonian original}, thereby restoring the validity of weak-coupling approaches \cite{Nick_PRXQ}. The RCPT mapping yields an effective Hermitian Hamiltonian of the form $\hat{H}^\text{eff} = \hat{H}^\text{eff}_S + \hat{H}^\text{eff}_I + \hat{H}^\text{eff}_B$, where the effects of strong SBC are absorbed into the effective Hamiltonian of the system, $\hat{H}^\text{eff}_S$, while the effective interaction term $\hat{H}^\text{eff}_I$ can be treated perturbatively. This mapping involves two successive transformations, followed by a projection onto a low-energy manifold, which is valid within specific parameter regimes to be discussed shortly.

The first of the two transformations is the \textit{reaction coordinate} (RC) mapping \cite{Nazir18}.
Here, we extract a collective degree of freedom from the bosonic reservoir and incorporate it into the system. The RC is essentially a single harmonic oscillator mode with frequency $\Omega$. It is coupled to the system with coupling strength $\lambda$. These parameters are defined based on the original spectral density function,  $ \lambda^2 = \frac{1}{\Omega}\int^\infty_0 \omega J(\omega)d\omega$ and $ \Omega^2  = \frac{\int^\infty_0 \omega^3 J(\omega)d\omega}{\int^\infty_0 \omega J(\omega)d\omega}$ \cite{Nazir18}.
With the RC transformation,
Eq.~\eqref{eq: total Hamiltonian original} is remolded into 
\begin{equation}
\begin{aligned}
    \hat{H}_\text{RC} = &\hat{H}_S+\lambda\hat{S}\left(\hat{a}^\dagger+\hat{a}\right)+\Omega\hat{a}^\dagger\hat{a}\\
    +&\left(\hat{a}^\dagger+\hat{a}\right)\sum_kf_k\left(\hat{b}^\dagger_k+\hat{b}_k\right)+\sum_k\omega_k\hat{b}^\dagger_k\hat{b}_k.
    \end{aligned}
    \label{eq:HRC}
\end{equation}
Here, $\hat{a}^\dagger~(\hat{a})$ is the creation (annihilation) operator for the RC mode and $\hat{b}^\dagger_k~(\hat{b}_k)$ are the creation (annihilation) operator for the residual bath modes. The RC transformation is essentially a \textit{Bogoliubov} transformation that achieves $\lambda(\hat{a}+\hat{a}^\dagger)=\sum_kt_k(\hat{c}_k+\hat{c}^\dagger_k)$. The coupling between the RC and the residual bath is described by a different spectral density function, $J_\text{RC}(\omega)=\sum_kf^2_k\delta(\omega-\omega_k)$, where $f_k$ are the new coupling parameters. This new spectral density function can be found from the original spectral function: $J_\text{RC}(\omega)=\frac{2\pi \lambda^2 J(\omega)}{[\mathcal{P}\int\frac{J(\omega')}{\omega'-\omega}d\omega']^2+\pi^2J(\omega)}$ with $\mathcal{P}$ indicating a principal-value integral \cite{Nazir18}.

The second transform is the unitary \textit{polaron} transformation given by $\hat{U}_P=\exp[\frac{\lambda}{\Omega}(\hat{a}^\dagger-\hat{a})\hat{S}]$, which is enacted on the system-RC Hilbert space while leaving the residual bath unaffected. This transform shifts the bosonic operator of the RC mode according to the state of the system (via the coupling to $\hat{S}$) as $\hat{a}^{(\dagger)}\rightarrow \hat{a}^{(\dagger)}-\frac{\lambda}{\Omega}\hat{S}$. The polaron transform 
seemingly complicates the Hamiltonian:
It imprints the RC coupling into the original system thereby weakening the coupling between the RC and the system. However, in addition, it generates a new coupling term between the original system and the residual bath. It may also generate bath-induced interactions between systems' degrees of freedom. Post polaron transform, the Hamiltonian 
Eq. (\ref{eq:HRC}) turns into
\bea
    \hat{H}_\text{RC-P} &=& 
    \hat{U}_P\hat{H}_\text{RC}\hat{U}^\dagger_P 
    \nonumber\\
    &=&
    \hat{U}_P\hat{H}_S\hat{U}^\dagger_P-\frac{\lambda^2}{\Omega}\hat{S}^2+\Omega\hat{a}^\dagger\hat{a}
    \nonumber\\
    &+&\left(\hat{a}^\dagger+\hat{a}-\frac{2\lambda}{\Omega}\hat{S}\right)\sum_kf_k\left(\hat{b}^\dagger_k+\hat{b}_k\right)+\sum_k\omega_k\hat{b}^\dagger_k\hat{b}_k,
    \nonumber\\
\eea
where the bath-induced interaction term is given by $-\frac{\lambda^2}{\Omega}\hat{S}^2$.

As the last step of the RCPT procedure, the RC-P Hamiltonian is truncated by projecting it onto the ground state of the polaron-transformed RC. Such a restriction to a sub-manifold of the Hamiltonian is justified under the assumption that $\Omega$, the characteristic frequency of the bath, is the largest energy scale in the problem. 
This truncation eliminates terms in $\hat{H}_\text{RC-P}$, those that are proportional to $\hat{a}^\dagger\hat{a}$, as well as $\hat{a}^\dagger+\hat{a}$. Hence, after the RCPT machinery, we completed the transition of the model Hamiltonian (\ref{eq: total Hamiltonian original}) into the final effective Hamiltonian, 
\begin{equation}
\label{eq: effective total Hamiltonian}
    \hat{H}^\text{eff} = \hat{H}^\text{eff}_S(\lambda,\Omega)-\hat{S}\sum_k\frac{2\lambda f_k}{\Omega}\left(\hat{b}^\dagger_k+\hat{b}_k\right)+\sum_k\omega_k\hat{b}^\dagger_k\hat{b}_k,
\end{equation}
where 
\bea 
\hat{H}^\text{eff}_S(\lambda,\Omega) = \bra{0}\hat{U}_P\hat{H}_S\hat{U}^\dagger_P\ket{0}-\frac{\lambda^2}{\Omega}\hat{S}^2.
\label{eq:HSeff}
\eea
Here, $\ket{0}$ is the ground state of the RC mode. Note that the system-bath coupling strength has been re-scaled $f_k\rightarrow 2\lambda f_k/\Omega$. As a result, the bath spectral density function after the RCPT machinery is $J^\text{eff}(\omega)=\frac{4\lambda^2}{\Omega^2}J_\text{RC}(\omega)$. 

We emphasize that under the RCPT mapping, the system-bath operator remains intact. Thus, the matrix $\hat S$ of Eq.~\eqref{eq:S} still describes the system bath coupling operator, with the ground state coupled to the two excited states.

For our model Eq.~\eqref{eq: total Hamiltonian original}, the RCPT mapping leads to the effective system's Hamiltonian,
\begin{widetext}
        \begin{equation}
        \label{eq: effective system Hamiltonian}
\hat{H}^\text{eff}_S(\lambda,\Omega) = \left(\begin{matrix}
     \frac{2v-\Delta}{4}(1-e^{-\frac{2\lambda^2}{\Omega^2}})-\frac{\lambda^2}{\Omega}& 0  & 0\cr 
     0   & \frac{-4\Delta e^{-\frac{\lambda^2}{2\Omega^2}}+(2v-\Delta)(3+e^{-\frac{2\lambda^2}{\Omega^2}})}{8}-\frac{\lambda^2}{2\Omega}& \frac{(-2v+\Delta)(1-e^{-\frac{2\lambda^2}{\Omega^2}})}{8}-\frac{\lambda^2}{2\Omega}\cr
   0   & \frac{(-2v+\Delta)(1-e^{-\frac{2\lambda^2}{\Omega^2}})}{8}-\frac{\lambda^2}{2\Omega}&  \frac{4\Delta e^{-\frac{\lambda^2}{2\Omega^2}}+(2v-\Delta)(3+e^{-\frac{2\lambda^2}{\Omega^2}})}{8}-\frac{\lambda^2}{2\Omega}\cr
    \end{matrix} \right),
\end{equation}
\end{widetext}
see Appendix~\ref{sec: Derivation for the effective system Hamiltonian} for a detailed derivation. The effective bath-spectral density takes the form \cite{NickNJP,Nazir18}
\bea
J^\text{eff}(\omega)=\frac{4\lambda^2}{\Omega^2}\Gamma \omega,
\label{eq:Jeff}
\eea
given the original spectral density is Brownian: 
\bea
J(\omega)=\frac{4\Gamma \Omega^2 \lambda^2 \omega}{(\omega^2-\Omega^2)^2+(2\pi\Gamma \Omega \omega)^2},
\label{eq:Joriginal}
\eea
which describes a peaked function centered at $\Omega$, with $\Gamma$ controlling the width \cite{Nazir18,NickNJP}. In the effective picture, the dimensionless width parameter $\Gamma$ becomes an overall prefactor in the spectral function. Consequently, an original Brownian spectral function with a narrow peak provides the most useful choice for the RCPT mapping, as it ensures that the effective system-bath interaction $\hat{H}^\text{eff}_I$ remains weak. 

Let us discuss some of the key features of $\hat{H}^\text{eff}_S(\lambda,\Omega)$:

(i) The ground state remains coherently decoupled from the two excited states. Its energy, given by
\begin{equation}
    E_0\equiv \frac{2v-\Delta}{4}(1-e^{-\frac{2\lambda^2}{\Omega^2}})-\frac{\lambda^2}{\Omega},
\end{equation}
is lowered quadratically as $\lambda$ increases.

(ii) The coupling to the bath introduces {\it off-diagonal elements} in the system's Hamiltonian connecting the excited states, $\ket{2}$ and $\ket{3}$. This matrix element, given by
\begin{equation}
\label{eq:h}
   h\equiv\frac{(-2v+\Delta)(1-e^{-\frac{2\lambda^2}{\Omega^2}})}{8}-\frac{\lambda^2}{2\Omega},
\end{equation}
increases in magnitude as the SBC increases.
Importantly, in our choice of parameters for $v$ and $\Delta$, at both weak coupling and strong coupling, the tunneling element $h$ is dominated by the second term, $h\approx -\lambda^2/(2\Omega)$.

(iii) The original energy separation $\Delta$ between $\ket{3}$ and $\ket{2}$ is exponentially suppressed as $\lambda$ increases. The midpoint between the two excited states in this effective picture is
\begin{equation}
    l\equiv\frac{(2v-\Delta)(3+e^{-\frac{2\lambda^2}{\Omega^2}})}{8}-\frac{\lambda^2}{2\Omega}.
\end{equation}
We label the corresponding energies  by $E_2=l+w$ and $E_3=l-w$, where
 \begin{equation}
     w=-\frac{\Delta e^{-\frac{\lambda^2}{2\Omega^2}}}{2},
 \end{equation}
is associated with the exponential suppression.

Accordingly, the matrix form of $\hat H_S^{\text{eff}}(\lambda,\Omega)$ may be expressed more concisely as
\begin{equation}\label{eq: effective system Hamiltonian simplified}
    \hat{H}^\text{eff}_S = \left(\begin{matrix}
        E_0 & 0 & 0 \cr
        0 & l+w & h \cr
        0 & h & l-w \cr
    \end{matrix} \right).
\end{equation}
Fig.~\ref{fig:figure 1}(c) presents a schematic diagram of the effective three-level system. 

What have we achieved? Starting from the model Hamiltonian Eq.~\eqref{eq: total Hamiltonian original}, with arbitrary coupling strength to the bath captured by $\lambda$, we have obtained a new effective Hamiltonian, Eq.~(\ref{eq: effective total Hamiltonian}), in which $\lambda$ renormalizes levels and generates bath-induced tunneling between excited states. The coupling of the system to the bath in the effective picture is assumed to be weak, controlled by the parameter $\Gamma$, see Eqs.~\eqref{eq:Jeff} and \eqref{eq:Joriginal}. 
The assumption of weak coupling for the effective Hamiltonian allows us to treat the dissipative dynamics of the effective system using weak coupling techniques, such as second-order QMEs. 
In fact, as we argue in the next section, we can use a secular QME over a broad range of system-bath coupling parameters, even if the excited states of the bare system Hamiltonian are close in energy. This is because, in the energy basis of the effective Hamiltonian, all three levels are well separated leading to a short bare oscillation period of the system, compared to the relaxation timescale. A more careful discussion is presented in Secs. \ref{sec: time scales}-\ref{sec: Numerical simulation}.

\section{Dissipative Time scales}
\label{sec: time scales}

Now that we have $\hat{H}^\text{eff}$, which describes an effective three-level system weakly coupled to a thermal bath, let us write a Lindblad QME associated with this model. To do so, we need to choose a basis in which to write the Lindblad jump operators. 
We employ the energy eigenstates of the system to construct the jump operators. Therefore, let us go to a diagonal basis of the system Hamiltonian to identify the Bohr frequencies and the associated jump operators. 

\begin{figure}[htbp]
\fontsize{6}{10}\selectfont 
\centering
\includegraphics[width=1.0\columnwidth]{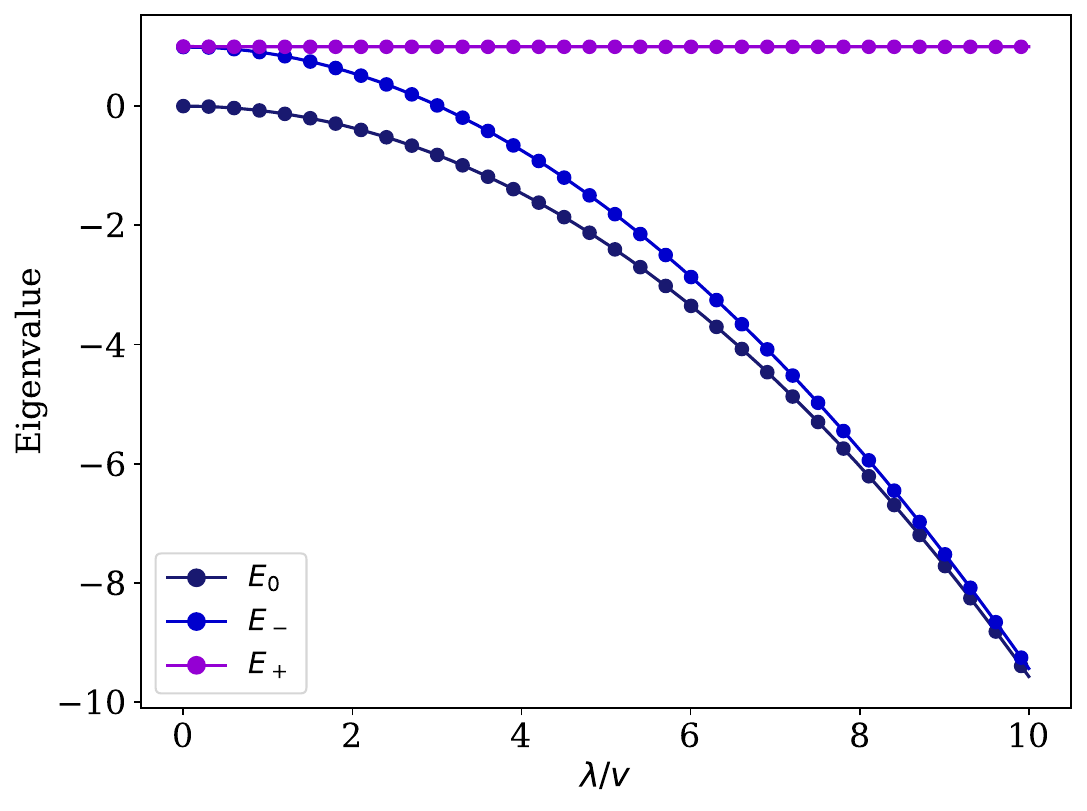}
\caption{Eigenvalues of $\hat{H}^\text{eff}_S$ as a function of $\lambda$. Parameters are $v=1$, $\Delta=0.01$, and $\Omega=10$. Stronger coupling pushes $E_0$ and $E_-$ down to lower energies, while suppressing the splitting between the two. $E_+$ remains relatively unchanged as a function of $\lambda$.}
\label{fig:figure 2}
\end{figure}

Note that $\hat{H}^\text{eff}_S$ in Eq.~\eqref{eq: effective system Hamiltonian simplified} is in a block-diagonal form. 
Therefore, the corresponding energy eigenvalues $\{E_0,E_\pm\}$ and the associated energy eigenstates $\{\ket{E_0},\ket{E_\pm}\}$ are easily identified as $E_\pm=l\pm\sqrt{w^2+h^2}$ and
\begin{equation}
\begin{aligned}
    \ket{E_0} =& (1,0,0)^T, \\
    \ket{E_+} =& (0,\cos(\phi/2),\sin(\phi/2))^T\\
    \ket{E_-} =&  (0,-\sin(\phi/2),\cos(\phi/2))^T,
    \end{aligned}
\end{equation}
where 
\begin{equation}
    \tan\phi=\frac{h}{w}.
\end{equation}
In Fig.~\ref{fig:figure 2}, we plot the energy eigenvalues as a function of the coupling strength $\lambda$ with the choice of the parameters $v=1$ and $\Delta=0.01$ for the system, and $\Omega=10$ for the characteristic frequency of the bath. We note that the value for $E_+$ remains relatively unchanged as a function of $\lambda$ while $E_0$ and $E_-$ are pushed to lower values, and their splitting becomes smaller.

\begin{figure}[htbp]
\fontsize{6}{10}\selectfont 
\centering
\includegraphics[width=1.0\columnwidth]{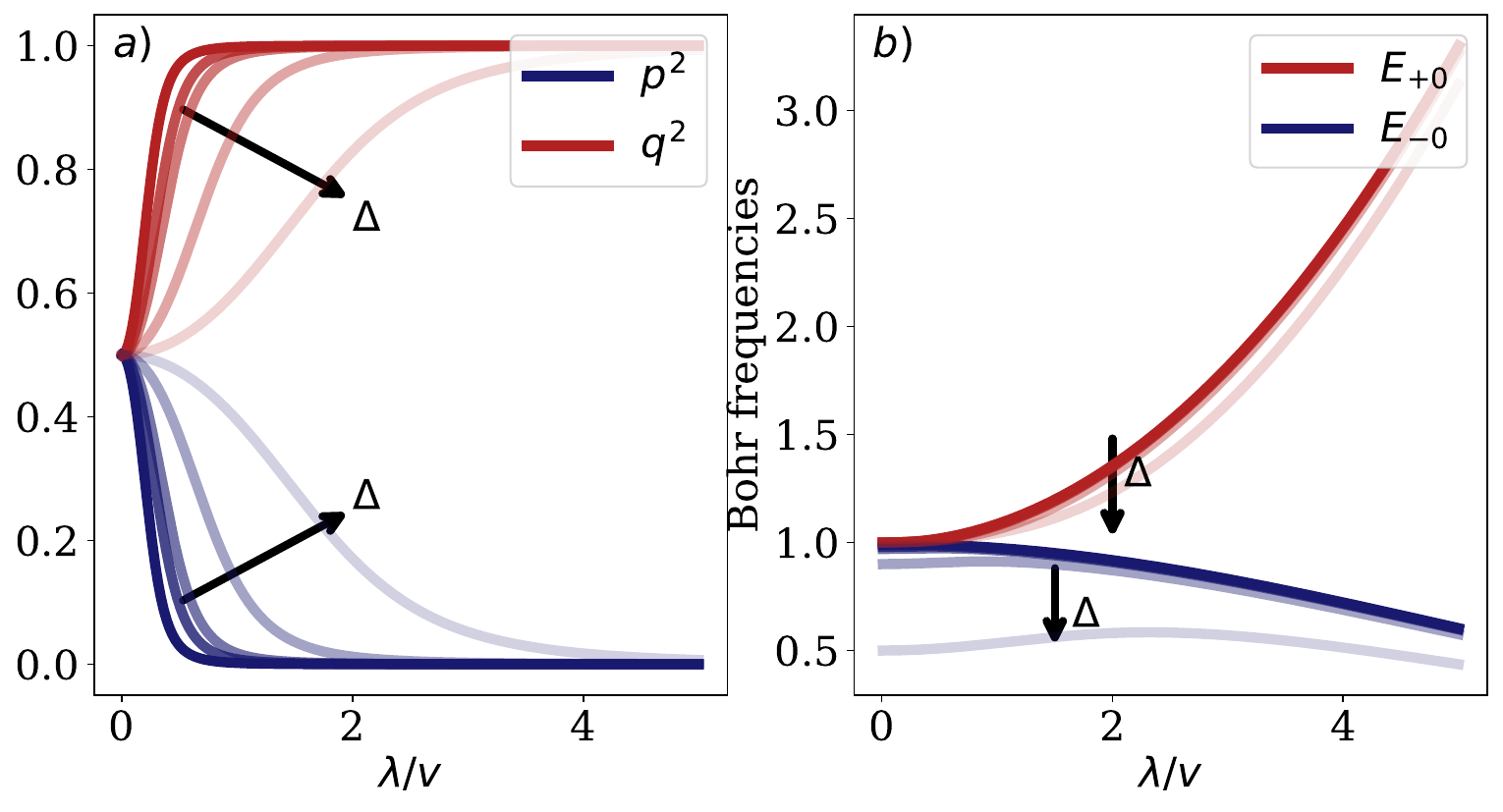}
\caption{(a) $p^2$ and $q^2$ as a function of $\lambda$ for  $\Delta=\{0.01,0.02,0.03,0.1,0.5\}$ in fading order indicated by the arrows. For large $\lambda$, $p^2\rightarrow 0$ while $q^2\rightarrow1$. 
(b) Bohr frequencies as a function of $\lambda$ at different values of $\Delta$ [identical to that of panel $(a)$]. For a wide range of coupling strength, the Bohr frequencies are of $\mathcal{O}(v)$. Here, $v=1$, $\Omega=10$, and $\Gamma=0.05$.}
\label{fig:figure 4}
\end{figure}

The effective total Hamiltonian $\hat{H}^\text{eff}$ in the diagonal basis ($E$) of the system is given as, 
\bea
\label{eq: effective Hamiltonian in the diagonal basis}
\hat{H}_E^\text{eff}
& =& 
 \hat{H}^\text{eff}_{S,E}-\left(\begin{matrix}
        0 & p & q \cr
        p & 0 & 0 \cr
        q & 0 & 0 \cr
    \end{matrix} \right)\sum_k\frac{2\lambda f_k}{\Omega}(\hat{b}^\dagger_k
    +\hat{b}_k)
    \nonumber\\
    &+&\sum_k\omega_k\hat{b}^\dagger_k\hat{b}_k,
\eea
where $\hat{H}^\text{eff}_{S,E}=\text{diag}(E_0,E_-,E_+)$.
The system-bath operator $\hat S$ maintained its general matrix form as in Eq. (\ref{eq:S}), but with the modified elements
\begin{equation}
\label{eq:pq}
    p=\sin\Big(-\frac{\phi}{2}+\frac{\pi}{4}\Big),\quad \text{and} \quad q = \sin\Big(\frac{\phi}{2}+\frac{\pi}{4}\Big),
\end{equation}
see Fig.~\ref{fig:figure 1}(c) for a schematic diagram
and Appendix \ref{AppB} for details.
These coefficients satisfy $p^2+q^2=1$. That is, in the diagonal basis for the system, the strong coupling induces an {\it anisotropy} between the jumps $\ket{E_0}\leftrightarrow\ket{E_-}$ and $\ket{E_0}\leftrightarrow\ket{E_+}$. As we argue shortly, this anisotropy will be the dominant reason behind the diverging timescales appearing in the relaxation dynamics at strong SBC. 

It is notable that the bath-induced tunneling energy $h$ is dominated by the second term in Eq.~\eqref{eq:h}, and that this approximation holds from weak to strong coupling,
$h\approx 
-\frac{\lambda^2}{2\Omega}$. 
As such, 
$\frac{h}{w} \approx \frac{\lambda^2 \exp(\frac{\lambda^2}{2\Omega^2}) }{\Omega\Delta}$.
%
This ratio grows exponentially with $\lambda$.
We now use the approximation, $\arctan x \approx  \pi/2 - 1/x $, valid for large $x$, leading to
\begin{equation}
    \frac{\phi}{2}=\frac{\arctan(h/w)}{2}\approx\frac{\pi}{4}-\frac{\Omega\Delta \exp(-\frac{\lambda^2}{2\Omega^2})
}{2\lambda^2}.
\label{eq:phi}
\end{equation}
 This approximation holds for our parameters $\Omega=10$, $\Delta =0.01$ 
once $\lambda>0.5$. 
For larger value of the excited state splitting, $\Delta=0.5$,
the approximation holds once $\lambda>4$. 


Plugging Eq.~\eqref{eq:phi} into Eq.~\eqref{eq:pq}, we 
find that, for large enough $\lambda$, $p\approx \frac{\Omega\Delta \exp(-\frac{\lambda^2}{2\Omega^2})
}{2\lambda^2}$ and $q \approx 1-
 \frac{\Omega^2\Delta^2 \exp(-\frac{\lambda^2}{\Omega^2})
}{8\lambda^4}$.
In words: at strong coupling, $p$ decreases, that is, $|E_-\rangle$ becomes a dark state. In contrast, $q\to 1$ at strong coupling.
Below we show that the relaxation timescales depend on $p^2$ and $q^2$, which are approximately given by
\begin{equation}
\label{eq: pandqss}
\begin{aligned}
    p^2 \approx& \frac{\Omega^2\Delta^2 \exp(-\frac{\lambda^2}{\Omega^2})
}{4\lambda^4},\\
q^2\approx&
1-  \frac{\Omega^2\Delta^2 \exp(-\frac{\lambda^2}{\Omega^2})
}{4\lambda^4}.
\end{aligned}
\end{equation}
These coefficients satisfy $p^2+q^2=1$, as required, even in their approximate form. Fig.~\ref{fig:figure 4} presents the behavior of $p^2$ and $q^2$ as a function of $\lambda$, 
as well as the Bohr frequencies in the eigenbasis of the effective Hamiltonian.

Watching the eigenenergies in Fig.~\ref{fig:figure 2} and the Bohr's frequencies in Fig. \ref{fig:figure 4}, it is clear that even when we start with a model where initially the two excited states are close in energy, in the energy basis of the effective Hamiltonian, $E_{\pm}$ separates as the interaction energy grows. As such, there is a broad regime of coupling parameters that allows making the secular approximation, where we neglect coherences between the three eigenstates of the system's Hamiltonian. 

We are now ready to write down the Lindblad QME and the associated Lindbladian $\hat{\mathcal{L}}$. The fully secular Lindblad QME is given by
\begin{equation}
    \frac{d}{dt}\hat{\rho} = -i\left[\hat{H}^\text{eff}_{S,E},\hat{\rho}\right]+\sum_n\gamma_n\left(\hat{L}_n\hat{\rho}\hat{L}^\dagger_n-\frac{1}{2}\{\hat{L}^\dagger_n\hat{L}_n,\hat{\rho}\} \right),
\end{equation}
where $n$ is an index for the possible jumps allowed by the system-bath coupling operator. We can read them off from Eq.~\eqref{eq: effective Hamiltonian in the diagonal basis} to be
\begin{equation}
\begin{aligned}
    \hat{L}_{+0}=&q\ket{E_+}\bra{E_0}, \quad \hat{L}_{-0} = p\ket{E_-}\bra{E_0}\\
    \hat{L}_{0+}=&q\ket{E_0}\bra{E_+},\quad \hat{L}_{0-}=p\ket{E_0}\bra{E_-}
    \end{aligned}
\end{equation}
with corresponding rates that satisfy local detailed balance relation given by
\begin{equation}
\begin{aligned}
    \gamma_{\pm 0} =& 2\pi J^\text{eff}(E_{\pm0})n(E_{\pm0})\\
\gamma_{0\pm} =& 2\pi J^\text{eff}(E_{\pm0})[n(E_{\pm0})+1]. \\
    \end{aligned}
\end{equation}
Here, $E_{ab}=E_a-E_b$, and $n(\omega)=(e^{\beta \omega}-1)^{-1}$ is the Bose-Einstein distribution with an inverse temperature $\beta=1/T$, setting $k_B\equiv 1$.

The fully secular Lindblad QME decouples here populations and coherences. Therefore, after eliminating the reduced density matrix element for the ground state population $\rho_{00}$ using the normalization condition, we can easily write down the following $2\times 2$ kinetic rate equation for the two populations $\rho_{--}$ and $\rho_{++}$:
\bea
     \left(\begin{matrix}
        \dot{\rho}_{--} \cr
        \dot{\rho}_{++} \cr
    \end{matrix} \right) & =&  \left(\begin{matrix}
        -p^2(\gamma_{-0}+\gamma_{0-}) & -p^2\gamma_{-0} \cr
        -q^2\gamma_{+0} & -q^2(\gamma_{+0}+\gamma_{0+})
    \end{matrix}\right)\left(\begin{matrix}
        \rho_{--} \cr
        \rho_{++} \cr
    \end{matrix}\right)
    \nonumber\\
    & +& \vec{d}
\eea
Here, $\vec{d}=(p^2\gamma_{-0},q^2\gamma_{+0})^T$, which leads to the steady-state, $\hat{\mathcal{L}}^{-1}\vec{d}$; 
we denote by $\hat{\mathcal{L}}$ the Liouvillian, which is the rate matrix in this case.
The eigenvalues of the $2\times 2$ Liouvillian matrix provide two timescales. They are given by the negative of the reciprocals of the eigenvalues themselves:

\begin{figure*}
\fontsize{6}{10}\selectfont 
\centering
\includegraphics[width=1.0\textwidth]{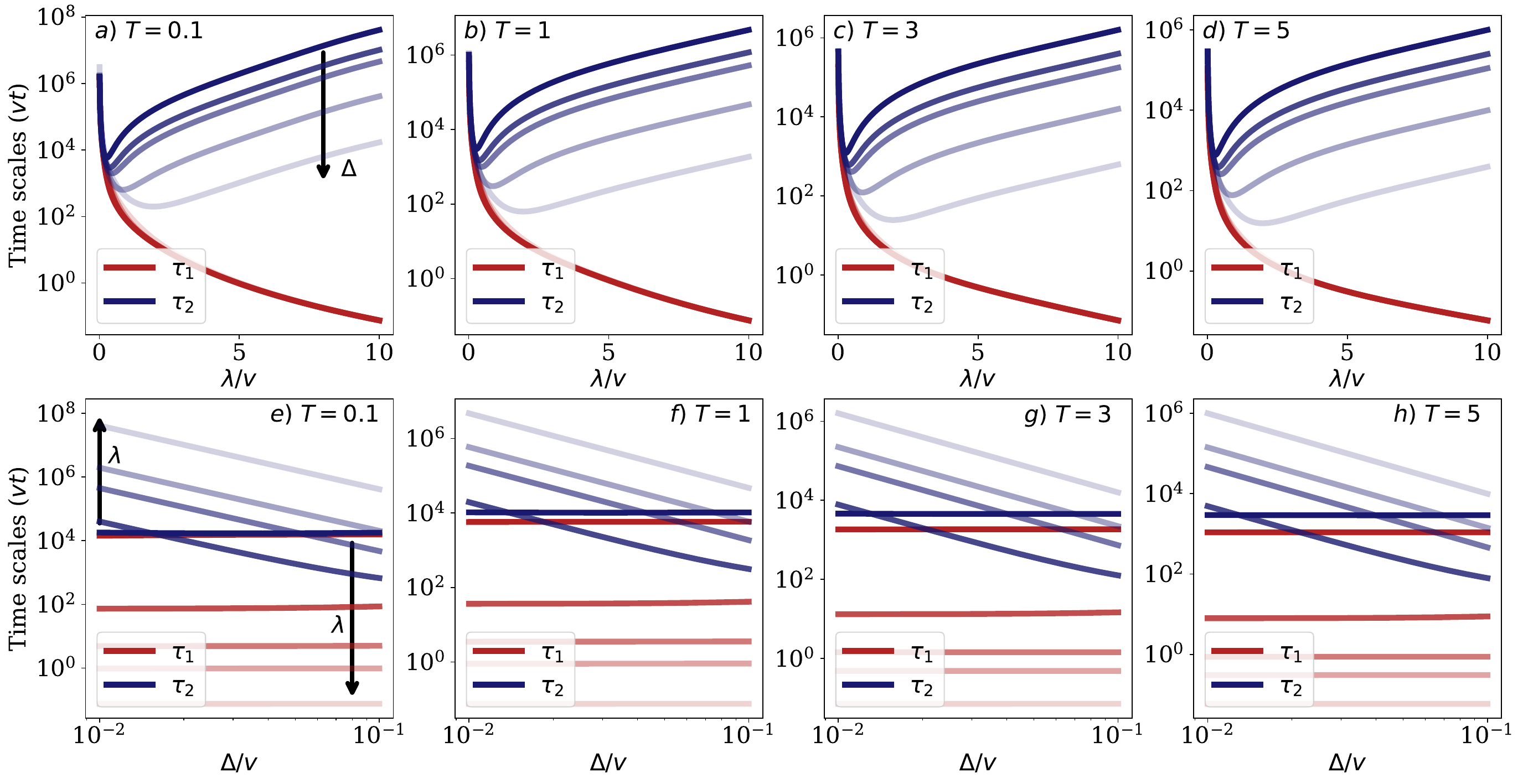}
\caption{[(a)-(d)] The relaxation timescales   $\tau_1$ and $\tau_2$ presented as a function of $\lambda$ at different temperatures and at different choices of $\Delta \in \{0.01,0.02,0.03,0.1,0.5\}$ in fading order indicated by the arrow. Note, $\tau_1$ is almost independent of $\Delta$ while $\tau_2$ is pulled towards shorter timescales upon increasing $\Delta$. It also depends nonmonotonically  on $\lambda$. [(e)-(h)] The dissipative timescales $\tau_1$ and $\tau_2$ presented as a function of $\Delta$ at different temperatures and for different choices of $\lambda\in\{0.1,1,3,5,10\}$ in fading order indicated by the arrows. As observed in [(a)-(d)], $\tau_1$ is almost independent of $\Delta$ while $\tau_2$ decreases as a function of $\Delta$ at large enough $\lambda$.}
\label{fig:figure 3}
\end{figure*}

\begin{widetext}
    \begin{equation}
    \label{eq: two timescales}
        \begin{aligned}
            \tau_1 = & \frac{2}{[p^2(\gamma_{-0}+\gamma_{0-})+q^2(\gamma_{+0}+\gamma_{0+})]+\sqrt{[p^2(\gamma_{-0}+\gamma_{0-})-q^2(\gamma_{+0}+\gamma_{0+})]^2+4p^2q^2\gamma_{+0}\gamma_{-0}}},\\
            \tau_2 = &\frac{2}{[p^2(\gamma_{-0}+\gamma_{0-})+q^2(\gamma_{+0}+\gamma_{0+})]-\sqrt{[p^2(\gamma_{-0}+\gamma_{0-})-q^2(\gamma_{+0}+\gamma_{0+})]^2+4p^2q^2\gamma_{+0}\gamma_{-0}}}.\\
        \end{aligned}
    \end{equation}
\end{widetext}
In what follows, we analyze these timescales numerically and analytically and show that, while at weak coupling they are of the same order (assuming the excited states are not close to degenerate to begin with),
at strong coupling they branch out substantially and scale differently with coupling strength.

First, in Fig.~\ref{fig:figure 3}, we plot $\tau_1$ and $\tau_2$ as a function of the coupling strength $\lambda$ (top panels) and excited state splitting $\Delta$ (bottom panels) at different temperatures. For the $\lambda$ dependence, we test different $\Delta$ values (0.01, 0.02, 0.03, 0.1, and 0.5 in fading order indicated by a black arrow) to showcase how this parameter alters the qualitative form of the curves. 

Let us first discuss some key features on the $\lambda$ dependence. We observe that the long timescale $\tau_2$ depends on $\lambda$ nonmonotonically. 
There exists a minimum for $\tau_2$ as a function of $\lambda$, which indicates an optimal coupling strength to reach the steady-state in the shortest time.
Furthermore, higher values of $\Delta$ push the curve to an overall shorter timescale. In contrast, the shorter time scale $\tau_1$ monotonically decreases as $\lambda$ increases, and it almost does not depend on the value of $\Delta$. At small $\lambda$, the two time scales $\tau_{1,2}$ are almost indistinguishable and are both very long, as it takes infinitely long time to fully thermalize with the bath with vanishingly small coupling to it. 

The $\Delta$ dependence of two time scales becomes starkly different for large $\lambda$ as  depicted in Fig.~\ref{fig:figure 3}. The short timescale $\tau_1$ does not depend on $\Delta$, while the long timescale $\tau_2$ develops an inverse dependence on $\Delta$, getting shorter with increasing $\Delta$.
We will substantiate all these observations next by simplifying Eq. (\ref{eq: two timescales}).

We examine the high- and low-temperature limits of the two timescales, giving us a better understanding of the effect of strong coupling.

(i) Taking a low temperature limit of the two timescales, Eq.~\eqref{eq: two timescales}, leads to 
\begin{equation}
\label{eq: low temperature limit}
\begin{aligned}
    \lim_{T\rightarrow0}\tau_1 &\simeq  \frac{1}{2\pi q^2 J^\text{eff}(E_{+0})}=\frac{1}{2\pi q^2(\frac{2\lambda}{\Omega})^2\Gamma E_{+0}}, \\
    \lim_{T\rightarrow0}\tau_2 &\simeq\frac{1}{2\pi p^2 J^\text{eff}(E_{-0})} =\frac{1}{2\pi p^2(\frac{2\lambda}{\Omega})^2\Gamma E_{-0}}.
    \end{aligned}
\end{equation}
That is, the separation of timescales at low temperature is attributed to the product of the SBC asymmetry ($p^2$ compared to $q^2$) with the Bohr frequencies $E_{\pm 0}$. In Fig.~\ref{fig:figure 4}(a), we show the $\lambda$ dependence of $p^2$ and $q^2$ at different $\Delta$ values (0.01, 0.02, 0.03, 0.1, and 0.5 in fading order). We observe that $q^2$ quickly reaches $1$ when increasing $\lambda$; $E_{+0}$ also increases as a function of $\lambda$. Together, these trends lead to a short timescale $\tau_1$. The opposite occurs for $\tau_2$ where $p^2$ quickly approaches zero when increasing $\lambda$. From Fig.~\ref{fig:figure 4}(b), we note 
that the $\lambda$ dependence of the two Bohr frequencies indicates that the splittings of the eigenenergy levels are not the dominant factor in determining the time scales.

(ii) Taking a high-temperature limit on the two timescales Eq.~\eqref{eq: two timescales} leads to 
\begin{equation}
\label{eq: high temperature limit}
    \begin{aligned}
        \lim_{T\rightarrow \infty}\tau_1 &\simeq \frac{1}{\frac{2\pi\Gamma}{\beta}(\frac{2\lambda}{\Omega})^2[1+\sqrt{(p^2-q^2)^2+p^2q^2}]},\\
         \lim_{T\rightarrow \infty}\tau_2 &\simeq \frac{1}{\frac{2\pi\Gamma}{\beta}(\frac{2\lambda}{\Omega})^2[1-\sqrt{(p^2-q^2)^2+p^2q^2}]}.
    \end{aligned}
\end{equation}
Let us further simplify the denominator based on Eq.~\eqref{eq: pandqss} to more clearly observe the $\lambda$ dependence. This leads to 
\begin{equation}
\label{eq: high T limit final}
\begin{aligned}
    \lim_{T\rightarrow \infty}\tau_1&\approx \frac{1}{\frac{2\pi \Gamma}{\beta}(\frac{2\lambda}{\Omega})^2\left(1+\left(q^2-\frac{p^2}{2}\right)\right)}\\
    \lim_{T\rightarrow \infty}\tau_2& \approx \frac{1}{\frac{2\pi \Gamma}{\beta}(\frac{2\lambda}{\Omega})^2\left(1-\left(q^2-\frac{p^2}{2}\right)\right)}
    \end{aligned}
\end{equation}
Since for large enough $\lambda$,
$q^2\gg p^2$ and $q^2\to 1$,
at high $T$, $\tau_1$ is dominated by $(\frac{2\lambda}{\Omega})^2$ in the denominator. Under the same high $T$ conditions, $\tau_2$  exhibits non-monotonic behavior as a function of $\lambda$ since $q^2\approx 1$ and therefore $p^2$ quickly dominates at large $\lambda$. 
Overall, we reach the following scaling of the timescales at high temperatures and at strong coupling,
\bea
\tau_1 \propto T\times\left(\frac{\Omega}{\lambda}\right)^2,  \,\,\,\,\ 
\tau_2 \propto T\times\left(\frac{\lambda}{\Delta}\right)^2 e^{\lambda^2/\Omega^2}.
\label{eq:t12}
\eea
These relations capture the fast rise of $\tau_2$ with growing interaction to the bath, and its decay with $\Delta$, as well as the suppression of $\tau_1$ with system-bath coupling, and its independence of $\Delta$.
The distinct trends for dissipative lifetimes, Eq.~\eqref{eq:t12},  is based on a microscopic derivation, and it constitutes our main result.

\begin{figure}[htbp]
\fontsize{6}{10}\selectfont 
\centering
\includegraphics[width=1.0\columnwidth]{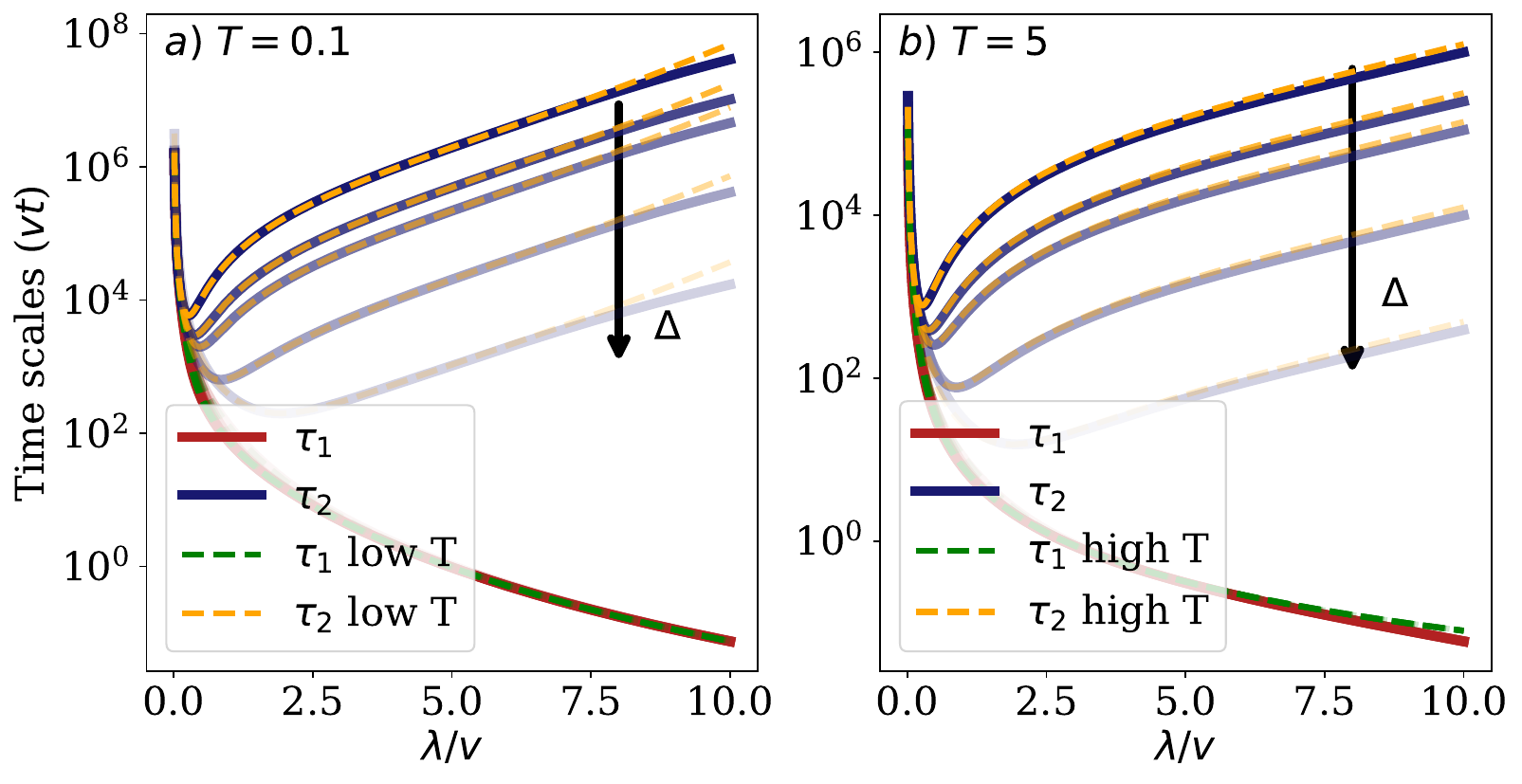}
\caption{(a) Comparison of two timescales described by the full expression Eq.~\eqref{eq: two timescales} to the approximated versions.
(a) Low temperature  comparison using Eq.~\eqref{eq: low temperature limit}. The arrows indicate the direction of increasing $\Delta$. At $T=0.1$, the low temperature limit accurately describes all the key features of the full expression. 
(b) Comparison at high $T$, $T=5$, given by Eq.~\eqref{eq: high T limit final}. For both limits [(a) and (b)], the ultrastrong coupling shows a slight deviation from the actual values. Parameters are identical to those of Fig.~\ref{fig:figure 3}.}
\label{fig:figure 5}
\end{figure}

In Fig.~\ref{fig:figure 5}, we showcase the accuracy of both the low and high temperature limits of the timescales as a function of $\lambda$, comparing the approximate expressions to the full formulae in Eq.~\eqref{eq: two timescales}. 
In Fig.~\ref{fig:figure 5}(a) we plot the expressions valid at low temperatures, Eq.~\eqref{eq: low temperature limit}, on top of the full description. 
In Fig.~\ref{fig:figure 5}(b), we plot the high-temperature limit, Eq.~\eqref{eq: high T limit final}, plotting it on top of the full expression. In both cases, simplified expressions accurately capture
the nonmonotonic dependence on $\lambda$ for $\tau_2$ as well as its suppression to shorter timescales when increasing $\Delta$. In addition, the negligible dependence on $\Delta$ for $\tau_1$ is also captured. In the ultrastrong coupling regime, the approximate expressions start to deviate slightly. 

We now go back to Eq. (\ref{eq: effective Hamiltonian in the diagonal basis}) and point out that there are two scenarios that lead to the observation of $p\to 0$ and $q\to 1$, thus leading to branching timescales. First, one could assume close-to-degeneracy excited state levels, $\Delta\ll \lambda^2/\Omega$.
In this case, $h/w\to \frac{\lambda^2}{\Omega\Delta}$, thus $\phi \to \pi/2$, and $p\to 0$, $q\to 1$.
At very weak SBC, this case leads to distinct lifetimes as shown in Refs. \cite{Brumer_2014,Gerry_2024}. 
Thus, even at very weak SBC, one can observe distinct timescales, short and long, so long as $\Delta$ is small enough below the characteristic decay rate.
In the other limit, which we focus on in this work, $\Delta $ is not necessarily small relative to other energy scales. However, once the system-bath interaction is made large enough, the environment creates the dark state with $p\to 0$ and $q\to 1$, and the two timescales depart from each other. 

Our work unifies seemingly unrelated observations of long-lived metastable dynamics: those occurring at weak coupling for nearly degenerate states (see, e.g., Refs.\cite{Brumer_2014, Gerry_2024}), and those emerging at strong coupling for more general system spectra (e.g., Ref. \cite{Zhou_2021}). We present a comprehensive microscopic framework that encompasses both scenarios in their respective limits.

\begin{figure*}[htbp]
\fontsize{6}{10}\selectfont 
\centering
\includegraphics[width=1.0\textwidth]{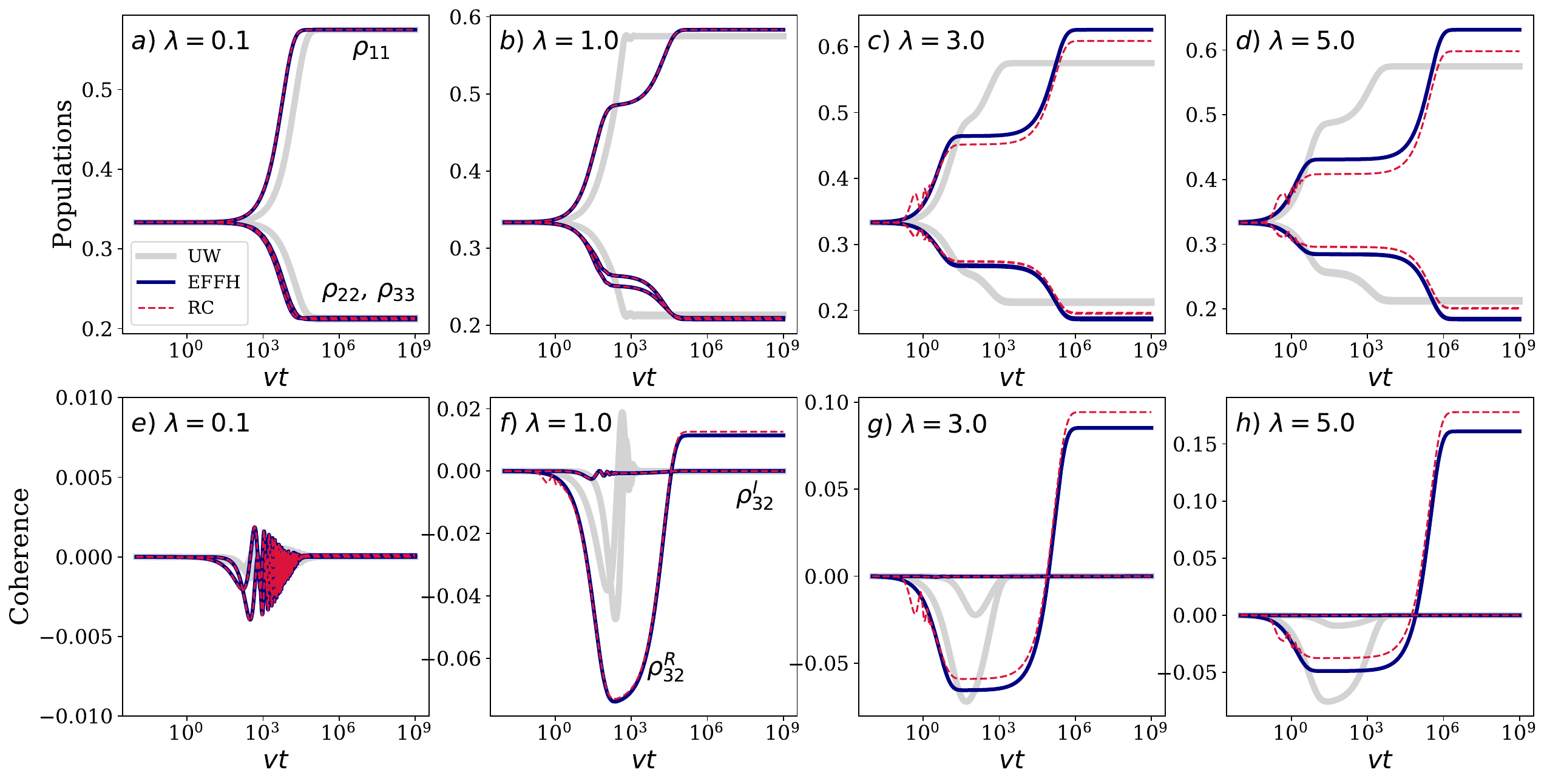}
\caption{Population (top row) and coherence (bottom row) dynamics of the dissipative three-level model at different coupling strengths $\lambda=0.1,1,3$ and $5$, left to right. Parameters are  $v=1$, $\Delta=0.01$, $T=1$, $\Omega=10$, $\Gamma=0.05$, and $\Lambda=1000$. For the RC simulation, 10 levels are included for convergence.}
\label{fig:figure 6}
\end{figure*}

\begin{table*}[hbt!]
\centering
\begin{tabular}{||c c c c||} 
 \hline
 Method & System Hamiltonian $\hat{H}_S$ &  System coupling operator $\hat{S}$& Spectral density function $J(\omega)$\\ [1.0ex] 
 \hline\hline
 Redfield QME (UW) & $\hat{H}_S$ & 
 Eq. (\ref{eq:S}) & $\frac{4\Gamma\Omega^2\lambda^2 \omega}{(\omega^2-\Omega^2)^2+(2\pi\Gamma\Omega\omega)^2}$ \\ 
 RC QME &  $\hat{H}_S+\lambda\hat{S}(\hat{a}^\dagger+\hat{a})+\Omega\hat{a}^\dagger\hat{a}$ & $\hat{a}^\dagger+\hat{a}$, \,\, see Eq. (\ref{eq:HRC}) & $\Gamma \omega e^{-\omega/\Lambda}$ \\
 EFFH QME & $\hat{H}_{S,E}^\text{eff}(\lambda,\Omega)$ & 
 see matrix in Eq. (\ref{eq: effective Hamiltonian in the diagonal basis}) & $\frac{4\lambda^2}{\Omega^2}\Gamma\omega e^{-\omega/\Lambda}$\\ [1ex] 
 \hline
\end{tabular}
\caption{System Hamiltonian, coupling operator, and the spectral density function used for three different numerical methods. These choices are fed into the numerical Redfield solver to obtain dissipative dynamics for the reduced density matrix. Here, $\Lambda$ is a large energy cut-off introduced. }
\label{table:table1}
\end{table*}

\section{Dynamics: Numerical Simulation}
\label{sec: Numerical simulation}

In this section, we pursue two  objectives. First, having established the presence of two distinct dynamical timescales, we examine their manifestations in both population and coherence dynamics. In particular, we highlight the emergence of long-lived transient coherences. We also analyze the generation of steady-state coherences. Notably, both the transient and steady-state coherences arise as a direct consequence of strong system-bath interactions.
Second, we aim to validate our analytical approach through numerical simulations. The timescales were derived from the effective Hamiltonian using the Lindblad QME. This procedure involves several approximations, both in the construction of the effective Hamiltonian and in the application of the secular approximation in its energy eigenbasis. To assess the accuracy of our predictions for the relaxation dynamics, we numerically simulate the dissipative evolution of the three-level system using the RC-QME, described in the following subsection. We then compare the behavior of the reduced density matrix elements with the analytical expectations given in Eq.~\eqref{eq: two timescales}, verifying whether the predicted decay or growth patterns are reproduced.

To this end, we write down the second order perturbative QME. Under the Born-Markov and weak-SBC approximations, one obtains the Redfield QME, which describes the dynamics of reduced density matrix (RDM) of the system, $\hat{\rho}(t)$. The matrix elements of the RDM, $\rho_{mn}=\bra{m}\hat{\rho}\ket{n}$, time evolve according to
\begin{equation}
\label{eq: Redfield QME numerical}
\begin{aligned}
    \frac{d\rho_{mn}}{dt} =& -i\omega_{mn}\rho_{mn}(t)\\
    -&\sum_{jl}\{ R_{mj,jl}(\omega_{lj})\rho_{ln}(t)+[R_{nl,lj}(\omega_{jl})]^*\rho_{mj}(t)\\
    -&[R_{ln,mj}(\omega_{jm})+[R_{jm,nl}(\omega_{ln})]^*]\rho_{jl}(t)\}.
    \end{aligned}
\end{equation}
The first line in Eq.~\eqref{eq: Redfield QME numerical} corresponds to the unitary dynamics of the system. The second and third lines correspond to the dissipative dynamics. Here, $\omega_m$ are the eigenvalues of the system's Hamiltonian, $\hat{H}_S$, and $\omega_{mn}=\omega_m-\omega_n$ are the Bohr frequencies. The Redfield tensor depends on the matrix elements of the system coupling operator $S_{mn}=\bra{m}\hat{S}\ket{n}$ and half Fourier transforms of the bath-bath correlation function. It is defined as 
\begin{equation}
    R_{mn,jl}(\omega) = S_{mn}S_{jl}\int^\infty_0d\tau e^{i\omega \tau}\langle\hat{B}(\tau)\hat{B}(0)\rangle.
\end{equation}
For the original model Eq.~\eqref{eq: total Hamiltonian original},
$\hat{B}(\tau) = \sum_kt_k(\hat{c}^\dagger_ke^{i\nu_k\tau}+\hat{c}_ke^{-i\nu_k\tau})$ and $\langle\hat{\mathcal{O}}\rangle=\Tr_B(\hat{\mathcal{O}}\hat{\rho}_B)$ with $\hat{\rho}_B$ being the Gibbs state with respect to the stationary bath Hamiltonian $\hat{H}_B$. Taking the continuum limit for the momentum degrees of freedom of the baths $k$ and splitting the Redfield tensor into its real and imaginary parts allows us to rewrite it as
\begin{equation}
    R_{mn,jl}(\omega)= S_{mn}S_{jl}[\Gamma(\omega)+i\delta(\omega)].
\end{equation}
Here, $\Gamma(\omega)$ is the symmetric part of the bath-bath correlation function and $\delta(\omega)$ is the Lamb shift. For the choice of our system and bath parameters, we are able to ignore the contribution of the Lamb shift \cite{Ivander_2022,Ivander_2023}. 

\begin{figure}[htbp]
\fontsize{6}{10}\selectfont 
\centering
\includegraphics[width=1.0\columnwidth]{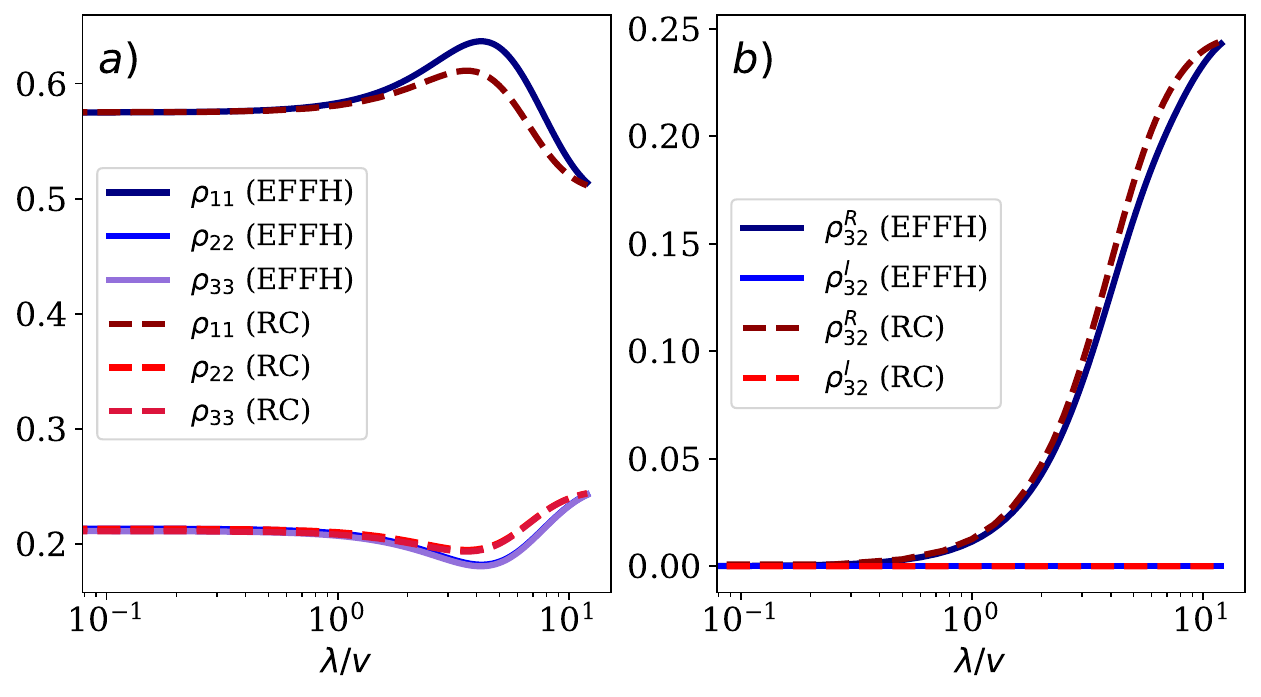}
\caption{Steady-state populations (a) and coherence (b) as a function of $\lambda$ obtained from the RC Hamiltonian (RC) and the effective Hamiltonian  (EFFH) simulations. At intermediate coupling strengths, we observe deviations between the methods in the steady-state RDM values. However, deviations disappear in the strong coupling limit. Parameters are  $v=1$, $\Delta=0.01$, $\Omega=10$. $\Gamma=0.05$, and $T=1$.}
\label{fig:figure 7}
\end{figure}

\begin{figure*}[htbp]
\fontsize{6}{10}\selectfont 
\centering
\includegraphics[width=1.0\textwidth]{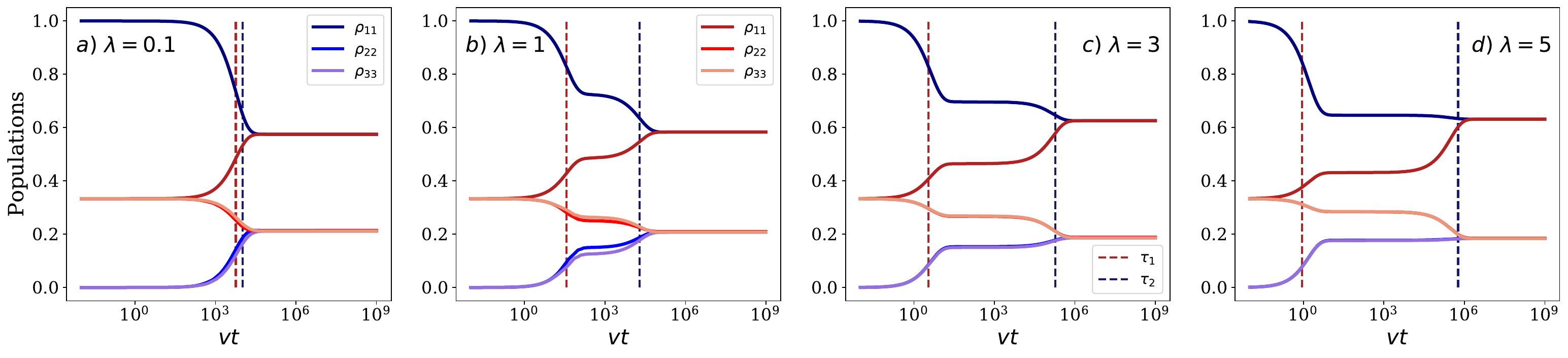}
\caption{Population dynamics obtained from the effective Hamiltonian QME for different values of $\lambda$ and with two different initial conditions. $\hat{\rho}(0)=\text{diag}(\frac{1}{3},\frac{1}{3},\frac{1}{3})$ (red) and $\hat{\rho}(0)=\text{diag}(1,0,0)$ (blue). Parameters are identical to that of Fig.~\ref{fig:figure 6}. Legends in panels (a)-(b) correspond to all four panels.}
\label{fig:figure 8}
\end{figure*}



We numerically simulate the dynamics of the RDM elements using the Redfield QME at three different stages of the mapping procedure:
(i) First, we directly simulate the dynamics based on the original Hamiltonian, assuming a Brownian bath spectral density for the bath, Eq.~\eqref{eq:Joriginal}. This approach accurately captures the dynamics only when the SBC is weak.
(ii) Second, we simulate the dynamics after applying the reaction coordinate transformation. In this case, the system Hamiltonian is extended to include the original system, the RC mode, and their interaction, see Eq.~\eqref{eq:HRC}. The SBC operator is taken to be the bosonic displacement operator of the RC mode, which now couples to a residual bath described by an Ohmic spectral density.
(iii) We simulate the dynamics using the effective system Hamiltonian coupled via the original system-bath coupling operator $\hat{S}$ to an effective Ohmic bath, see Eq.~\eqref{eq:Jeff}.
Table~\ref{table:table1} summarizes the system Hamiltonians, system-bath coupling operators, and bath spectral density functions used in each case.

In Fig.~\ref{fig:figure 6}, we plot the dynamics of the populations [(a)-(d)] and the coherence [(e)-(h)] for an initial state $\hat{\rho}(0)=\text{diag}(\frac{1}{3},\frac{1}{3},\frac{1}{3})$ using different coupling strengths $\lambda=\{0.1,1,3,5\}$. The parameters for the system and bath are chosen to be $v=1$, $\Delta=0.01$, $\Omega=10$, $\Gamma=0.05$, and $T=1$. Dynamics with the different levels of approximations (i)-(iii) are color-coded with gray, dotted red, and blue, which corresponds to direct simulation of the original Hamiltonian valid for ultraweak (UW) SBC strength, post-RC mapping, and effective Hamiltonian simulations, respectively. 

Let us first discuss some key features of the dissipative dynamics. At relatively weak SBC ($\lambda=0.1$) corresponding to Fig.~\ref{fig:figure 6}(a), the relaxation dynamics is characterized by a single timescale. This is expected, as seen in Fig.~\ref{fig:figure 3}(b), where the two timescales are almost indistinguishable at weak coupling for our set of parameters (the key is that $\Delta$ is not made too small to lead to long timescales due to the quasidegenerate nature of the excited states \cite{Gerry_2024}). As the coupling increases, we observe the emergence of a metastable state bounded by two distinct timescales, with the duration of this state increasing as the coupling strengthens. 
Furthermore, the strong-coupling-induced metastable state is accompanied by long-lived coherence, and even steady-state coherence at strong coupling. In Fig.~\ref{fig:figure 6} [(e)-(h)], we plot the real and imaginary parts of the coherences between the excited state, $\rho_{32}$. The generation of long-lived transient coherences in the system {\it due to the strong coupling with the environment} is a nontrivial prediction of our work.

The RC simulation can be considered numerically accurate for our parameters, since we use small $\Gamma$ \cite{Ahsan16}; we kept 10 levels of extracted RC mode for convergence. We note that although some deviations emerge between the RC dynamics and the effective Hamiltonian dynamics, the timescales over which the populations and coherences change still agree. From previous studies \cite{Nick_PRXQ,Nick_PRB}, we know that the RCPT method leads to deviations from exact results at intermediate coupling, but that these deviations vanish in the ultrastrong coupling limit. 
To support this argument, we plot the steady-state values of the system's density matrix as a function of the coupling strength in Fig.~\ref{fig:figure 7}. Once the ultrastrong coupling limit is reached ($\lambda=10$ in our parameters), the steady-state values predicted by the RC Hamiltonian and the effective Hamiltonian match. 

Back to Fig.~\ref{fig:figure 6},
as expected, ignoring strong-coupling effects and using the Redfield equation for the original Hamiltonian leads to incorrect dynamics in the intermediate and strong coupling regimes. Although two timescales do emerge at stronger coupling, their precise values are off from the correct values by orders of magnitude. These are most easily in the third and fourth columns of Fig.~\ref{fig:figure 6}, corresponding to population and coherence dynamics at strong SBC.

We end this section by studying the dynamics with different initial conditions and plotting the two timescales described in Eq.~\eqref{eq: two timescales}. In Fig.~\ref{fig:figure 8}, we present the population dynamics using the effective Hamiltonian QME for $\lambda\in\{0.1,1,3,5\}$ using two different initial conditions: (i) $\hat{\rho}(0)=\text{diag}(\frac{1}{3},\frac{1}{3},\frac{1}{3})$, corresponding to blue curves and (ii) $\hat{\rho}(0)=\text{diag}(1,0,0)$, corresponding to red curves. Regardless of the initial conditions, we observe that the dynamics relax to the same steady-state as it should. The predicted timescales from Eq.~\eqref{eq: two timescales} are indicated by vertical lines, which accurately correspond to the actual dynamics, and where the metastable state emerges and ends.

\section{Conclusion} 
\label{sec: conclusion}

We discussed the rich dynamics that emerge under strong system-bath coupling. At weak coupling, the relaxation dynamics of the system is slow. A natural expectation is that stronger interactions with the environment should accelerate relaxation dynamics.
Rather than adopting phenomenological methods, 
we use here an approach grounded in a microscopic description of the system-bath interaction. 
Focusing on a three-level system as a case study, we analytically demonstrated that strong coupling induced two distinct dynamical timescales. The first corresponded to relaxation that indeed accelerated upon increasing coupling strength to the bath, in line with intuition. The second timescale, however, exhibited a nonintuitive behavior: it slowed down exponentially with the coupling energy, giving rise to a bath-induced metastable state that typically involved quantum coherences.
To the best of our knowledge, this is the first rigorous demonstration of such timescale branching and the associated slowdown of relaxation dynamics in the strong-coupling regime.

In future studies, we are interested to explore this effect in other impurity models (Lambda model, for example) and many-body models (Bose-Hubbard model) as analyzed in Refs. using the Lindblad equation \cite{Luitz_2020,Luitz_2022,Luitz_2024}. 
Considering, for example, a spin chain globally coupled to an environment, we showed in Ref. \cite{Min_PRL} that the environment created an effective coupling between the spins [the analogue of $h$ in Eq.~\eqref{eq: effective system Hamiltonian simplified}] and that it suppressed the spin splitting [the analogue of $w$ in Eq.~\eqref{eq: effective system Hamiltonian simplified}]. 
As such, we expect similar physics of diverging timescales to develop in the many-body spin chain case. In future work, a rigorous study of this behavior will be done.

Our work predicts that strong system-bath coupling generates
long-lived metastable states. The effect could be tested in atomic systems, NV centers, and superconducting junctions. 
We anticipate applications of our study in the field of many-body thermalization and bath engineering, as we identify both conditions for fast thermalization with an optimally chosen coupling strength and the genetic creation of metastable states with coherences, due to coupling to thermal environments.  We also expect generalizations of our analysis to account for the creation of correlations between systems at strong system-bath coupling.
We anticipate that in analogy to the present results, bath-induced transient quantum correlations could persist for long times as one increases the coupling to the bath, decaying eventually to their steady-state value after a slow dynamics.

Open quantum systems may exhibit rich and nonintuitive dynamics, especially beyond the regime of asymptotically weak system-bath coupling. Our work, based on the Reaction Coordinate Polaron Transform semi-analytical approach, provides transparent analytical insights into the strong-coupling regime. Our study highlights the critical importance of developing analytical tools along with the ongoing advancement of numerical techniques to accurately capture strong system-bath interactions. The development of mapping approaches for strong coupling, e.g., \cite{Ashida_2021,Nick_PRXQ,Chain15,MD22} 
remains a significant and rewarding challenge, with broad applications across condensed matter physics, quantum optics, and quantum information science.





\section*{Acknowledgments}
The work of B.M. was supported by the Ontario Graduate Scholarship (OGS). The work of M.G. was supported by the NSERC Canada Graduate Scholarship-Doctoral. D.S. acknowledges the NSERC Discovery Grant.
We acknowledge Paul Brumer for discussions that suggested this research direction and Jakub Garwoła for useful discussions.

\appendix
\begin{widetext}
\section{Details on the derivation for the effective system Hamiltonian}
\label{sec: Derivation for the effective system Hamiltonian}

Here, we go through the details of deriving the effective model Hamiltonian 
\begin{equation}
\label{eq: effective H definition}
    \hat{H}^\text{eff}_S(\lambda,\Omega) = \bra{0}\hat{U}_p\hat{H}_S\hat{U}^\dagger_p\ket{0}-\frac{\lambda^2}{\Omega}\hat{S}^2.
\end{equation}
Let us first go through the polaron unitary transform of the system Hamiltonian. Since $\hat{S}$ is a traceless $3\times 3$ Hermitian matrix, the most straightforward approach in carrying out the polaron unitary transform is to utilize the following identity for any $SU(3)$ group element generated by a traceless $3\times 3$ Hermitian matrix $\hat{S}$ \cite{Curtright_2015}:
\begin{equation}
\label{eqn: SU3 unitary}
    \exp(i\theta\hat{S}) = \sum_{k=0,1,2}\left[\hat{S}^2+\frac{2}{\sqrt{3}}\hat{S}\sin(\Phi+\frac{2\pi k}{3})-\frac{1}{3}\hat{I}\left(1+2\cos(2(\Phi+\frac{2\pi k}{3}))\right)\right]\frac{\exp(\frac{2}{\sqrt{3}}i\theta\sin(\Phi+\frac{2\pi k}{3}))}{1-2\cos(2(\Phi+\frac{2\pi k}{3}))}
\end{equation}
where
\begin{equation}
    \Phi = \frac{1}{3}\left(\arccos(\frac{3}{2}\sqrt{3}\det(\hat{S}))-\frac{\pi}{2}\right).
\end{equation}
With $\hat{S}=\frac{1}{\sqrt{2}}\left(\ket{1}\bra{2}+\ket{1}\bra{3}+\text{H.c.}\right)$, we have $\det(\hat{S})=0$ and therefore $\Phi=0$. The polaron unitary can then be written as 
\begin{equation}
\begin{cases}    \hat{U}_p=&\exp(i\theta\hat{S})=\exp(i\left[-i\left(\frac{\lambda}{\Omega}\hat{A}\right)\right]\hat{S})= \hat{I}-\hat{S}^2+\cos\theta\hat{S}^2+i\sin\theta\hat{S}\\
\hat{U}^\dagger_p=&\exp(-i\theta\hat{S})=\exp(i\left[+i\left(\frac{\lambda}{\Omega}\hat{A}\right)\right]\hat{S})= \hat{I}-\hat{S}^2+\cos\theta\hat{S}^2-i\sin\theta\hat{S}\\
\end{cases}
\end{equation}
Here, we denoted $\hat{A}=\hat{a}^\dagger-\hat{a}$ and defined $\theta=-i(\frac{\lambda}{\Omega}\hat{A})$. Transforming $\hat{H}_S$ under the polaron unitary then gives us
\begin{equation}
    \hat{U}_p\hat{H}_S\hat{U}^\dagger_p  = \left(\begin{matrix}
     \frac{2v-\Delta}{2}\sin^2\theta   &-\frac{i(\Delta+(-2v+\Delta)\cos\theta)\sin\theta}{2\sqrt{2}} & \frac{i(\Delta+(2v-\Delta)\cos\theta)\sin\theta}{2\sqrt{2}} \cr 
     \frac{i(\Delta+(-2v+\Delta)\cos\theta)\sin\theta}{2\sqrt{2}}   & \frac{-4\Delta\cos\theta+(2v-\Delta)(3+\cos(2\theta))}{8}& \frac{(-2v+\Delta)}{4}\sin^2\theta\cr
     -\frac{i(\Delta+(2v-\Delta)\cos\theta)\sin\theta}{2\sqrt{2}}    & \frac{(-2v+\Delta)}{4}\sin^2\theta& \frac{4\Delta\cos\theta+(2v-\Delta)(3+\cos(2\theta))}{8}\cr
    \end{matrix} \right).
\end{equation}
Projecting on to the ground state of the RC mode will get rid of matrix elements that are linear in $\sin\theta$ as it is odd power in $\hat{A}$. We are then left with
\begin{equation}
    \bra{0}\hat{U}_p\hat{H}_S\hat{U}^\dagger_p\ket{0} = \left(\begin{matrix}
     \frac{2v-\Delta}{2}\bra{0}\sin^2\theta\ket{0}   & 0  & 0\cr 
     0   & \frac{-4\Delta\bra{0}\cos\theta\ket{0}+(2v-\Delta)(3+\bra{0}\cos(2\theta)\ket{0})}{8}& \frac{(-2v+\Delta)}{4}\bra{0}\sin^2\theta\ket{0}\cr
   0   & \frac{(-2v+\Delta)}{4}\bra{0}\sin^2\theta\ket{0}& \frac{4\Delta\bra{0}\cos\theta\ket{0}+(2v-\Delta)(3+\bra{0}\cos(2\theta)\ket{0})}{8}\cr
    \end{matrix} \right).
\end{equation}
Note that 
\begin{equation}
    \bra{0}\cos(\theta)\ket{0} = \frac{1}{2}\bra{0}e^{\frac{\lambda}{\Omega}(\hat{a}^\dagger-\hat{a})}+e^{-\frac{\lambda}{\Omega}}(\hat{a}^\dagger-\hat{a})\ket{0}.
\end{equation}
Defining a displacement operator, $D(\alpha)=e^{\alpha(\hat{a}^\dagger-\hat{a})}$, where $\alpha=\frac{\lambda}{\Omega}\hat{I}$, we get 
\begin{equation}
\begin{aligned}
    \bra{0}\cos(-i\frac{\lambda}{\Omega}(\hat{a}^\dagger-\hat{a}))\ket{0} =& \frac{1}{2}\bra{0}D(\alpha)+D(-\alpha)\ket{0}\\
    =&\frac{1}{2}\bra{0}\left(e^{-\frac{|\lambda/\Omega|^2}{2}}\sum^\infty_{n=0}\frac{(\frac{\lambda}{\Omega})^n}{\sqrt{n!}}\ket{n}+e^{-\frac{|\lambda/\Omega|^2}{2}}\sum^\infty_{n=0}\frac{(-\frac{\lambda}{\Omega})^n}{\sqrt{n!}}\ket{n}\right)\\
    =&\frac{1}{2}\bra{0}e^{-\frac{\lambda^2}{2\Omega^2}}+e^{-\frac{\lambda^2}{2\Omega^2}}\ket{0} = e^{-\frac{\lambda^2}{2\Omega^2}}
    \end{aligned}
\end{equation}
This implies $\bra{0}\cos(2\theta)\ket{0}=e^{-\frac{2\lambda^2}{\Omega^2}}$ and $\bra{0}\sin^2\theta\ket{0}=\bra{0}\frac{1-\cos(2\theta)}{2}\ket{0}=\frac{1-e^{-\frac{2\lambda^2}{\Omega^2}}}{2}$. Adding the $-\frac{\lambda^2}{\Omega}\hat{S}^2$ contribution from Eq.~\eqref{eq: effective H definition}, we finally obtain
\begin{equation}
\label{eq: effective system Hamiltonian app}
\hat{H}^\text{eff}_S(\lambda,\Omega) = \left(\begin{matrix}
     \frac{2v-\Delta}{4}(1-e^{-\frac{2\lambda^2}{\Omega^2}})-\frac{\lambda^2}{\Omega}& 0  & 0\cr 
     0   & \frac{-4\Delta e^{-\frac{\lambda^2}{2\Omega^2}}+(2v-\Delta)(3+e^{-\frac{2\lambda^2}{\Omega^2}})}{8}-\frac{\lambda^2}{2\Omega}& \frac{(-2v+\Delta)(1-e^{-\frac{2\lambda^2}{\Omega^2}})}{8}-\frac{\lambda^2}{2\Omega}\cr
   0   & \frac{(-2v+\Delta)(1-e^{-\frac{2\lambda^2}{\Omega^2}})}{8}-\frac{\lambda^2}{2\Omega}&  \frac{4\Delta e^{-\frac{\lambda^2}{2\Omega^2}}+(2v-\Delta)(3+e^{-\frac{2\lambda^2}{\Omega^2}})}{8}-\frac{\lambda^2}{2\Omega}\cr
    \end{matrix} \right),
\end{equation}
which is Eq. (\ref{eq: effective system Hamiltonian}) in the main text.
One can easily confirm that in the $\lambda\rightarrow 0$ limit, we recover the correct $\hat{H}_S=\text{diag}(0,v-\Delta,v)$.

\section{Effective system Hamiltonian in the energy eigenbasis}
\label{AppB}

In this appendix, we diagonalize the effective system Hamiltonian as described in Eq.~\eqref{eq: effective system Hamiltonian app} to identify the Bohr frequencies and corresponding eigenstates that are needed to build the Lindblad jump operators for the QME. Note that $\hat{H}^\text{eff}_S$ is in a block diagonal form of 
\begin{equation}
    \hat{H}^\text{eff}_S = \left(\begin{matrix}
        E_0 & 0 & 0 \cr
        0 & l+w & h \cr
        0 & h & l-w \cr
    \end{matrix} \right)
\end{equation}
with $E_0 = \frac{(2v-\Delta)(1-e^{-\frac{2\lambda^2}{\Omega^2}})}{4}-\frac{\lambda^2}{\Omega}$, and
\begin{equation}
\begin{aligned}
    l= \frac{(2v-\Delta)(3+e^{-\frac{2\lambda^2}{\Omega^2}})}{8}-\frac{\lambda^2}{2\Omega},\quad
    h=\frac{(-2v+\Delta)(1-e^{-\frac{2\lambda^2}{\Omega^2}})}{8}-\frac{\lambda^2}{2\Omega},\quad \text{and} \quad 
    w=-\frac{\Delta e^{-\frac{\lambda^2}{2\Omega^2}}}{2}.
    \end{aligned}
\end{equation}
Let us look at the eigenenergies and states of this Hamiltonian. Since it is in a block diagonal form, these are $\{E_0,E_+,E_-\}$ with $E_\pm=l\pm\sqrt{w^2+h^2}$ and the corresponding eigenvectors $\{\ket{E_0},\ket{E_+},\ket{E_-}\}$ which are given as
\begin{equation}
    \ket{E_0} = \left(\begin{matrix}
        1,
        0,
        0
    \end{matrix} \right)^T,\quad \ket{E_+} =\left(\begin{matrix}
        0,
        \cos(\frac{\phi}{2}),
        \sin(\frac{\phi}{2})
    \end{matrix} \right)^T,\quad \ket{E_-} = \left(\begin{matrix}
        0,
        -\sin(\frac{\phi}{2}),
        \cos(\frac{\phi}{2})
    \end{matrix} \right)^T
\end{equation}
with $\tan(\phi)=h/w$. Again, we remind the reader that in the small coupling limit, $\lim_{\lambda\rightarrow 0}h= 0 $ hence, $\phi=0$. The effective system Hamiltonian in the energy eigenbasis $\hat{H}^\text{eff}_{S,E}=\text{diag}(E_0,E_-,E_+)$ is related to the original site basis $\hat{H}^\text{eff}_S$ as
\begin{equation}
    \hat{H}^\text{eff}_S = \hat{P}\hat{H}^\text{eff}_{S,E}\hat{P}^{-1}
\end{equation}
with $\hat{P}=(\ket{E_0},\ket{E_-},\ket{E_+})=\hat{P}^{-1}$. Then, we are able to write $\hat{S}$ in the energy eigenbasis as
\begin{equation}
\begin{aligned}
    \hat{S}_E = \hat{P}^{-1}\hat{S}\hat{P} = &
    \left(\begin{matrix}
       1 & 0 & 0 \cr
       0 & -\sin(\frac{\phi}{2})&\cos(\frac{\phi}{2}) \cr
       0 &\cos(\frac{\phi}{2}) &\sin(\frac{\phi}{2}) \cr
    \end{matrix} \right)\left(\begin{matrix}
        0 & \frac{1}{\sqrt{2}} & \frac{1}{\sqrt{2}} \cr
        \frac{1}{\sqrt{2}} & 0 & 0 \cr
        \frac{1}{\sqrt{2}} & 0 & 0\cr
    \end{matrix} \right)\left(\begin{matrix}
        1 & 0 & 0 \cr
        0 & -\sin(\frac{\phi}{2}) & \cos(\frac{\phi}{2}) \cr
        0 &\cos(\frac{\phi}{2}) &\sin(\frac{\phi}{2})
    \end{matrix} \right)\\
    =&\left(\begin{matrix}
        0 & \sin(-\frac{\phi}{2}+\frac{\pi}{4}) & \sin(\frac{\phi}{2}+\frac{\pi}{4})\\
        \sin(-\frac{\phi}{2}+\frac{\pi}{4}) & 0 & 0 \cr
        \sin(\frac{\phi}{2}+\frac{\pi}{4}) & 0 & 0 \cr
    \end{matrix} \right).
    \end{aligned}
\end{equation}
Let us denote by $p=\sin(-\frac{\phi}{2}+\frac{\pi}{4})$ and $q= \sin(\frac{\phi}{2}+\frac{\pi}{4})$, where $p^2+q^2=1$. That is, strong coupling induces anisotropy between $\ket{E_0}\leftrightarrow\ket{E_-}$ and $\ket{E_0}\leftrightarrow\ket{E_+}$. We therefore get Eq.~\eqref{eq: effective Hamiltonian in the diagonal basis}, 
\begin{equation}
     \hat{H}^\text{eff}_E = \left(\begin{matrix}
         E_0 & 0 & 0 \cr
         0 & E_- & 0 \cr
         0 & 0 & E_+ \cr
     \end{matrix} \right)-\left(\begin{matrix}
    0 & p & q \cr
    p & 0 & 0 \cr
    q & 0 & 0 \cr
    \end{matrix}\right)\sum_k\frac{2\lambda f_k}{\Omega}\left(\hat{b}^\dagger_k+\hat{b}_k\right)+\sum_k\omega_k\hat{b}^\dagger_k\hat{b}_k
\end{equation}
which will be our starting point in constructing master equations. Note that the structure of the Hamiltonian is identical to the original system Hamiltonian, Eq. (\ref{eq: total Hamiltonian original}), except for the asymmetry induced in the system operator that is coupled to the bath, $\hat{S}_E$.

\end{widetext}

\bibliography{bibliography}

\begin{thebibliography}{100}%
\makeatletter
\providecommand \@ifxundefined [1]{%
 \@ifx{#1\undefined}
}%
\providecommand \@ifnum [1]{%
 \ifnum #1\expandafter \@firstoftwo
 \else \expandafter \@secondoftwo
 \fi
}%
\providecommand \@ifx [1]{%
 \ifx #1\expandafter \@firstoftwo
 \else \expandafter \@secondoftwo
 \fi
}%
\providecommand \natexlab [1]{#1}%
\providecommand \enquote  [1]{``#1''}%
\providecommand \bibnamefont  [1]{#1}%
\providecommand \bibfnamefont [1]{#1}%
\providecommand \citenamefont [1]{#1}%
\providecommand \href@noop [0]{\@secondoftwo}%
\providecommand \href [0]{\begingroup \@sanitize@url \@href}%
\providecommand \@href[1]{\@@startlink{#1}\@@href}%
\providecommand \@@href[1]{\endgroup#1\@@endlink}%
\providecommand \@sanitize@url [0]{\catcode `\\12\catcode `\$12\catcode `\&12\catcode `\#12\catcode `\^12\catcode `\_12\catcode `\%12\relax}%
\providecommand \@@startlink[1]{}%
\providecommand \@@endlink[0]{}%
\providecommand \url  [0]{\begingroup\@sanitize@url \@url }%
\providecommand \@url [1]{\endgroup\@href {#1}{\urlprefix }}%
\providecommand \urlprefix  [0]{URL }%
\providecommand \Eprint [0]{\href }%
\providecommand \doibase [0]{http://dx.doi.org/}%
\providecommand \selectlanguage [0]{\@gobble}%
\providecommand \bibinfo  [0]{\@secondoftwo}%
\providecommand \bibfield  [0]{\@secondoftwo}%
\providecommand \translation [1]{[#1]}%
\providecommand \BibitemOpen [0]{}%
\providecommand \bibitemStop [0]{}%
\providecommand \bibitemNoStop [0]{.\EOS\space}%
\providecommand \EOS [0]{\spacefactor3000\relax}%
\providecommand \BibitemShut  [1]{\csname bibitem#1\endcsname}%
\let\auto@bib@innerbib\@empty
\bibitem [{\citenamefont {Barreiro}\ \emph {et~al.}(2011)\citenamefont {Barreiro}, \citenamefont {M{\"u}ller}, \citenamefont {Schindler}, \citenamefont {Nigg}, \citenamefont {Monz}, \citenamefont {Chwalla}, \citenamefont {Hennrich}, \citenamefont {Roos}, \citenamefont {Zoller},\ and\ \citenamefont {Blatt}}]{Julio_2011}%
  \BibitemOpen
  \bibfield  {author} {\bibinfo {author} {\bibfnamefont {Julio~T.}\ \bibnamefont {Barreiro}}, \bibinfo {author} {\bibfnamefont {Markus}\ \bibnamefont {M{\"u}ller}}, \bibinfo {author} {\bibfnamefont {Philipp}\ \bibnamefont {Schindler}}, \bibinfo {author} {\bibfnamefont {Daniel}\ \bibnamefont {Nigg}}, \bibinfo {author} {\bibfnamefont {Thomas}\ \bibnamefont {Monz}}, \bibinfo {author} {\bibfnamefont {Michael}\ \bibnamefont {Chwalla}}, \bibinfo {author} {\bibfnamefont {Markus}\ \bibnamefont {Hennrich}}, \bibinfo {author} {\bibfnamefont {Christian~F.}\ \bibnamefont {Roos}}, \bibinfo {author} {\bibfnamefont {Peter}\ \bibnamefont {Zoller}}, \ and\ \bibinfo {author} {\bibfnamefont {Rainer}\ \bibnamefont {Blatt}},\ }\bibfield  {title} {\enquote {\bibinfo {title} {An open-system quantum simulator with trapped ions},}\ }\href {\doibase 10.1038/nature09801} {\bibfield  {journal} {\bibinfo  {journal} {Nature}\ }\textbf {\bibinfo {volume} {470}},\ \bibinfo {pages} {486--491} (\bibinfo {year} {2011})}\BibitemShut {NoStop}%
\bibitem [{\citenamefont {Barontini}\ \emph {et~al.}(2013)\citenamefont {Barontini}, \citenamefont {Labouvie}, \citenamefont {Stubenrauch}, \citenamefont {Vogler}, \citenamefont {Guarrera},\ and\ \citenamefont {Ott}}]{Barontini_2013}%
  \BibitemOpen
  \bibfield  {author} {\bibinfo {author} {\bibfnamefont {G.}~\bibnamefont {Barontini}}, \bibinfo {author} {\bibfnamefont {R.}~\bibnamefont {Labouvie}}, \bibinfo {author} {\bibfnamefont {F.}~\bibnamefont {Stubenrauch}}, \bibinfo {author} {\bibfnamefont {A.}~\bibnamefont {Vogler}}, \bibinfo {author} {\bibfnamefont {V.}~\bibnamefont {Guarrera}}, \ and\ \bibinfo {author} {\bibfnamefont {H.}~\bibnamefont {Ott}},\ }\bibfield  {title} {\enquote {\bibinfo {title} {Controlling the dynamics of an open many-body quantum system with localized dissipation},}\ }\href {\doibase 10.1103/PhysRevLett.110.035302} {\bibfield  {journal} {\bibinfo  {journal} {Phys. Rev. Lett.}\ }\textbf {\bibinfo {volume} {110}},\ \bibinfo {pages} {035302} (\bibinfo {year} {2013})}\BibitemShut {NoStop}%
\bibitem [{\citenamefont {Tomita}\ \emph {et~al.}(2017)\citenamefont {Tomita}, \citenamefont {Nakajima}, \citenamefont {Danshita}, \citenamefont {Takasu},\ and\ \citenamefont {Takahashi}}]{Tomita_2017}%
  \BibitemOpen
  \bibfield  {author} {\bibinfo {author} {\bibfnamefont {Takafumi}\ \bibnamefont {Tomita}}, \bibinfo {author} {\bibfnamefont {Shuta}\ \bibnamefont {Nakajima}}, \bibinfo {author} {\bibfnamefont {Ippei}\ \bibnamefont {Danshita}}, \bibinfo {author} {\bibfnamefont {Yosuke}\ \bibnamefont {Takasu}}, \ and\ \bibinfo {author} {\bibfnamefont {Yoshiro}\ \bibnamefont {Takahashi}},\ }\bibfield  {title} {\enquote {\bibinfo {title} {Observation of the mott insulator to superfluid crossover of a driven-dissipative bose-hubbard system},}\ }\href {\doibase 10.1126/sciadv.1701513} {\bibfield  {journal} {\bibinfo  {journal} {Science Advances}\ }\textbf {\bibinfo {volume} {3}},\ \bibinfo {pages} {e1701513} (\bibinfo {year} {2017})}\BibitemShut {NoStop}%
\bibitem [{\citenamefont {Kouzelis}\ \emph {et~al.}(2020)\citenamefont {Kouzelis}, \citenamefont {Macieszczak}, \citenamefont {Min\'a\ifmmode~\check{r}\else \v{r}\fi{}},\ and\ \citenamefont {Lesanovsky}}]{Kouzelis_2020}%
  \BibitemOpen
  \bibfield  {author} {\bibinfo {author} {\bibfnamefont {Andreas}\ \bibnamefont {Kouzelis}}, \bibinfo {author} {\bibfnamefont {Katarzyna}\ \bibnamefont {Macieszczak}}, \bibinfo {author} {\bibfnamefont {Ji\ifmmode \check{r}\else~\v{r}\fi{}\'{\i}}\ \bibnamefont {Min\'a\ifmmode~\check{r}\else \v{r}\fi{}}}, \ and\ \bibinfo {author} {\bibfnamefont {Igor}\ \bibnamefont {Lesanovsky}},\ }\bibfield  {title} {\enquote {\bibinfo {title} {Dissipative quantum state preparation and metastability in two-photon micromasers},}\ }\href {\doibase 10.1103/PhysRevA.101.043847} {\bibfield  {journal} {\bibinfo  {journal} {Phys. Rev. A}\ }\textbf {\bibinfo {volume} {101}},\ \bibinfo {pages} {043847} (\bibinfo {year} {2020})}\BibitemShut {NoStop}%
\bibitem [{\citenamefont {Lim}\ \emph {et~al.}(2024)\citenamefont {Lim}, \citenamefont {Mok}, \citenamefont {You}, \citenamefont {Kong},\ and\ \citenamefont {Aghamalyan}}]{Lim_2024}%
  \BibitemOpen
  \bibfield  {author} {\bibinfo {author} {\bibfnamefont {Kian~Hwee}\ \bibnamefont {Lim}}, \bibinfo {author} {\bibfnamefont {Wai-Keong}\ \bibnamefont {Mok}}, \bibinfo {author} {\bibfnamefont {Jia-Bin}\ \bibnamefont {You}}, \bibinfo {author} {\bibfnamefont {Jian~Feng}\ \bibnamefont {Kong}}, \ and\ \bibinfo {author} {\bibfnamefont {Davit}\ \bibnamefont {Aghamalyan}},\ }\bibfield  {title} {\enquote {\bibinfo {title} {Exponentially faster preparation of quantum dimers via driven-dissipative stabilization},}\ }\href {\doibase 10.1103/PhysRevResearch.6.L032047} {\bibfield  {journal} {\bibinfo  {journal} {Phys. Rev. Res.}\ }\textbf {\bibinfo {volume} {6}},\ \bibinfo {pages} {L032047} (\bibinfo {year} {2024})}\BibitemShut {NoStop}%
\bibitem [{\citenamefont {Neri}\ \emph {et~al.}(2025)\citenamefont {Neri}, \citenamefont {Damanet}, \citenamefont {Daley}, \citenamefont {Chiofalo},\ and\ \citenamefont {Malo}}]{Neri_2025}%
  \BibitemOpen
  \bibfield  {author} {\bibinfo {author} {\bibfnamefont {Silvia}\ \bibnamefont {Neri}}, \bibinfo {author} {\bibfnamefont {François}\ \bibnamefont {Damanet}}, \bibinfo {author} {\bibfnamefont {Andrew~J.}\ \bibnamefont {Daley}}, \bibinfo {author} {\bibfnamefont {Marialuisa}\ \bibnamefont {Chiofalo}}, \ and\ \bibinfo {author} {\bibfnamefont {Jorga~Yago}\ \bibnamefont {Malo}},\ }\href {https://arxiv.org/abs/2507.00553} {\enquote {\bibinfo {title} {Dissipation engineering of fermionic long-range order beyond lindblad},}\ } (\bibinfo {year} {2025}),\ \Eprint {http://arxiv.org/abs/2507.00553} {arXiv:2507.00553 [cond-mat.supr-con]} \BibitemShut {NoStop}%
\bibitem [{\citenamefont {Raghunandan}\ \emph {et~al.}(2020)\citenamefont {Raghunandan}, \citenamefont {Wolf}, \citenamefont {Ospelkaus}, \citenamefont {Schmidt},\ and\ \citenamefont {Weimer}}]{Raghunandan_2020}%
  \BibitemOpen
  \bibfield  {author} {\bibinfo {author} {\bibfnamefont {Meghana}\ \bibnamefont {Raghunandan}}, \bibinfo {author} {\bibfnamefont {Fabian}\ \bibnamefont {Wolf}}, \bibinfo {author} {\bibfnamefont {Christian}\ \bibnamefont {Ospelkaus}}, \bibinfo {author} {\bibfnamefont {Piet~O.}\ \bibnamefont {Schmidt}}, \ and\ \bibinfo {author} {\bibfnamefont {Hendrik}\ \bibnamefont {Weimer}},\ }\bibfield  {title} {\enquote {\bibinfo {title} {Initialization of quantum simulators by sympathetic cooling},}\ }\href {\doibase 10.1126/sciadv.aaw9268} {\bibfield  {journal} {\bibinfo  {journal} {Science Advances}\ }\textbf {\bibinfo {volume} {6}},\ \bibinfo {pages} {eaaw9268} (\bibinfo {year} {2020})}\BibitemShut {NoStop}%
\bibitem [{\citenamefont {Wu}\ \emph {et~al.}(2025)\citenamefont {Wu}, \citenamefont {Ma}, \citenamefont {Wang}, \citenamefont {Brumer},\ and\ \citenamefont {Wu}}]{Wu_2025}%
  \BibitemOpen
  \bibfield  {author} {\bibinfo {author} {\bibfnamefont {S.~L.}\ \bibnamefont {Wu}}, \bibinfo {author} {\bibfnamefont {W.}~\bibnamefont {Ma}}, \bibinfo {author} {\bibfnamefont {Zhao-Ming}\ \bibnamefont {Wang}}, \bibinfo {author} {\bibfnamefont {P.}~\bibnamefont {Brumer}}, \ and\ \bibinfo {author} {\bibfnamefont {Lian-Ao}\ \bibnamefont {Wu}},\ }\href {https://arxiv.org/abs/2507.02557} {\enquote {\bibinfo {title} {Steady-state coherences under partial collective non-markovian decoherence},}\ } (\bibinfo {year} {2025}),\ \Eprint {http://arxiv.org/abs/2507.02557} {arXiv:2507.02557 [quant-ph]} \BibitemShut {NoStop}%
\bibitem [{\citenamefont {Min}\ \emph {et~al.}(2024{\natexlab{a}})\citenamefont {Min}, \citenamefont {Anto-Sztrikacs}, \citenamefont {Brenes},\ and\ \citenamefont {Segal}}]{Min_PRL}%
  \BibitemOpen
  \bibfield  {author} {\bibinfo {author} {\bibfnamefont {Brett}\ \bibnamefont {Min}}, \bibinfo {author} {\bibfnamefont {Nicholas}\ \bibnamefont {Anto-Sztrikacs}}, \bibinfo {author} {\bibfnamefont {Marlon}\ \bibnamefont {Brenes}}, \ and\ \bibinfo {author} {\bibfnamefont {Dvira}\ \bibnamefont {Segal}},\ }\bibfield  {title} {\enquote {\bibinfo {title} {Bath-engineering magnetic order in quantum spin chains: An analytic mapping approach},}\ }\href {\doibase 10.1103/PhysRevLett.132.266701} {\bibfield  {journal} {\bibinfo  {journal} {Phys. Rev. Lett.}\ }\textbf {\bibinfo {volume} {132}},\ \bibinfo {pages} {266701} (\bibinfo {year} {2024}{\natexlab{a}})}\BibitemShut {NoStop}%
\bibitem [{\citenamefont {Min}\ \emph {et~al.}(2024{\natexlab{b}})\citenamefont {Min}, \citenamefont {Agarwal},\ and\ \citenamefont {Segal}}]{Min_PRB}%
  \BibitemOpen
  \bibfield  {author} {\bibinfo {author} {\bibfnamefont {Brett}\ \bibnamefont {Min}}, \bibinfo {author} {\bibfnamefont {Kartiek}\ \bibnamefont {Agarwal}}, \ and\ \bibinfo {author} {\bibfnamefont {Dvira}\ \bibnamefont {Segal}},\ }\bibfield  {title} {\enquote {\bibinfo {title} {Role of bath-induced many-body interactions in the dissipative phases of the su-schrieffer-heeger model},}\ }\href {\doibase 10.1103/PhysRevB.110.125415} {\bibfield  {journal} {\bibinfo  {journal} {Phys. Rev. B}\ }\textbf {\bibinfo {volume} {110}},\ \bibinfo {pages} {125415} (\bibinfo {year} {2024}{\natexlab{b}})}\BibitemShut {NoStop}%
\bibitem [{\citenamefont {Macieszczak}\ \emph {et~al.}(2016)\citenamefont {Macieszczak}, \citenamefont {Gu\ifmmode \mbox{\c{t}}\else \c{t}\fi{}\ifmmode~\u{a}\else \u{a}\fi{}}, \citenamefont {Lesanovsky},\ and\ \citenamefont {Garrahan}}]{Macieszczak_2016}%
  \BibitemOpen
  \bibfield  {author} {\bibinfo {author} {\bibfnamefont {Katarzyna}\ \bibnamefont {Macieszczak}}, \bibinfo {author} {\bibfnamefont {M\ifmmode \u{a}\else \u{a}\fi{}d\ifmmode \u{a}\else~\u{a}\fi{}lin}\ \bibnamefont {Gu\ifmmode \mbox{\c{t}}\else \c{t}\fi{}\ifmmode~\u{a}\else \u{a}\fi{}}}, \bibinfo {author} {\bibfnamefont {Igor}\ \bibnamefont {Lesanovsky}}, \ and\ \bibinfo {author} {\bibfnamefont {Juan~P.}\ \bibnamefont {Garrahan}},\ }\bibfield  {title} {\enquote {\bibinfo {title} {Towards a theory of metastability in open quantum dynamics},}\ }\href {\doibase 10.1103/PhysRevLett.116.240404} {\bibfield  {journal} {\bibinfo  {journal} {Phys. Rev. Lett.}\ }\textbf {\bibinfo {volume} {116}},\ \bibinfo {pages} {240404} (\bibinfo {year} {2016})}\BibitemShut {NoStop}%
\bibitem [{\citenamefont {Rose}\ \emph {et~al.}(2016)\citenamefont {Rose}, \citenamefont {Macieszczak}, \citenamefont {Lesanovsky},\ and\ \citenamefont {Garrahan}}]{Rose_2016}%
  \BibitemOpen
  \bibfield  {author} {\bibinfo {author} {\bibfnamefont {Dominic~C.}\ \bibnamefont {Rose}}, \bibinfo {author} {\bibfnamefont {Katarzyna}\ \bibnamefont {Macieszczak}}, \bibinfo {author} {\bibfnamefont {Igor}\ \bibnamefont {Lesanovsky}}, \ and\ \bibinfo {author} {\bibfnamefont {Juan~P.}\ \bibnamefont {Garrahan}},\ }\bibfield  {title} {\enquote {\bibinfo {title} {Metastability in an open quantum ising model},}\ }\href {\doibase 10.1103/PhysRevE.94.052132} {\bibfield  {journal} {\bibinfo  {journal} {Phys. Rev. E}\ }\textbf {\bibinfo {volume} {94}},\ \bibinfo {pages} {052132} (\bibinfo {year} {2016})}\BibitemShut {NoStop}%
\bibitem [{\citenamefont {Le~Boit\'e}\ \emph {et~al.}(2017)\citenamefont {Le~Boit\'e}, \citenamefont {Hwang},\ and\ \citenamefont {Plenio}}]{Hwang_2017}%
  \BibitemOpen
  \bibfield  {author} {\bibinfo {author} {\bibfnamefont {Alexandre}\ \bibnamefont {Le~Boit\'e}}, \bibinfo {author} {\bibfnamefont {Myung-Joong}\ \bibnamefont {Hwang}}, \ and\ \bibinfo {author} {\bibfnamefont {Martin~B.}\ \bibnamefont {Plenio}},\ }\bibfield  {title} {\enquote {\bibinfo {title} {Metastability in the driven-dissipative rabi model},}\ }\href {\doibase 10.1103/PhysRevA.95.023829} {\bibfield  {journal} {\bibinfo  {journal} {Phys. Rev. A}\ }\textbf {\bibinfo {volume} {95}},\ \bibinfo {pages} {023829} (\bibinfo {year} {2017})}\BibitemShut {NoStop}%
\bibitem [{\citenamefont {Macieszczak}\ \emph {et~al.}(2021)\citenamefont {Macieszczak}, \citenamefont {Rose}, \citenamefont {Lesanovsky},\ and\ \citenamefont {Garrahan}}]{Maciezczak_2021}%
  \BibitemOpen
  \bibfield  {author} {\bibinfo {author} {\bibfnamefont {Katarzyna}\ \bibnamefont {Macieszczak}}, \bibinfo {author} {\bibfnamefont {Dominic~C.}\ \bibnamefont {Rose}}, \bibinfo {author} {\bibfnamefont {Igor}\ \bibnamefont {Lesanovsky}}, \ and\ \bibinfo {author} {\bibfnamefont {Juan~P.}\ \bibnamefont {Garrahan}},\ }\bibfield  {title} {\enquote {\bibinfo {title} {Theory of classical metastability in open quantum systems},}\ }\href {\doibase 10.1103/PhysRevResearch.3.033047} {\bibfield  {journal} {\bibinfo  {journal} {Phys. Rev. Res.}\ }\textbf {\bibinfo {volume} {3}},\ \bibinfo {pages} {033047} (\bibinfo {year} {2021})}\BibitemShut {NoStop}%
\bibitem [{\citenamefont {Anto-Sztrikacs}\ \emph {et~al.}(2024)\citenamefont {Anto-Sztrikacs}, \citenamefont {Miller}, \citenamefont {Nazir},\ and\ \citenamefont {Segal}}]{Nick_PRA}%
  \BibitemOpen
  \bibfield  {author} {\bibinfo {author} {\bibfnamefont {Nicholas}\ \bibnamefont {Anto-Sztrikacs}}, \bibinfo {author} {\bibfnamefont {Harry J.~D.}\ \bibnamefont {Miller}}, \bibinfo {author} {\bibfnamefont {Ahsan}\ \bibnamefont {Nazir}}, \ and\ \bibinfo {author} {\bibfnamefont {Dvira}\ \bibnamefont {Segal}},\ }\bibfield  {title} {\enquote {\bibinfo {title} {Bypassing thermalization timescales in temperature estimation using prethermal probes},}\ }\href {\doibase 10.1103/PhysRevA.109.L060201} {\bibfield  {journal} {\bibinfo  {journal} {Phys. Rev. A}\ }\textbf {\bibinfo {volume} {109}},\ \bibinfo {pages} {L060201} (\bibinfo {year} {2024})}\BibitemShut {NoStop}%
\bibitem [{\citenamefont {Mori}(2021)}]{Mori_2021}%
  \BibitemOpen
  \bibfield  {author} {\bibinfo {author} {\bibfnamefont {Takashi}\ \bibnamefont {Mori}},\ }\bibfield  {title} {\enquote {\bibinfo {title} {Metastability associated with many-body explosion of eigenmode expansion coefficients},}\ }\href {\doibase 10.1103/PhysRevResearch.3.043137} {\bibfield  {journal} {\bibinfo  {journal} {Phys. Rev. Res.}\ }\textbf {\bibinfo {volume} {3}},\ \bibinfo {pages} {043137} (\bibinfo {year} {2021})}\BibitemShut {NoStop}%
\bibitem [{\citenamefont {Cabot}\ \emph {et~al.}(2021)\citenamefont {Cabot}, \citenamefont {Giorgi},\ and\ \citenamefont {Zambrini}}]{Cabot_2021}%
  \BibitemOpen
  \bibfield  {author} {\bibinfo {author} {\bibfnamefont {Albert}\ \bibnamefont {Cabot}}, \bibinfo {author} {\bibfnamefont {Gian~Luca}\ \bibnamefont {Giorgi}}, \ and\ \bibinfo {author} {\bibfnamefont {Roberta}\ \bibnamefont {Zambrini}},\ }\bibfield  {title} {\enquote {\bibinfo {title} {Metastable quantum entrainment},}\ }\href {\doibase 10.1088/1367-2630/ac29fe} {\bibfield  {journal} {\bibinfo  {journal} {New Journal of Physics}\ }\textbf {\bibinfo {volume} {23}},\ \bibinfo {pages} {103017} (\bibinfo {year} {2021})}\BibitemShut {NoStop}%
\bibitem [{\citenamefont {Cabot}\ \emph {et~al.}(2022)\citenamefont {Cabot}, \citenamefont {Carollo},\ and\ \citenamefont {Lesanovsky}}]{Cabot_2022}%
  \BibitemOpen
  \bibfield  {author} {\bibinfo {author} {\bibfnamefont {Albert}\ \bibnamefont {Cabot}}, \bibinfo {author} {\bibfnamefont {Federico}\ \bibnamefont {Carollo}}, \ and\ \bibinfo {author} {\bibfnamefont {Igor}\ \bibnamefont {Lesanovsky}},\ }\bibfield  {title} {\enquote {\bibinfo {title} {Metastable discrete time-crystal resonances in a dissipative central spin system},}\ }\href {\doibase 10.1103/PhysRevB.106.134311} {\bibfield  {journal} {\bibinfo  {journal} {Phys. Rev. B}\ }\textbf {\bibinfo {volume} {106}},\ \bibinfo {pages} {134311} (\bibinfo {year} {2022})}\BibitemShut {NoStop}%
\bibitem [{\citenamefont {Wolff}\ \emph {et~al.}(2020)\citenamefont {Wolff}, \citenamefont {Sheikhan}, \citenamefont {Diehl},\ and\ \citenamefont {Kollath}}]{Wolff_2020}%
  \BibitemOpen
  \bibfield  {author} {\bibinfo {author} {\bibfnamefont {Stefan}\ \bibnamefont {Wolff}}, \bibinfo {author} {\bibfnamefont {Ameneh}\ \bibnamefont {Sheikhan}}, \bibinfo {author} {\bibfnamefont {Sebastian}\ \bibnamefont {Diehl}}, \ and\ \bibinfo {author} {\bibfnamefont {Corinna}\ \bibnamefont {Kollath}},\ }\bibfield  {title} {\enquote {\bibinfo {title} {Nonequilibrium metastable state in a chain of interacting spinless fermions with localized loss},}\ }\href {\doibase 10.1103/PhysRevB.101.075139} {\bibfield  {journal} {\bibinfo  {journal} {Phys. Rev. B}\ }\textbf {\bibinfo {volume} {101}},\ \bibinfo {pages} {075139} (\bibinfo {year} {2020})}\BibitemShut {NoStop}%
\bibitem [{\citenamefont {Labay-Mora}\ \emph {et~al.}(2023)\citenamefont {Labay-Mora}, \citenamefont {Zambrini},\ and\ \citenamefont {Giorgi}}]{Giorgi_2023}%
  \BibitemOpen
  \bibfield  {author} {\bibinfo {author} {\bibfnamefont {Adri\`a}\ \bibnamefont {Labay-Mora}}, \bibinfo {author} {\bibfnamefont {Roberta}\ \bibnamefont {Zambrini}}, \ and\ \bibinfo {author} {\bibfnamefont {Gian~Luca}\ \bibnamefont {Giorgi}},\ }\bibfield  {title} {\enquote {\bibinfo {title} {Quantum associative memory with a single driven-dissipative nonlinear oscillator},}\ }\href {\doibase 10.1103/PhysRevLett.130.190602} {\bibfield  {journal} {\bibinfo  {journal} {Phys. Rev. Lett.}\ }\textbf {\bibinfo {volume} {130}},\ \bibinfo {pages} {190602} (\bibinfo {year} {2023})}\BibitemShut {NoStop}%
\bibitem [{\citenamefont {Gerry}\ \emph {et~al.}(2024)\citenamefont {Gerry}, \citenamefont {Kewming},\ and\ \citenamefont {Segal}}]{Gerry_2024}%
  \BibitemOpen
  \bibfield  {author} {\bibinfo {author} {\bibfnamefont {Matthew}\ \bibnamefont {Gerry}}, \bibinfo {author} {\bibfnamefont {Michael~J.}\ \bibnamefont {Kewming}}, \ and\ \bibinfo {author} {\bibfnamefont {Dvira}\ \bibnamefont {Segal}},\ }\bibfield  {title} {\enquote {\bibinfo {title} {Understanding multiple timescales in quantum dissipative dynamics: Insights from quantum trajectories},}\ }\href {\doibase 10.1103/PhysRevResearch.6.033106} {\bibfield  {journal} {\bibinfo  {journal} {Phys. Rev. Res.}\ }\textbf {\bibinfo {volume} {6}},\ \bibinfo {pages} {033106} (\bibinfo {year} {2024})}\BibitemShut {NoStop}%
\bibitem [{\citenamefont {Tscherbul}\ and\ \citenamefont {Brumer}(2014)}]{Brumer_2014}%
  \BibitemOpen
  \bibfield  {author} {\bibinfo {author} {\bibfnamefont {Timur~V.}\ \bibnamefont {Tscherbul}}\ and\ \bibinfo {author} {\bibfnamefont {Paul}\ \bibnamefont {Brumer}},\ }\bibfield  {title} {\enquote {\bibinfo {title} {Long-lived quasistationary coherences in a $v$-type system driven by incoherent light},}\ }\href {\doibase 10.1103/PhysRevLett.113.113601} {\bibfield  {journal} {\bibinfo  {journal} {Phys. Rev. Lett.}\ }\textbf {\bibinfo {volume} {113}},\ \bibinfo {pages} {113601} (\bibinfo {year} {2014})}\BibitemShut {NoStop}%
\bibitem [{\citenamefont {Zhou}\ \emph {et~al.}(2021)\citenamefont {Zhou}, \citenamefont {Mao},\ and\ \citenamefont {Zhai}}]{Zhou_2021}%
  \BibitemOpen
  \bibfield  {author} {\bibinfo {author} {\bibfnamefont {Yi-Neng}\ \bibnamefont {Zhou}}, \bibinfo {author} {\bibfnamefont {Liang}\ \bibnamefont {Mao}}, \ and\ \bibinfo {author} {\bibfnamefont {Hui}\ \bibnamefont {Zhai}},\ }\bibfield  {title} {\enquote {\bibinfo {title} {R\'enyi entropy dynamics and lindblad spectrum for open quantum systems},}\ }\href {\doibase 10.1103/PhysRevResearch.3.043060} {\bibfield  {journal} {\bibinfo  {journal} {Phys. Rev. Res.}\ }\textbf {\bibinfo {volume} {3}},\ \bibinfo {pages} {043060} (\bibinfo {year} {2021})}\BibitemShut {NoStop}%
\bibitem [{\citenamefont {Wang}\ \emph {et~al.}(2020)\citenamefont {Wang}, \citenamefont {Piazza},\ and\ \citenamefont {Luitz}}]{Luitz_2020}%
  \BibitemOpen
  \bibfield  {author} {\bibinfo {author} {\bibfnamefont {Kevin}\ \bibnamefont {Wang}}, \bibinfo {author} {\bibfnamefont {Francesco}\ \bibnamefont {Piazza}}, \ and\ \bibinfo {author} {\bibfnamefont {David~J.}\ \bibnamefont {Luitz}},\ }\bibfield  {title} {\enquote {\bibinfo {title} {Hierarchy of relaxation timescales in local random liouvillians},}\ }\href {\doibase 10.1103/PhysRevLett.124.100604} {\bibfield  {journal} {\bibinfo  {journal} {Phys. Rev. Lett.}\ }\textbf {\bibinfo {volume} {124}},\ \bibinfo {pages} {100604} (\bibinfo {year} {2020})}\BibitemShut {NoStop}%
\bibitem [{\citenamefont {Li}\ \emph {et~al.}(2022)\citenamefont {Li}, \citenamefont {Rose}, \citenamefont {Garrahan},\ and\ \citenamefont {Luitz}}]{Luitz_2022}%
  \BibitemOpen
  \bibfield  {author} {\bibinfo {author} {\bibfnamefont {Jimin~L.}\ \bibnamefont {Li}}, \bibinfo {author} {\bibfnamefont {Dominic~C.}\ \bibnamefont {Rose}}, \bibinfo {author} {\bibfnamefont {Juan~P.}\ \bibnamefont {Garrahan}}, \ and\ \bibinfo {author} {\bibfnamefont {David~J.}\ \bibnamefont {Luitz}},\ }\bibfield  {title} {\enquote {\bibinfo {title} {Random matrix theory for quantum and classical metastability in local liouvillians},}\ }\href {\doibase 10.1103/PhysRevB.105.L180201} {\bibfield  {journal} {\bibinfo  {journal} {Phys. Rev. B}\ }\textbf {\bibinfo {volume} {105}},\ \bibinfo {pages} {L180201} (\bibinfo {year} {2022})}\BibitemShut {NoStop}%
\bibitem [{\citenamefont {Hartmann}\ \emph {et~al.}(2024)\citenamefont {Hartmann}, \citenamefont {Li},\ and\ \citenamefont {Luitz}}]{Luitz_2024}%
  \BibitemOpen
  \bibfield  {author} {\bibinfo {author} {\bibfnamefont {Nick~D.}\ \bibnamefont {Hartmann}}, \bibinfo {author} {\bibfnamefont {Jimin~L.}\ \bibnamefont {Li}}, \ and\ \bibinfo {author} {\bibfnamefont {David~J.}\ \bibnamefont {Luitz}},\ }\bibfield  {title} {\enquote {\bibinfo {title} {Fate of dissipative hierarchy of timescales in the presence of unitary dynamics},}\ }\href {\doibase 10.1103/PhysRevB.109.054203} {\bibfield  {journal} {\bibinfo  {journal} {Phys. Rev. B}\ }\textbf {\bibinfo {volume} {109}},\ \bibinfo {pages} {054203} (\bibinfo {year} {2024})}\BibitemShut {NoStop}%
\bibitem [{\citenamefont {Popkov}\ and\ \citenamefont {Presilla}(2021)}]{Presilla_2021}%
  \BibitemOpen
  \bibfield  {author} {\bibinfo {author} {\bibfnamefont {Vladislav}\ \bibnamefont {Popkov}}\ and\ \bibinfo {author} {\bibfnamefont {Carlo}\ \bibnamefont {Presilla}},\ }\bibfield  {title} {\enquote {\bibinfo {title} {Full spectrum of the liouvillian of open dissipative quantum systems in the zeno limit},}\ }\href {\doibase 10.1103/PhysRevLett.126.190402} {\bibfield  {journal} {\bibinfo  {journal} {Phys. Rev. Lett.}\ }\textbf {\bibinfo {volume} {126}},\ \bibinfo {pages} {190402} (\bibinfo {year} {2021})}\BibitemShut {NoStop}%
\bibitem [{\citenamefont {\ifmmode \check{Z}\else \v{Z}\fi{}nidari\ifmmode~\check{c}\else \v{c}\fi{}}(2015)}]{Marko_2015}%
  \BibitemOpen
  \bibfield  {author} {\bibinfo {author} {\bibfnamefont {Marko}\ \bibnamefont {\ifmmode \check{Z}\else \v{Z}\fi{}nidari\ifmmode~\check{c}\else \v{c}\fi{}}},\ }\bibfield  {title} {\enquote {\bibinfo {title} {Relaxation times of dissipative many-body quantum systems},}\ }\href {\doibase 10.1103/PhysRevE.92.042143} {\bibfield  {journal} {\bibinfo  {journal} {Phys. Rev. E}\ }\textbf {\bibinfo {volume} {92}},\ \bibinfo {pages} {042143} (\bibinfo {year} {2015})}\BibitemShut {NoStop}%
\bibitem [{\citenamefont {Sommer}\ \emph {et~al.}(2021)\citenamefont {Sommer}, \citenamefont {Piazza},\ and\ \citenamefont {Luitz}}]{Sommer_2021}%
  \BibitemOpen
  \bibfield  {author} {\bibinfo {author} {\bibfnamefont {Ophelia~Evelyn}\ \bibnamefont {Sommer}}, \bibinfo {author} {\bibfnamefont {Francesco}\ \bibnamefont {Piazza}}, \ and\ \bibinfo {author} {\bibfnamefont {David~J.}\ \bibnamefont {Luitz}},\ }\bibfield  {title} {\enquote {\bibinfo {title} {Many-body hierarchy of dissipative timescales in a quantum computer},}\ }\href {\doibase 10.1103/PhysRevResearch.3.023190} {\bibfield  {journal} {\bibinfo  {journal} {Phys. Rev. Res.}\ }\textbf {\bibinfo {volume} {3}},\ \bibinfo {pages} {023190} (\bibinfo {year} {2021})}\BibitemShut {NoStop}%
\bibitem [{\citenamefont {Rose}\ \emph {et~al.}(2022)\citenamefont {Rose}, \citenamefont {Macieszczak}, \citenamefont {Lesanovsky},\ and\ \citenamefont {Garrahan}}]{Rose_2022}%
  \BibitemOpen
  \bibfield  {author} {\bibinfo {author} {\bibfnamefont {Dominic~C.}\ \bibnamefont {Rose}}, \bibinfo {author} {\bibfnamefont {Katarzyna}\ \bibnamefont {Macieszczak}}, \bibinfo {author} {\bibfnamefont {Igor}\ \bibnamefont {Lesanovsky}}, \ and\ \bibinfo {author} {\bibfnamefont {Juan~P.}\ \bibnamefont {Garrahan}},\ }\bibfield  {title} {\enquote {\bibinfo {title} {Hierarchical classical metastability in an open quantum east model},}\ }\href {\doibase 10.1103/PhysRevE.105.044121} {\bibfield  {journal} {\bibinfo  {journal} {Phys. Rev. E}\ }\textbf {\bibinfo {volume} {105}},\ \bibinfo {pages} {044121} (\bibinfo {year} {2022})}\BibitemShut {NoStop}%
\bibitem [{\citenamefont {Zheng}\ \emph {et~al.}(2023)\citenamefont {Zheng}, \citenamefont {Wang},\ and\ \citenamefont {Chen}}]{Zheng_2023}%
  \BibitemOpen
  \bibfield  {author} {\bibinfo {author} {\bibfnamefont {Zhen-Yu}\ \bibnamefont {Zheng}}, \bibinfo {author} {\bibfnamefont {Xueliang}\ \bibnamefont {Wang}}, \ and\ \bibinfo {author} {\bibfnamefont {Shu}\ \bibnamefont {Chen}},\ }\bibfield  {title} {\enquote {\bibinfo {title} {Exact solution of the boundary-dissipated transverse field ising model: Structure of the liouvillian spectrum and dynamical duality},}\ }\href {\doibase 10.1103/PhysRevB.108.024404} {\bibfield  {journal} {\bibinfo  {journal} {Phys. Rev. B}\ }\textbf {\bibinfo {volume} {108}},\ \bibinfo {pages} {024404} (\bibinfo {year} {2023})}\BibitemShut {NoStop}%
\bibitem [{\citenamefont {Liu}\ and\ \citenamefont {Chen}(2023)}]{Shu_2023}%
  \BibitemOpen
  \bibfield  {author} {\bibinfo {author} {\bibfnamefont {Yu-Guo}\ \bibnamefont {Liu}}\ and\ \bibinfo {author} {\bibfnamefont {Shu}\ \bibnamefont {Chen}},\ }\bibfield  {title} {\enquote {\bibinfo {title} {Dynamical signatures of the liouvillian flat band},}\ }\href {\doibase 10.1103/PhysRevB.107.134307} {\bibfield  {journal} {\bibinfo  {journal} {Phys. Rev. B}\ }\textbf {\bibinfo {volume} {107}},\ \bibinfo {pages} {134307} (\bibinfo {year} {2023})}\BibitemShut {NoStop}%
\bibitem [{\citenamefont {Yang}\ \emph {et~al.}(2024)\citenamefont {Yang}, \citenamefont {Xu},\ and\ \citenamefont {del Campo}}]{Yang_2024}%
  \BibitemOpen
  \bibfield  {author} {\bibinfo {author} {\bibfnamefont {Yifeng}\ \bibnamefont {Yang}}, \bibinfo {author} {\bibfnamefont {Zhenyu}\ \bibnamefont {Xu}}, \ and\ \bibinfo {author} {\bibfnamefont {Adolfo}\ \bibnamefont {del Campo}},\ }\bibfield  {title} {\enquote {\bibinfo {title} {Decoherence rate in random lindblad dynamics},}\ }\href {\doibase 10.1103/PhysRevResearch.6.023229} {\bibfield  {journal} {\bibinfo  {journal} {Phys. Rev. Res.}\ }\textbf {\bibinfo {volume} {6}},\ \bibinfo {pages} {023229} (\bibinfo {year} {2024})}\BibitemShut {NoStop}%
\bibitem [{\citenamefont {Zündel}(2025)}]{Martina_2025}%
  \BibitemOpen
  \bibfield  {author} {\bibinfo {author} {\bibfnamefont {Martina}\ \bibnamefont {Zündel}},\ }\href {https://arxiv.org/abs/2503.21381} {\enquote {\bibinfo {title} {On the non-integrability of driven-dissipative one-dimensional hard-core bosons},}\ } (\bibinfo {year} {2025}),\ \Eprint {http://arxiv.org/abs/2503.21381} {arXiv:2503.21381 [math-ph]} \BibitemShut {NoStop}%
\bibitem [{\citenamefont {Zhou}\ \emph {et~al.}(2023{\natexlab{a}})\citenamefont {Zhou}, \citenamefont {Yu}, \citenamefont {Wu}, \citenamefont {Li}, \citenamefont {Zhang}, \citenamefont {Li},\ and\ \citenamefont {Chen}}]{Zhou_2023}%
  \BibitemOpen
  \bibfield  {author} {\bibinfo {author} {\bibfnamefont {Yan-Li}\ \bibnamefont {Zhou}}, \bibinfo {author} {\bibfnamefont {Xiao-Die}\ \bibnamefont {Yu}}, \bibinfo {author} {\bibfnamefont {Chun-Wang}\ \bibnamefont {Wu}}, \bibinfo {author} {\bibfnamefont {Xie-Qian}\ \bibnamefont {Li}}, \bibinfo {author} {\bibfnamefont {Jie}\ \bibnamefont {Zhang}}, \bibinfo {author} {\bibfnamefont {Weibin}\ \bibnamefont {Li}}, \ and\ \bibinfo {author} {\bibfnamefont {Ping-Xing}\ \bibnamefont {Chen}},\ }\bibfield  {title} {\enquote {\bibinfo {title} {Accelerating relaxation through liouvillian exceptional point},}\ }\href {\doibase 10.1103/PhysRevResearch.5.043036} {\bibfield  {journal} {\bibinfo  {journal} {Phys. Rev. Res.}\ }\textbf {\bibinfo {volume} {5}},\ \bibinfo {pages} {043036} (\bibinfo {year} {2023}{\natexlab{a}})}\BibitemShut {NoStop}%
\bibitem [{\citenamefont {Rossatto}\ and\ \citenamefont {Villas-Boas}(2016)}]{Rossatto_2016}%
  \BibitemOpen
  \bibfield  {author} {\bibinfo {author} {\bibfnamefont {D.~Z.}\ \bibnamefont {Rossatto}}\ and\ \bibinfo {author} {\bibfnamefont {C.~J.}\ \bibnamefont {Villas-Boas}},\ }\bibfield  {title} {\enquote {\bibinfo {title} {Relaxation time for monitoring the quantumness of an intense cavity field},}\ }\href {\doibase 10.1103/PhysRevA.94.033819} {\bibfield  {journal} {\bibinfo  {journal} {Phys. Rev. A}\ }\textbf {\bibinfo {volume} {94}},\ \bibinfo {pages} {033819} (\bibinfo {year} {2016})}\BibitemShut {NoStop}%
\bibitem [{\citenamefont {Zhou}\ \emph {et~al.}(2023{\natexlab{b}})\citenamefont {Zhou}, \citenamefont {Zou},\ and\ \citenamefont {Shao}}]{Shao_2023}%
  \BibitemOpen
  \bibfield  {author} {\bibinfo {author} {\bibfnamefont {Kun-Jie}\ \bibnamefont {Zhou}}, \bibinfo {author} {\bibfnamefont {Jian}\ \bibnamefont {Zou}}, \ and\ \bibinfo {author} {\bibfnamefont {Bin}\ \bibnamefont {Shao}},\ }\bibfield  {title} {\enquote {\bibinfo {title} {Dynamical transition between synchronization and antisynchronization with exceptional points},}\ }\href {\doibase 10.1103/PhysRevA.108.042206} {\bibfield  {journal} {\bibinfo  {journal} {Phys. Rev. A}\ }\textbf {\bibinfo {volume} {108}},\ \bibinfo {pages} {042206} (\bibinfo {year} {2023}{\natexlab{b}})}\BibitemShut {NoStop}%
\bibitem [{\citenamefont {Nemeth}\ \emph {et~al.}(2025)\citenamefont {Nemeth}, \citenamefont {Principi},\ and\ \citenamefont {Nazir}}]{Nazir_2025}%
  \BibitemOpen
  \bibfield  {author} {\bibinfo {author} {\bibfnamefont {Dominik}\ \bibnamefont {Nemeth}}, \bibinfo {author} {\bibfnamefont {Alessandro}\ \bibnamefont {Principi}}, \ and\ \bibinfo {author} {\bibfnamefont {Ahsan}\ \bibnamefont {Nazir}},\ }\href {https://arxiv.org/abs/2507.06998} {\enquote {\bibinfo {title} {Solving boundary time crystals via the superspin method},}\ } (\bibinfo {year} {2025}),\ \Eprint {http://arxiv.org/abs/2507.06998} {arXiv:2507.06998 [quant-ph]} \BibitemShut {NoStop}%
\bibitem [{\citenamefont {Mori}\ and\ \citenamefont {Shirai}(2020)}]{Mori_2020}%
  \BibitemOpen
  \bibfield  {author} {\bibinfo {author} {\bibfnamefont {Takashi}\ \bibnamefont {Mori}}\ and\ \bibinfo {author} {\bibfnamefont {Tatsuhiko}\ \bibnamefont {Shirai}},\ }\bibfield  {title} {\enquote {\bibinfo {title} {Resolving a discrepancy between liouvillian gap and relaxation time in boundary-dissipated quantum many-body systems},}\ }\href {\doibase 10.1103/PhysRevLett.125.230604} {\bibfield  {journal} {\bibinfo  {journal} {Phys. Rev. Lett.}\ }\textbf {\bibinfo {volume} {125}},\ \bibinfo {pages} {230604} (\bibinfo {year} {2020})}\BibitemShut {NoStop}%
\bibitem [{\citenamefont {Mori}\ and\ \citenamefont {Shirai}(2023)}]{Mori_2023}%
  \BibitemOpen
  \bibfield  {author} {\bibinfo {author} {\bibfnamefont {Takashi}\ \bibnamefont {Mori}}\ and\ \bibinfo {author} {\bibfnamefont {Tatsuhiko}\ \bibnamefont {Shirai}},\ }\bibfield  {title} {\enquote {\bibinfo {title} {Symmetrized liouvillian gap in markovian open quantum systems},}\ }\href {\doibase 10.1103/PhysRevLett.130.230404} {\bibfield  {journal} {\bibinfo  {journal} {Phys. Rev. Lett.}\ }\textbf {\bibinfo {volume} {130}},\ \bibinfo {pages} {230404} (\bibinfo {year} {2023})}\BibitemShut {NoStop}%
\bibitem [{\citenamefont {Mori}(2024)}]{Mori_2024}%
  \BibitemOpen
  \bibfield  {author} {\bibinfo {author} {\bibfnamefont {Takashi}\ \bibnamefont {Mori}},\ }\bibfield  {title} {\enquote {\bibinfo {title} {Liouvillian-gap analysis of open quantum many-body systems in the weak dissipation limit},}\ }\href {\doibase 10.1103/PhysRevB.109.064311} {\bibfield  {journal} {\bibinfo  {journal} {Phys. Rev. B}\ }\textbf {\bibinfo {volume} {109}},\ \bibinfo {pages} {064311} (\bibinfo {year} {2024})}\BibitemShut {NoStop}%
\bibitem [{\citenamefont {Yuan}\ \emph {et~al.}(2021)\citenamefont {Yuan}, \citenamefont {Wang}, \citenamefont {Wang},\ and\ \citenamefont {Deng}}]{Yuan_2021}%
  \BibitemOpen
  \bibfield  {author} {\bibinfo {author} {\bibfnamefont {Dong}\ \bibnamefont {Yuan}}, \bibinfo {author} {\bibfnamefont {He-Ran}\ \bibnamefont {Wang}}, \bibinfo {author} {\bibfnamefont {Zhong}\ \bibnamefont {Wang}}, \ and\ \bibinfo {author} {\bibfnamefont {Dong-Ling}\ \bibnamefont {Deng}},\ }\bibfield  {title} {\enquote {\bibinfo {title} {Solving the liouvillian gap with artificial neural networks},}\ }\href {\doibase 10.1103/PhysRevLett.126.160401} {\bibfield  {journal} {\bibinfo  {journal} {Phys. Rev. Lett.}\ }\textbf {\bibinfo {volume} {126}},\ \bibinfo {pages} {160401} (\bibinfo {year} {2021})}\BibitemShut {NoStop}%
\bibitem [{\citenamefont {Kessler}\ \emph {et~al.}(2012)\citenamefont {Kessler}, \citenamefont {Giedke}, \citenamefont {Imamoglu}, \citenamefont {Yelin}, \citenamefont {Lukin},\ and\ \citenamefont {Cirac}}]{Kessler_2012}%
  \BibitemOpen
  \bibfield  {author} {\bibinfo {author} {\bibfnamefont {E.~M.}\ \bibnamefont {Kessler}}, \bibinfo {author} {\bibfnamefont {G.}~\bibnamefont {Giedke}}, \bibinfo {author} {\bibfnamefont {A.}~\bibnamefont {Imamoglu}}, \bibinfo {author} {\bibfnamefont {S.~F.}\ \bibnamefont {Yelin}}, \bibinfo {author} {\bibfnamefont {M.~D.}\ \bibnamefont {Lukin}}, \ and\ \bibinfo {author} {\bibfnamefont {J.~I.}\ \bibnamefont {Cirac}},\ }\bibfield  {title} {\enquote {\bibinfo {title} {Dissipative phase transition in a central spin system},}\ }\href {\doibase 10.1103/PhysRevA.86.012116} {\bibfield  {journal} {\bibinfo  {journal} {Phys. Rev. A}\ }\textbf {\bibinfo {volume} {86}},\ \bibinfo {pages} {012116} (\bibinfo {year} {2012})}\BibitemShut {NoStop}%
\bibitem [{\citenamefont {H\"oning}\ \emph {et~al.}(2012)\citenamefont {H\"oning}, \citenamefont {Moos},\ and\ \citenamefont {Fleischhauer}}]{Moos_2012}%
  \BibitemOpen
  \bibfield  {author} {\bibinfo {author} {\bibfnamefont {M.}~\bibnamefont {H\"oning}}, \bibinfo {author} {\bibfnamefont {M.}~\bibnamefont {Moos}}, \ and\ \bibinfo {author} {\bibfnamefont {M.}~\bibnamefont {Fleischhauer}},\ }\bibfield  {title} {\enquote {\bibinfo {title} {Critical exponents of steady-state phase transitions in fermionic lattice models},}\ }\href {\doibase 10.1103/PhysRevA.86.013606} {\bibfield  {journal} {\bibinfo  {journal} {Phys. Rev. A}\ }\textbf {\bibinfo {volume} {86}},\ \bibinfo {pages} {013606} (\bibinfo {year} {2012})}\BibitemShut {NoStop}%
\bibitem [{\citenamefont {Horstmann}\ \emph {et~al.}(2013)\citenamefont {Horstmann}, \citenamefont {Cirac},\ and\ \citenamefont {Giedke}}]{Horstmann_2013}%
  \BibitemOpen
  \bibfield  {author} {\bibinfo {author} {\bibfnamefont {Birger}\ \bibnamefont {Horstmann}}, \bibinfo {author} {\bibfnamefont {J.~Ignacio}\ \bibnamefont {Cirac}}, \ and\ \bibinfo {author} {\bibfnamefont {G\'eza}\ \bibnamefont {Giedke}},\ }\bibfield  {title} {\enquote {\bibinfo {title} {Noise-driven dynamics and phase transitions in fermionic systems},}\ }\href {\doibase 10.1103/PhysRevA.87.012108} {\bibfield  {journal} {\bibinfo  {journal} {Phys. Rev. A}\ }\textbf {\bibinfo {volume} {87}},\ \bibinfo {pages} {012108} (\bibinfo {year} {2013})}\BibitemShut {NoStop}%
\bibitem [{\citenamefont {Minganti}\ \emph {et~al.}(2018)\citenamefont {Minganti}, \citenamefont {Biella}, \citenamefont {Bartolo},\ and\ \citenamefont {Ciuti}}]{Minganti_2018}%
  \BibitemOpen
  \bibfield  {author} {\bibinfo {author} {\bibfnamefont {Fabrizio}\ \bibnamefont {Minganti}}, \bibinfo {author} {\bibfnamefont {Alberto}\ \bibnamefont {Biella}}, \bibinfo {author} {\bibfnamefont {Nicola}\ \bibnamefont {Bartolo}}, \ and\ \bibinfo {author} {\bibfnamefont {Cristiano}\ \bibnamefont {Ciuti}},\ }\bibfield  {title} {\enquote {\bibinfo {title} {Spectral theory of liouvillians for dissipative phase transitions},}\ }\href {\doibase 10.1103/PhysRevA.98.042118} {\bibfield  {journal} {\bibinfo  {journal} {Phys. Rev. A}\ }\textbf {\bibinfo {volume} {98}},\ \bibinfo {pages} {042118} (\bibinfo {year} {2018})}\BibitemShut {NoStop}%
\bibitem [{\citenamefont {Haga}\ \emph {et~al.}(2021)\citenamefont {Haga}, \citenamefont {Nakagawa}, \citenamefont {Hamazaki},\ and\ \citenamefont {Ueda}}]{Haga_2021}%
  \BibitemOpen
  \bibfield  {author} {\bibinfo {author} {\bibfnamefont {Taiki}\ \bibnamefont {Haga}}, \bibinfo {author} {\bibfnamefont {Masaya}\ \bibnamefont {Nakagawa}}, \bibinfo {author} {\bibfnamefont {Ryusuke}\ \bibnamefont {Hamazaki}}, \ and\ \bibinfo {author} {\bibfnamefont {Masahito}\ \bibnamefont {Ueda}},\ }\bibfield  {title} {\enquote {\bibinfo {title} {Liouvillian skin effect: Slowing down of relaxation processes without gap closing},}\ }\href {\doibase 10.1103/PhysRevLett.127.070402} {\bibfield  {journal} {\bibinfo  {journal} {Phys. Rev. Lett.}\ }\textbf {\bibinfo {volume} {127}},\ \bibinfo {pages} {070402} (\bibinfo {year} {2021})}\BibitemShut {NoStop}%
\bibitem [{\citenamefont {Zhou}\ \emph {et~al.}(2023{\natexlab{c}})\citenamefont {Zhou}, \citenamefont {Yu}, \citenamefont {Wu}, \citenamefont {Li}, \citenamefont {Zhang}, \citenamefont {Li},\ and\ \citenamefont {Chen}}]{Weibin_2023}%
  \BibitemOpen
  \bibfield  {author} {\bibinfo {author} {\bibfnamefont {Yan-Li}\ \bibnamefont {Zhou}}, \bibinfo {author} {\bibfnamefont {Xiao-Die}\ \bibnamefont {Yu}}, \bibinfo {author} {\bibfnamefont {Chun-Wang}\ \bibnamefont {Wu}}, \bibinfo {author} {\bibfnamefont {Xie-Qian}\ \bibnamefont {Li}}, \bibinfo {author} {\bibfnamefont {Jie}\ \bibnamefont {Zhang}}, \bibinfo {author} {\bibfnamefont {Weibin}\ \bibnamefont {Li}}, \ and\ \bibinfo {author} {\bibfnamefont {Ping-Xing}\ \bibnamefont {Chen}},\ }\bibfield  {title} {\enquote {\bibinfo {title} {Accelerating relaxation through liouvillian exceptional point},}\ }\href {\doibase 10.1103/PhysRevResearch.5.043036} {\bibfield  {journal} {\bibinfo  {journal} {Phys. Rev. Res.}\ }\textbf {\bibinfo {volume} {5}},\ \bibinfo {pages} {043036} (\bibinfo {year} {2023}{\natexlab{c}})}\BibitemShut {NoStop}%
\bibitem [{\citenamefont {Zhang}\ \emph {et~al.}(2025)\citenamefont {Zhang}, \citenamefont {Xia}, \citenamefont {Wu}, \citenamefont {Chen}, \citenamefont {Zhang}, \citenamefont {Xie}, \citenamefont {Su}, \citenamefont {Wu}, \citenamefont {Qiu}, \citenamefont {Chen}, \citenamefont {Li}, \citenamefont {Jing},\ and\ \citenamefont {Zhou}}]{Zhang2025}%
  \BibitemOpen
  \bibfield  {author} {\bibinfo {author} {\bibfnamefont {Jie}\ \bibnamefont {Zhang}}, \bibinfo {author} {\bibfnamefont {Gang}\ \bibnamefont {Xia}}, \bibinfo {author} {\bibfnamefont {Chun-Wang}\ \bibnamefont {Wu}}, \bibinfo {author} {\bibfnamefont {Ting}\ \bibnamefont {Chen}}, \bibinfo {author} {\bibfnamefont {Qian}\ \bibnamefont {Zhang}}, \bibinfo {author} {\bibfnamefont {Yi}~\bibnamefont {Xie}}, \bibinfo {author} {\bibfnamefont {Wen-Bo}\ \bibnamefont {Su}}, \bibinfo {author} {\bibfnamefont {Wei}\ \bibnamefont {Wu}}, \bibinfo {author} {\bibfnamefont {Cheng-Wei}\ \bibnamefont {Qiu}}, \bibinfo {author} {\bibfnamefont {Ping-Xing}\ \bibnamefont {Chen}}, \bibinfo {author} {\bibfnamefont {Weibin}\ \bibnamefont {Li}}, \bibinfo {author} {\bibfnamefont {Hui}\ \bibnamefont {Jing}}, \ and\ \bibinfo {author} {\bibfnamefont {Yan-Li}\ \bibnamefont {Zhou}},\ }\bibfield  {title} {\enquote {\bibinfo {title} {Observation of quantum strong mpemba effect},}\ }\href {\doibase 10.1038/s41467-024-54303-0} {\bibfield  {journal}
  {\bibinfo  {journal} {Nature Communications}\ }\textbf {\bibinfo {volume} {16}},\ \bibinfo {pages} {301} (\bibinfo {year} {2025})}\BibitemShut {NoStop}%
\bibitem [{\citenamefont {Pino}\ \emph {et~al.}(2015)\citenamefont {Pino}, \citenamefont {Feist},\ and\ \citenamefont {Garcia-Vidal}}]{Pino_2015}%
  \BibitemOpen
  \bibfield  {author} {\bibinfo {author} {\bibfnamefont {Javier~del}\ \bibnamefont {Pino}}, \bibinfo {author} {\bibfnamefont {Johannes}\ \bibnamefont {Feist}}, \ and\ \bibinfo {author} {\bibfnamefont {Francisco~J}\ \bibnamefont {Garcia-Vidal}},\ }\bibfield  {title} {\enquote {\bibinfo {title} {Quantum theory of collective strong coupling of molecular vibrations with a microcavity mode},}\ }\href {\doibase 10.1088/1367-2630/17/5/053040} {\bibfield  {journal} {\bibinfo  {journal} {New Journal of Physics}\ }\textbf {\bibinfo {volume} {17}},\ \bibinfo {pages} {053040} (\bibinfo {year} {2015})}\BibitemShut {NoStop}%
\bibitem [{\citenamefont {Damanet}\ \emph {et~al.}(2019)\citenamefont {Damanet}, \citenamefont {Daley},\ and\ \citenamefont {Keeling}}]{Damanet_2019}%
  \BibitemOpen
  \bibfield  {author} {\bibinfo {author} {\bibfnamefont {Fran\ifmmode \mbox{\c{c}}\else~\c{c}\fi{}ois}\ \bibnamefont {Damanet}}, \bibinfo {author} {\bibfnamefont {Andrew~J.}\ \bibnamefont {Daley}}, \ and\ \bibinfo {author} {\bibfnamefont {Jonathan}\ \bibnamefont {Keeling}},\ }\bibfield  {title} {\enquote {\bibinfo {title} {Atom-only descriptions of the driven-dissipative dicke model},}\ }\href {\doibase 10.1103/PhysRevA.99.033845} {\bibfield  {journal} {\bibinfo  {journal} {Phys. Rev. A}\ }\textbf {\bibinfo {volume} {99}},\ \bibinfo {pages} {033845} (\bibinfo {year} {2019})}\BibitemShut {NoStop}%
\bibitem [{\citenamefont {Qu}\ \emph {et~al.}(2025)\citenamefont {Qu}, \citenamefont {Stefanini}, \citenamefont {Shi}, \citenamefont {Esslinger}, \citenamefont {Gopalakrishnan}, \citenamefont {Marino},\ and\ \citenamefont {Demler}}]{Qu_2025s}%
  \BibitemOpen
  \bibfield  {author} {\bibinfo {author} {\bibfnamefont {Yi-Fan}\ \bibnamefont {Qu}}, \bibinfo {author} {\bibfnamefont {Martino}\ \bibnamefont {Stefanini}}, \bibinfo {author} {\bibfnamefont {Tao}\ \bibnamefont {Shi}}, \bibinfo {author} {\bibfnamefont {Tilman}\ \bibnamefont {Esslinger}}, \bibinfo {author} {\bibfnamefont {Sarang}\ \bibnamefont {Gopalakrishnan}}, \bibinfo {author} {\bibfnamefont {Jamir}\ \bibnamefont {Marino}}, \ and\ \bibinfo {author} {\bibfnamefont {Eugene}\ \bibnamefont {Demler}},\ }\bibfield  {title} {\enquote {\bibinfo {title} {Variational approach to the dynamics of dissipative quantum impurity models},}\ }\href {\doibase 10.1103/PhysRevB.111.155113} {\bibfield  {journal} {\bibinfo  {journal} {Phys. Rev. B}\ }\textbf {\bibinfo {volume} {111}},\ \bibinfo {pages} {155113} (\bibinfo {year} {2025})}\BibitemShut {NoStop}%
\bibitem [{\citenamefont {Poletti}\ \emph {et~al.}(2012)\citenamefont {Poletti}, \citenamefont {Bernier}, \citenamefont {Georges},\ and\ \citenamefont {Kollath}}]{Poletti_2012}%
  \BibitemOpen
  \bibfield  {author} {\bibinfo {author} {\bibfnamefont {Dario}\ \bibnamefont {Poletti}}, \bibinfo {author} {\bibfnamefont {Jean-S\'ebastien}\ \bibnamefont {Bernier}}, \bibinfo {author} {\bibfnamefont {Antoine}\ \bibnamefont {Georges}}, \ and\ \bibinfo {author} {\bibfnamefont {Corinna}\ \bibnamefont {Kollath}},\ }\bibfield  {title} {\enquote {\bibinfo {title} {Interaction-induced impeding of decoherence and anomalous diffusion},}\ }\href {\doibase 10.1103/PhysRevLett.109.045302} {\bibfield  {journal} {\bibinfo  {journal} {Phys. Rev. Lett.}\ }\textbf {\bibinfo {volume} {109}},\ \bibinfo {pages} {045302} (\bibinfo {year} {2012})}\BibitemShut {NoStop}%
\bibitem [{\citenamefont {Rosso}\ \emph {et~al.}(2021)\citenamefont {Rosso}, \citenamefont {Rossini}, \citenamefont {Biella},\ and\ \citenamefont {Mazza}}]{Rosso_2021}%
  \BibitemOpen
  \bibfield  {author} {\bibinfo {author} {\bibfnamefont {Lorenzo}\ \bibnamefont {Rosso}}, \bibinfo {author} {\bibfnamefont {Davide}\ \bibnamefont {Rossini}}, \bibinfo {author} {\bibfnamefont {Alberto}\ \bibnamefont {Biella}}, \ and\ \bibinfo {author} {\bibfnamefont {Leonardo}\ \bibnamefont {Mazza}},\ }\bibfield  {title} {\enquote {\bibinfo {title} {One-dimensional spin-1/2 fermionic gases with two-body losses: Weak dissipation and spin conservation},}\ }\href {\doibase 10.1103/PhysRevA.104.053305} {\bibfield  {journal} {\bibinfo  {journal} {Phys. Rev. A}\ }\textbf {\bibinfo {volume} {104}},\ \bibinfo {pages} {053305} (\bibinfo {year} {2021})}\BibitemShut {NoStop}%
\bibitem [{\citenamefont {qiang Liu}\ \emph {et~al.}(2025)\citenamefont {qiang Liu}, \citenamefont {Long}, \citenamefont {jia Yang}, \citenamefont {Liu}, \citenamefont {Ting-ting}, \citenamefont {Ma}, \citenamefont {Bao-qing}, \citenamefont {Guo}, \citenamefont {Xingdong}, \citenamefont {Zhao}, \citenamefont {Zunlue}, \citenamefont {Zhu}, \citenamefont {Wuming}, \citenamefont {Liu},\ and\ \citenamefont {shui Yu}}]{Liu_2025}%
  \BibitemOpen
  \bibfield  {author} {\bibinfo {author} {\bibfnamefont {Yu}~\bibnamefont {qiang Liu}}, \bibinfo {author} {\bibfnamefont {Qiulin}\ \bibnamefont {Long}}, \bibinfo {author} {\bibfnamefont {Yi}~\bibnamefont {jia Yang}}, \bibinfo {author} {\bibfnamefont {Zheng}\ \bibnamefont {Liu}}, \bibinfo {author} {\bibnamefont {Ting-ting}}, \bibinfo {author} {\bibnamefont {Ma}}, \bibinfo {author} {\bibnamefont {Bao-qing}}, \bibinfo {author} {\bibnamefont {Guo}}, \bibinfo {author} {\bibnamefont {Xingdong}}, \bibinfo {author} {\bibnamefont {Zhao}}, \bibinfo {author} {\bibnamefont {Zunlue}}, \bibinfo {author} {\bibnamefont {Zhu}}, \bibinfo {author} {\bibnamefont {Wuming}}, \bibinfo {author} {\bibnamefont {Liu}}, \ and\ \bibinfo {author} {\bibfnamefont {Chang}\ \bibnamefont {shui Yu}},\ }\href {https://arxiv.org/abs/2507.00638} {\enquote {\bibinfo {title} {Behavior of quantum coherence in the ultrastrong and deep strong coupling regimes of light-matter system},}\ } (\bibinfo {year} {2025}),\ \Eprint
  {http://arxiv.org/abs/2507.00638} {arXiv:2507.00638 [quant-ph]} \BibitemShut {NoStop}%
\bibitem [{\citenamefont {Beaudoin}\ \emph {et~al.}(2011)\citenamefont {Beaudoin}, \citenamefont {Gambetta},\ and\ \citenamefont {Blais}}]{Blais_2011}%
  \BibitemOpen
  \bibfield  {author} {\bibinfo {author} {\bibfnamefont {F\'elix}\ \bibnamefont {Beaudoin}}, \bibinfo {author} {\bibfnamefont {Jay~M.}\ \bibnamefont {Gambetta}}, \ and\ \bibinfo {author} {\bibfnamefont {A.}~\bibnamefont {Blais}},\ }\bibfield  {title} {\enquote {\bibinfo {title} {Dissipation and ultrastrong coupling in circuit qed},}\ }\href {\doibase 10.1103/PhysRevA.84.043832} {\bibfield  {journal} {\bibinfo  {journal} {Phys. Rev. A}\ }\textbf {\bibinfo {volume} {84}},\ \bibinfo {pages} {043832} (\bibinfo {year} {2011})}\BibitemShut {NoStop}%
\bibitem [{\citenamefont {Anto-Sztrikacs}\ \emph {et~al.}(2023{\natexlab{a}})\citenamefont {Anto-Sztrikacs}, \citenamefont {Nazir},\ and\ \citenamefont {Segal}}]{Nick_PRXQ}%
  \BibitemOpen
  \bibfield  {author} {\bibinfo {author} {\bibfnamefont {Nicholas}\ \bibnamefont {Anto-Sztrikacs}}, \bibinfo {author} {\bibfnamefont {Ahsan}\ \bibnamefont {Nazir}}, \ and\ \bibinfo {author} {\bibfnamefont {Dvira}\ \bibnamefont {Segal}},\ }\bibfield  {title} {\enquote {\bibinfo {title} {Effective-hamiltonian theory of open quantum systems at strong coupling},}\ }\href {\doibase 10.1103/PRXQuantum.4.020307} {\bibfield  {journal} {\bibinfo  {journal} {PRX Quantum}\ }\textbf {\bibinfo {volume} {4}},\ \bibinfo {pages} {020307} (\bibinfo {year} {2023}{\natexlab{a}})}\BibitemShut {NoStop}%
\bibitem [{\citenamefont {Anto-Sztrikacs}\ \emph {et~al.}(2023{\natexlab{b}})\citenamefont {Anto-Sztrikacs}, \citenamefont {Min}, \citenamefont {Brenes},\ and\ \citenamefont {Segal}}]{Nick_PRB}%
  \BibitemOpen
  \bibfield  {author} {\bibinfo {author} {\bibfnamefont {Nicholas}\ \bibnamefont {Anto-Sztrikacs}}, \bibinfo {author} {\bibfnamefont {Brett}\ \bibnamefont {Min}}, \bibinfo {author} {\bibfnamefont {Marlon}\ \bibnamefont {Brenes}}, \ and\ \bibinfo {author} {\bibfnamefont {Dvira}\ \bibnamefont {Segal}},\ }\bibfield  {title} {\enquote {\bibinfo {title} {Effective hamiltonian theory: An approximation to the equilibrium state of open quantum systems},}\ }\href {\doibase 10.1103/PhysRevB.108.115437} {\bibfield  {journal} {\bibinfo  {journal} {Phys. Rev. B}\ }\textbf {\bibinfo {volume} {108}},\ \bibinfo {pages} {115437} (\bibinfo {year} {2023}{\natexlab{b}})}\BibitemShut {NoStop}%
\bibitem [{\citenamefont {Chen}\ \emph {et~al.}(2025)\citenamefont {Chen}, \citenamefont {Garwoła},\ and\ \citenamefont {Segal}}]{Jitian_2025}%
  \BibitemOpen
  \bibfield  {author} {\bibinfo {author} {\bibfnamefont {Jitian}\ \bibnamefont {Chen}}, \bibinfo {author} {\bibfnamefont {Jakub}\ \bibnamefont {Garwoła}}, \ and\ \bibinfo {author} {\bibfnamefont {Dvira}\ \bibnamefont {Segal}},\ }\href {https://arxiv.org/abs/2504.02796} {\enquote {\bibinfo {title} {Suppression of decoherence dynamics by a dissipative bath at strong coupling},}\ } (\bibinfo {year} {2025}),\ \Eprint {http://arxiv.org/abs/2504.02796} {arXiv:2504.02796 [quant-ph]} \BibitemShut {NoStop}%
\bibitem [{\citenamefont {Garwo\l{}a}\ and\ \citenamefont {Segal}(2024)}]{Garwola_PRB}%
  \BibitemOpen
  \bibfield  {author} {\bibinfo {author} {\bibfnamefont {Jakub}\ \bibnamefont {Garwo\l{}a}}\ and\ \bibinfo {author} {\bibfnamefont {Dvira}\ \bibnamefont {Segal}},\ }\bibfield  {title} {\enquote {\bibinfo {title} {Open quantum systems with noncommuting coupling operators: An analytic approach},}\ }\href {\doibase 10.1103/PhysRevB.110.174304} {\bibfield  {journal} {\bibinfo  {journal} {Phys. Rev. B}\ }\textbf {\bibinfo {volume} {110}},\ \bibinfo {pages} {174304} (\bibinfo {year} {2024})}\BibitemShut {NoStop}%
\bibitem [{\citenamefont {Brenes}\ \emph {et~al.}(2025)\citenamefont {Brenes}, \citenamefont {Garwo\l{}a},\ and\ \citenamefont {Segal}}]{Brenes_2025}%
  \BibitemOpen
  \bibfield  {author} {\bibinfo {author} {\bibfnamefont {Marlon}\ \bibnamefont {Brenes}}, \bibinfo {author} {\bibfnamefont {Jakub}\ \bibnamefont {Garwo\l{}a}}, \ and\ \bibinfo {author} {\bibfnamefont {Dvira}\ \bibnamefont {Segal}},\ }\bibfield  {title} {\enquote {\bibinfo {title} {Optimal qubit-mediated quantum heat transfer via noncommuting operators and strong coupling effects},}\ }\href {\doibase 10.1103/m18l-t1hk} {\bibfield  {journal} {\bibinfo  {journal} {Phys. Rev. B}\ }\textbf {\bibinfo {volume} {111}},\ \bibinfo {pages} {235440} (\bibinfo {year} {2025})}\BibitemShut {NoStop}%
\bibitem [{\citenamefont {Brenes}\ \emph {et~al.}(2024)\citenamefont {Brenes}, \citenamefont {Min}, \citenamefont {Anto-Sztrikacs}, \citenamefont {Bar-Gill},\ and\ \citenamefont {Segal}}]{Brenes_JCP}%
  \BibitemOpen
  \bibfield  {author} {\bibinfo {author} {\bibfnamefont {Marlon}\ \bibnamefont {Brenes}}, \bibinfo {author} {\bibfnamefont {Brett}\ \bibnamefont {Min}}, \bibinfo {author} {\bibfnamefont {Nicholas}\ \bibnamefont {Anto-Sztrikacs}}, \bibinfo {author} {\bibfnamefont {Nir}\ \bibnamefont {Bar-Gill}}, \ and\ \bibinfo {author} {\bibfnamefont {Dvira}\ \bibnamefont {Segal}},\ }\bibfield  {title} {\enquote {\bibinfo {title} {Bath-induced interactions and transient dynamics in open quantum systems at strong coupling: Effective hamiltonian approach},}\ }\href {\doibase 10.1063/5.0207028} {\bibfield  {journal} {\bibinfo  {journal} {The Journal of Chemical Physics}\ }\textbf {\bibinfo {volume} {160}},\ \bibinfo {pages} {244106} (\bibinfo {year} {2024})}\BibitemShut {NoStop}%
\bibitem [{\citenamefont {Nazir}\ and\ \citenamefont {Schaller}(2018)}]{Nazir18}%
  \BibitemOpen
  \bibfield  {author} {\bibinfo {author} {\bibfnamefont {Ahsan}\ \bibnamefont {Nazir}}\ and\ \bibinfo {author} {\bibfnamefont {Gernot}\ \bibnamefont {Schaller}},\ }\enquote {\bibinfo {title} {The reaction coordinate mapping in quantum thermodynamics},}\ in\ \href@noop {} {\emph {\bibinfo {booktitle} {Thermodynamics in the Quantum Regime: Fundamental Aspects and New Directions}}}\ (\bibinfo  {publisher} {Springer International Publishing},\ \bibinfo {year} {2018})\ pp.\ \bibinfo {pages} {551--577}\BibitemShut {NoStop}%
\bibitem [{\citenamefont {Xu}\ and\ \citenamefont {Cao}(2016)}]{Cao_2016}%
  \BibitemOpen
  \bibfield  {author} {\bibinfo {author} {\bibfnamefont {Dazhi}\ \bibnamefont {Xu}}\ and\ \bibinfo {author} {\bibfnamefont {Jianshu}\ \bibnamefont {Cao}},\ }\bibfield  {title} {\enquote {\bibinfo {title} {Non-canonical distribution and non-equilibrium transport beyond weak system-bath coupling regime: A polaron transformation approach},}\ }\href {\doibase 10.1007/s11467-016-0540-2} {\bibfield  {journal} {\bibinfo  {journal} {Front. Phys.}\ }\textbf {\bibinfo {volume} {11}},\ \bibinfo {pages} {110308} (\bibinfo {year} {2016})}\BibitemShut {NoStop}%
\bibitem [{\citenamefont {Napoli}\ \emph {et~al.}(2025)\citenamefont {Napoli}, \citenamefont {Mercurio}, \citenamefont {Lamberto}, \citenamefont {Zappal\`a}, \citenamefont {Di~Stefano},\ and\ \citenamefont {Savasta}}]{Napoli_2025}%
  \BibitemOpen
  \bibfield  {author} {\bibinfo {author} {\bibfnamefont {Samuel}\ \bibnamefont {Napoli}}, \bibinfo {author} {\bibfnamefont {Alberto}\ \bibnamefont {Mercurio}}, \bibinfo {author} {\bibfnamefont {Daniele}\ \bibnamefont {Lamberto}}, \bibinfo {author} {\bibfnamefont {Andrea}\ \bibnamefont {Zappal\`a}}, \bibinfo {author} {\bibfnamefont {Omar}\ \bibnamefont {Di~Stefano}}, \ and\ \bibinfo {author} {\bibfnamefont {Salvatore}\ \bibnamefont {Savasta}},\ }\bibfield  {title} {\enquote {\bibinfo {title} {Circuit qed spectra in the ultrastrong coupling regime: How they differ from cavity qed},}\ }\href {\doibase 10.1103/9zl7-31f3} {\bibfield  {journal} {\bibinfo  {journal} {Phys. Rev. Res.}\ }\textbf {\bibinfo {volume} {7}},\ \bibinfo {pages} {033037} (\bibinfo {year} {2025})}\BibitemShut {NoStop}%
\bibitem [{\citenamefont {Mochida}\ and\ \citenamefont {Ashida}(2024)}]{Mochida_2024}%
  \BibitemOpen
  \bibfield  {author} {\bibinfo {author} {\bibfnamefont {Jun}\ \bibnamefont {Mochida}}\ and\ \bibinfo {author} {\bibfnamefont {Yuto}\ \bibnamefont {Ashida}},\ }\bibfield  {title} {\enquote {\bibinfo {title} {Cavity-enhanced kondo effect},}\ }\href {\doibase 10.1103/PhysRevB.110.035158} {\bibfield  {journal} {\bibinfo  {journal} {Phys. Rev. B}\ }\textbf {\bibinfo {volume} {110}},\ \bibinfo {pages} {035158} (\bibinfo {year} {2024})}\BibitemShut {NoStop}%
\bibitem [{\citenamefont {Masuki}\ and\ \citenamefont {Ashida}(2024)}]{Masuki_2024}%
  \BibitemOpen
  \bibfield  {author} {\bibinfo {author} {\bibfnamefont {Kanta}\ \bibnamefont {Masuki}}\ and\ \bibinfo {author} {\bibfnamefont {Yuto}\ \bibnamefont {Ashida}},\ }\bibfield  {title} {\enquote {\bibinfo {title} {Cavity moir\'e materials: Controlling magnetic frustration with quantum light-matter interaction},}\ }\href {\doibase 10.1103/PhysRevB.109.195173} {\bibfield  {journal} {\bibinfo  {journal} {Phys. Rev. B}\ }\textbf {\bibinfo {volume} {109}},\ \bibinfo {pages} {195173} (\bibinfo {year} {2024})}\BibitemShut {NoStop}%
\bibitem [{\citenamefont {Ashida}\ \emph {et~al.}(2023)\citenamefont {Ashida}, \citenamefont {\ifmmode \dot{I}\else \.{I}\fi{}mamo\ifmmode~\breve{g}\else \u{g}\fi{}lu},\ and\ \citenamefont {Demler}}]{Ashida_2023}%
  \BibitemOpen
  \bibfield  {author} {\bibinfo {author} {\bibfnamefont {Yuto}\ \bibnamefont {Ashida}}, \bibinfo {author} {\bibfnamefont {Ata\ifmmode \mbox{\c{c}}\else~\c{c}\fi{}}\ \bibnamefont {\ifmmode \dot{I}\else \.{I}\fi{}mamo\ifmmode~\breve{g}\else \u{g}\fi{}lu}}, \ and\ \bibinfo {author} {\bibfnamefont {Eugene}\ \bibnamefont {Demler}},\ }\bibfield  {title} {\enquote {\bibinfo {title} {Cavity quantum electrodynamics with hyperbolic van der waals materials},}\ }\href {\doibase 10.1103/PhysRevLett.130.216901} {\bibfield  {journal} {\bibinfo  {journal} {Phys. Rev. Lett.}\ }\textbf {\bibinfo {volume} {130}},\ \bibinfo {pages} {216901} (\bibinfo {year} {2023})}\BibitemShut {NoStop}%
\bibitem [{\citenamefont {Masuki}\ and\ \citenamefont {Ashida}(2023)}]{Masuki_2023}%
  \BibitemOpen
  \bibfield  {author} {\bibinfo {author} {\bibfnamefont {Kanta}\ \bibnamefont {Masuki}}\ and\ \bibinfo {author} {\bibfnamefont {Yuto}\ \bibnamefont {Ashida}},\ }\bibfield  {title} {\enquote {\bibinfo {title} {Berry phase and topology in ultrastrongly coupled quantum light-matter systems},}\ }\href {\doibase 10.1103/PhysRevB.107.195104} {\bibfield  {journal} {\bibinfo  {journal} {Phys. Rev. B}\ }\textbf {\bibinfo {volume} {107}},\ \bibinfo {pages} {195104} (\bibinfo {year} {2023})}\BibitemShut {NoStop}%
\bibitem [{\citenamefont {Masuki}\ \emph {et~al.}(2022)\citenamefont {Masuki}, \citenamefont {Sudo}, \citenamefont {Oshikawa},\ and\ \citenamefont {Ashida}}]{Masuki_2022}%
  \BibitemOpen
  \bibfield  {author} {\bibinfo {author} {\bibfnamefont {Kanta}\ \bibnamefont {Masuki}}, \bibinfo {author} {\bibfnamefont {Hiroyuki}\ \bibnamefont {Sudo}}, \bibinfo {author} {\bibfnamefont {Masaki}\ \bibnamefont {Oshikawa}}, \ and\ \bibinfo {author} {\bibfnamefont {Yuto}\ \bibnamefont {Ashida}},\ }\bibfield  {title} {\enquote {\bibinfo {title} {Absence versus presence of dissipative quantum phase transition in josephson junctions},}\ }\href {\doibase 10.1103/PhysRevLett.129.087001} {\bibfield  {journal} {\bibinfo  {journal} {Phys. Rev. Lett.}\ }\textbf {\bibinfo {volume} {129}},\ \bibinfo {pages} {087001} (\bibinfo {year} {2022})}\BibitemShut {NoStop}%
\bibitem [{\citenamefont {Ashida}\ \emph {et~al.}(2022)\citenamefont {Ashida}, \citenamefont {Yokota}, \citenamefont {\ifmmode \dot{I}\else \.{I}\fi{}mamo\ifmmode~\breve{g}\else \u{g}\fi{}lu},\ and\ \citenamefont {Demler}}]{Ashida_2022}%
  \BibitemOpen
  \bibfield  {author} {\bibinfo {author} {\bibfnamefont {Yuto}\ \bibnamefont {Ashida}}, \bibinfo {author} {\bibfnamefont {Takeru}\ \bibnamefont {Yokota}}, \bibinfo {author} {\bibfnamefont {Ata\ifmmode \mbox{\c{c}}\else~\c{c}\fi{}}\ \bibnamefont {\ifmmode \dot{I}\else \.{I}\fi{}mamo\ifmmode~\breve{g}\else \u{g}\fi{}lu}}, \ and\ \bibinfo {author} {\bibfnamefont {Eugene}\ \bibnamefont {Demler}},\ }\bibfield  {title} {\enquote {\bibinfo {title} {Nonperturbative waveguide quantum electrodynamics},}\ }\href {\doibase 10.1103/PhysRevResearch.4.023194} {\bibfield  {journal} {\bibinfo  {journal} {Phys. Rev. Res.}\ }\textbf {\bibinfo {volume} {4}},\ \bibinfo {pages} {023194} (\bibinfo {year} {2022})}\BibitemShut {NoStop}%
\bibitem [{\citenamefont {Ashida}\ \emph {et~al.}(2021)\citenamefont {Ashida}, \citenamefont {\ifmmode \dot{I}\else \.{I}\fi{}mamo\ifmmode~\breve{g}\else \u{g}\fi{}lu},\ and\ \citenamefont {Demler}}]{Ashida_2021}%
  \BibitemOpen
  \bibfield  {author} {\bibinfo {author} {\bibfnamefont {Yuto}\ \bibnamefont {Ashida}}, \bibinfo {author} {\bibfnamefont {Ata\ifmmode \mbox{\c{c}}\else~\c{c}\fi{}}\ \bibnamefont {\ifmmode \dot{I}\else \.{I}\fi{}mamo\ifmmode~\breve{g}\else \u{g}\fi{}lu}}, \ and\ \bibinfo {author} {\bibfnamefont {Eugene}\ \bibnamefont {Demler}},\ }\bibfield  {title} {\enquote {\bibinfo {title} {Cavity quantum electrodynamics at arbitrary light-matter coupling strengths},}\ }\href {\doibase 10.1103/PhysRevLett.126.153603} {\bibfield  {journal} {\bibinfo  {journal} {Phys. Rev. Lett.}\ }\textbf {\bibinfo {volume} {126}},\ \bibinfo {pages} {153603} (\bibinfo {year} {2021})}\BibitemShut {NoStop}%
\bibitem [{\citenamefont {Manzano}(2020)}]{Manzano_2020}%
  \BibitemOpen
  \bibfield  {author} {\bibinfo {author} {\bibfnamefont {Daniel}\ \bibnamefont {Manzano}},\ }\bibfield  {title} {\enquote {\bibinfo {title} {A short introduction to the lindblad master equation},}\ }\href {\doibase 10.1063/1.5115323} {\bibfield  {journal} {\bibinfo  {journal} {AIP Advances}\ }\textbf {\bibinfo {volume} {10}},\ \bibinfo {pages} {025106} (\bibinfo {year} {2020})}\BibitemShut {NoStop}%
\bibitem [{\citenamefont {Kast}\ and\ \citenamefont {Ankerhold}(2013)}]{Ankerhold_2013}%
  \BibitemOpen
  \bibfield  {author} {\bibinfo {author} {\bibfnamefont {Denis}\ \bibnamefont {Kast}}\ and\ \bibinfo {author} {\bibfnamefont {Joachim}\ \bibnamefont {Ankerhold}},\ }\bibfield  {title} {\enquote {\bibinfo {title} {Dynamics of quantum coherences at strong coupling to a heat bath},}\ }\href {\doibase 10.1103/PhysRevB.87.134301} {\bibfield  {journal} {\bibinfo  {journal} {Phys. Rev. B}\ }\textbf {\bibinfo {volume} {87}},\ \bibinfo {pages} {134301} (\bibinfo {year} {2013})}\BibitemShut {NoStop}%
\bibitem [{\citenamefont {Becker}\ \emph {et~al.}(2021)\citenamefont {Becker}, \citenamefont {Wu},\ and\ \citenamefont {Eckardt}}]{Becker_2021}%
  \BibitemOpen
  \bibfield  {author} {\bibinfo {author} {\bibfnamefont {Tobias}\ \bibnamefont {Becker}}, \bibinfo {author} {\bibfnamefont {Ling-Na}\ \bibnamefont {Wu}}, \ and\ \bibinfo {author} {\bibfnamefont {Andr\'e}\ \bibnamefont {Eckardt}},\ }\bibfield  {title} {\enquote {\bibinfo {title} {Lindbladian approximation beyond ultraweak coupling},}\ }\href {\doibase 10.1103/PhysRevE.104.014110} {\bibfield  {journal} {\bibinfo  {journal} {Phys. Rev. E}\ }\textbf {\bibinfo {volume} {104}},\ \bibinfo {pages} {014110} (\bibinfo {year} {2021})}\BibitemShut {NoStop}%
\bibitem [{\citenamefont {Xu}\ \emph {et~al.}(2019)\citenamefont {Xu}, \citenamefont {Thingna}, \citenamefont {Guo},\ and\ \citenamefont {Poletti}}]{Poletti_2019}%
  \BibitemOpen
  \bibfield  {author} {\bibinfo {author} {\bibfnamefont {Xiansong}\ \bibnamefont {Xu}}, \bibinfo {author} {\bibfnamefont {Juzar}\ \bibnamefont {Thingna}}, \bibinfo {author} {\bibfnamefont {Chu}\ \bibnamefont {Guo}}, \ and\ \bibinfo {author} {\bibfnamefont {Dario}\ \bibnamefont {Poletti}},\ }\bibfield  {title} {\enquote {\bibinfo {title} {Many-body open quantum systems beyond lindblad master equations},}\ }\href {\doibase 10.1103/PhysRevA.99.012106} {\bibfield  {journal} {\bibinfo  {journal} {Phys. Rev. A}\ }\textbf {\bibinfo {volume} {99}},\ \bibinfo {pages} {012106} (\bibinfo {year} {2019})}\BibitemShut {NoStop}%
\bibitem [{\citenamefont {Nathan}\ and\ \citenamefont {Rudner}(2020)}]{Rudner_2020}%
  \BibitemOpen
  \bibfield  {author} {\bibinfo {author} {\bibfnamefont {Frederik}\ \bibnamefont {Nathan}}\ and\ \bibinfo {author} {\bibfnamefont {Mark~S.}\ \bibnamefont {Rudner}},\ }\bibfield  {title} {\enquote {\bibinfo {title} {Universal lindblad equation for open quantum systems},}\ }\href {\doibase 10.1103/PhysRevB.102.115109} {\bibfield  {journal} {\bibinfo  {journal} {Phys. Rev. B}\ }\textbf {\bibinfo {volume} {102}},\ \bibinfo {pages} {115109} (\bibinfo {year} {2020})}\BibitemShut {NoStop}%
\bibitem [{\citenamefont {J\"ager}\ \emph {et~al.}(2022)\citenamefont {J\"ager}, \citenamefont {Schmit}, \citenamefont {Morigi}, \citenamefont {Holland},\ and\ \citenamefont {Betzholz}}]{Betzholz_2022}%
  \BibitemOpen
  \bibfield  {author} {\bibinfo {author} {\bibfnamefont {Simon~B.}\ \bibnamefont {J\"ager}}, \bibinfo {author} {\bibfnamefont {Tom}\ \bibnamefont {Schmit}}, \bibinfo {author} {\bibfnamefont {Giovanna}\ \bibnamefont {Morigi}}, \bibinfo {author} {\bibfnamefont {Murray~J.}\ \bibnamefont {Holland}}, \ and\ \bibinfo {author} {\bibfnamefont {Ralf}\ \bibnamefont {Betzholz}},\ }\bibfield  {title} {\enquote {\bibinfo {title} {Lindblad master equations for quantum systems coupled to dissipative bosonic modes},}\ }\href {\doibase 10.1103/PhysRevLett.129.063601} {\bibfield  {journal} {\bibinfo  {journal} {Phys. Rev. Lett.}\ }\textbf {\bibinfo {volume} {129}},\ \bibinfo {pages} {063601} (\bibinfo {year} {2022})}\BibitemShut {NoStop}%
\bibitem [{\citenamefont {Trushechkin}(2022)}]{Trushechkin_2022}%
  \BibitemOpen
  \bibfield  {author} {\bibinfo {author} {\bibfnamefont {Anton}\ \bibnamefont {Trushechkin}},\ }\bibfield  {title} {\enquote {\bibinfo {title} {Quantum master equations and steady states for the ultrastrong-coupling limit and the strong-decoherence limit},}\ }\href@noop {} {\bibfield  {journal} {\bibinfo  {journal} {Phys. Rev. A}\ }\textbf {\bibinfo {volume} {106}},\ \bibinfo {pages} {042209} (\bibinfo {year} {2022})}\BibitemShut {NoStop}%
\bibitem [{\citenamefont {Ivander}\ \emph {et~al.}(2023)\citenamefont {Ivander}, \citenamefont {Anto-Sztrikacs},\ and\ \citenamefont {Segal}}]{Ivander_2023}%
  \BibitemOpen
  \bibfield  {author} {\bibinfo {author} {\bibfnamefont {Felix}\ \bibnamefont {Ivander}}, \bibinfo {author} {\bibfnamefont {Nicholas}\ \bibnamefont {Anto-Sztrikacs}}, \ and\ \bibinfo {author} {\bibfnamefont {Dvira}\ \bibnamefont {Segal}},\ }\bibfield  {title} {\enquote {\bibinfo {title} {Hyperacceleration of quantum thermalization dynamics by bypassing long-lived coherences: An analytical treatment},}\ }\href@noop {} {\bibfield  {journal} {\bibinfo  {journal} {Phys. Rev. E}\ }\textbf {\bibinfo {volume} {108}},\ \bibinfo {pages} {014130} (\bibinfo {year} {2023})}\BibitemShut {NoStop}%
\bibitem [{\citenamefont {Ivander}\ \emph {et~al.}(2022)\citenamefont {Ivander}, \citenamefont {Anto-Sztrikacs},\ and\ \citenamefont {Segal}}]{Ivander_2022}%
  \BibitemOpen
  \bibfield  {author} {\bibinfo {author} {\bibfnamefont {Felix}\ \bibnamefont {Ivander}}, \bibinfo {author} {\bibfnamefont {Nicholas}\ \bibnamefont {Anto-Sztrikacs}}, \ and\ \bibinfo {author} {\bibfnamefont {Dvira}\ \bibnamefont {Segal}},\ }\bibfield  {title} {\enquote {\bibinfo {title} {Quantum coherence-control of thermal energy transport: the v model as a case study},}\ }\href@noop {} {\bibfield  {journal} {\bibinfo  {journal} {New J. Phys.}\ }\textbf {\bibinfo {volume} {24}},\ \bibinfo {pages} {103010} (\bibinfo {year} {2022})}\BibitemShut {NoStop}%
\bibitem [{\citenamefont {Gerry}\ and\ \citenamefont {Segal}(2023)}]{Gerry_2023}%
  \BibitemOpen
  \bibfield  {author} {\bibinfo {author} {\bibfnamefont {Matthew}\ \bibnamefont {Gerry}}\ and\ \bibinfo {author} {\bibfnamefont {Dvira}\ \bibnamefont {Segal}},\ }\bibfield  {title} {\enquote {\bibinfo {title} {Full counting statistics and coherences: Fluctuation symmetry in heat transport with the unified quantum master equation},}\ }\href@noop {} {\bibfield  {journal} {\bibinfo  {journal} {Phys. Rev. E}\ }\textbf {\bibinfo {volume} {107}},\ \bibinfo {pages} {054115} (\bibinfo {year} {2023})}\BibitemShut {NoStop}%
\bibitem [{\citenamefont {Koyu}\ \emph {et~al.}(2021)\citenamefont {Koyu}, \citenamefont {Dodin}, \citenamefont {Brumer},\ and\ \citenamefont {Tscherbul}}]{Koyu_2021}%
  \BibitemOpen
  \bibfield  {author} {\bibinfo {author} {\bibfnamefont {Suyesh}\ \bibnamefont {Koyu}}, \bibinfo {author} {\bibfnamefont {Amro}\ \bibnamefont {Dodin}}, \bibinfo {author} {\bibfnamefont {Paul}\ \bibnamefont {Brumer}}, \ and\ \bibinfo {author} {\bibfnamefont {Timur}\ \bibnamefont {Tscherbul}},\ }\bibfield  {title} {\enquote {\bibinfo {title} {Steady-state fano coherences in a v-type system driven by polarized incoherent light},}\ }\href@noop {} {\bibfield  {journal} {\bibinfo  {journal} {Phys. Rev. Research}\ }\textbf {\bibinfo {volume} {3}},\ \bibinfo {pages} {013295} (\bibinfo {year} {2021})}\BibitemShut {NoStop}%
\bibitem [{\citenamefont {Kundu}\ and\ \citenamefont {Makri}(2023)}]{Kundu23}%
  \BibitemOpen
  \bibfield  {author} {\bibinfo {author} {\bibfnamefont {Sohang}\ \bibnamefont {Kundu}}\ and\ \bibinfo {author} {\bibfnamefont {Nancy}\ \bibnamefont {Makri}},\ }\bibfield  {title} {\enquote {\bibinfo {title} {{PathSum: A C++ and Fortran suite of fully quantum mechanical real-time path integral methods for (multi-)system + bath dynamics}},}\ }\href {\doibase 10.1063/5.0151748} {\bibfield  {journal} {\bibinfo  {journal} {The Journal of Chemical Physics}\ }\textbf {\bibinfo {volume} {158}},\ \bibinfo {pages} {224801} (\bibinfo {year} {2023})}\BibitemShut {NoStop}%
\bibitem [{\citenamefont {Segal}\ \emph {et~al.}(2010)\citenamefont {Segal}, \citenamefont {Millis},\ and\ \citenamefont {Reichman}}]{Segal_2010}%
  \BibitemOpen
  \bibfield  {author} {\bibinfo {author} {\bibfnamefont {Dvira}\ \bibnamefont {Segal}}, \bibinfo {author} {\bibfnamefont {Andrew~J.}\ \bibnamefont {Millis}}, \ and\ \bibinfo {author} {\bibfnamefont {David~R.}\ \bibnamefont {Reichman}},\ }\bibfield  {title} {\enquote {\bibinfo {title} {Numerically exact path-integral simulation of nonequilibrium quantum transport and dissipation},}\ }\href {\doibase 10.1103/PhysRevB.82.205323} {\bibfield  {journal} {\bibinfo  {journal} {Phys. Rev. B}\ }\textbf {\bibinfo {volume} {82}},\ \bibinfo {pages} {205323} (\bibinfo {year} {2010})}\BibitemShut {NoStop}%
\bibitem [{\citenamefont {Chen}\ \emph {et~al.}(2017{\natexlab{a}})\citenamefont {Chen}, \citenamefont {Cohen},\ and\ \citenamefont {Reichman}}]{Theta1}%
  \BibitemOpen
  \bibfield  {author} {\bibinfo {author} {\bibfnamefont {Hsing-Ta}\ \bibnamefont {Chen}}, \bibinfo {author} {\bibfnamefont {Guy}\ \bibnamefont {Cohen}}, \ and\ \bibinfo {author} {\bibfnamefont {David~R.}\ \bibnamefont {Reichman}},\ }\bibfield  {title} {\enquote {\bibinfo {title} {Inchworm monte carlo for exact non-adiabatic dynamics i. theory and algorithms},}\ }\href {\doibase 10.1063/1.4974727} {\bibfield  {journal} {\bibinfo  {journal} {The Journal of Chemical Physics}\ }\textbf {\bibinfo {volume} {146}},\ \bibinfo {pages} {054105} (\bibinfo {year} {2017}{\natexlab{a}})}\BibitemShut {NoStop}%
\bibitem [{\citenamefont {Chen}\ \emph {et~al.}(2017{\natexlab{b}})\citenamefont {Chen}, \citenamefont {Cohen},\ and\ \citenamefont {Reichman}}]{Theta2}%
  \BibitemOpen
  \bibfield  {author} {\bibinfo {author} {\bibfnamefont {Hsing-Ta}\ \bibnamefont {Chen}}, \bibinfo {author} {\bibfnamefont {Guy}\ \bibnamefont {Cohen}}, \ and\ \bibinfo {author} {\bibfnamefont {David~R.}\ \bibnamefont {Reichman}},\ }\bibfield  {title} {\enquote {\bibinfo {title} {Inchworm monte carlo for exact non-adiabatic dynamics ii. benchmarks and comparison with established methods},}\ }\href {\doibase 10.1063/1.4974728} {\bibfield  {journal} {\bibinfo  {journal} {The Journal of Chemical Physics}\ }\textbf {\bibinfo {volume} {146}},\ \bibinfo {pages} {054106} (\bibinfo {year} {2017}{\natexlab{b}})}\BibitemShut {NoStop}%
\bibitem [{\citenamefont {Gribben}\ \emph {et~al.}(2020)\citenamefont {Gribben}, \citenamefont {Strathearn}, \citenamefont {Iles-Smith}, \citenamefont {Kilda}, \citenamefont {Nazir}, \citenamefont {Lovett},\ and\ \citenamefont {Kirton}}]{Gribben_2020}%
  \BibitemOpen
  \bibfield  {author} {\bibinfo {author} {\bibfnamefont {Dominic}\ \bibnamefont {Gribben}}, \bibinfo {author} {\bibfnamefont {Aidan}\ \bibnamefont {Strathearn}}, \bibinfo {author} {\bibfnamefont {Jake}\ \bibnamefont {Iles-Smith}}, \bibinfo {author} {\bibfnamefont {Dainius}\ \bibnamefont {Kilda}}, \bibinfo {author} {\bibfnamefont {Ahsan}\ \bibnamefont {Nazir}}, \bibinfo {author} {\bibfnamefont {Brendon~W.}\ \bibnamefont {Lovett}}, \ and\ \bibinfo {author} {\bibfnamefont {Peter}\ \bibnamefont {Kirton}},\ }\bibfield  {title} {\enquote {\bibinfo {title} {Exact quantum dynamics in structured environments},}\ }\href {\doibase 10.1103/PhysRevResearch.2.013265} {\bibfield  {journal} {\bibinfo  {journal} {Phys. Rev. Res.}\ }\textbf {\bibinfo {volume} {2}},\ \bibinfo {pages} {013265} (\bibinfo {year} {2020})}\BibitemShut {NoStop}%
\bibitem [{\citenamefont {Tanimura}(2020)}]{Tanimura_2020}%
  \BibitemOpen
  \bibfield  {author} {\bibinfo {author} {\bibfnamefont {Yoshitaka}\ \bibnamefont {Tanimura}},\ }\bibfield  {title} {\enquote {\bibinfo {title} {Numerically “exact” approach to open quantum dynamics: The hierarchical equations of motion (heom)},}\ }\href {\doibase 10.1063/5.0011599} {\bibfield  {journal} {\bibinfo  {journal} {The Journal of Chemical Physics}\ }\textbf {\bibinfo {volume} {153}},\ \bibinfo {pages} {020901} (\bibinfo {year} {2020})}\BibitemShut {NoStop}%
\bibitem [{\citenamefont {Lambert}\ \emph {et~al.}(2023)\citenamefont {Lambert}, \citenamefont {Raheja}, \citenamefont {Cross}, \citenamefont {Menczel}, \citenamefont {Ahmed}, \citenamefont {Pitchford}, \citenamefont {Burgarth},\ and\ \citenamefont {Nori}}]{Lambert_2023}%
  \BibitemOpen
  \bibfield  {author} {\bibinfo {author} {\bibfnamefont {Neill}\ \bibnamefont {Lambert}}, \bibinfo {author} {\bibfnamefont {Tarun}\ \bibnamefont {Raheja}}, \bibinfo {author} {\bibfnamefont {Simon}\ \bibnamefont {Cross}}, \bibinfo {author} {\bibfnamefont {Paul}\ \bibnamefont {Menczel}}, \bibinfo {author} {\bibfnamefont {Shahnawaz}\ \bibnamefont {Ahmed}}, \bibinfo {author} {\bibfnamefont {Alexander}\ \bibnamefont {Pitchford}}, \bibinfo {author} {\bibfnamefont {Daniel}\ \bibnamefont {Burgarth}}, \ and\ \bibinfo {author} {\bibfnamefont {Franco}\ \bibnamefont {Nori}},\ }\bibfield  {title} {\enquote {\bibinfo {title} {Qutip-bofin: A bosonic and fermionic numerical hierarchical-equations-of-motion library with applications in light-harvesting, quantum control, and single-molecule electronics},}\ }\href {\doibase 10.1103/PhysRevResearch.5.013181} {\bibfield  {journal} {\bibinfo  {journal} {Phys. Rev. Res.}\ }\textbf {\bibinfo {volume} {5}},\ \bibinfo {pages} {013181} (\bibinfo {year} {2023})}\BibitemShut {NoStop}%
\bibitem [{\citenamefont {Segal}\ and\ \citenamefont {Nitzan}(2002)}]{Segal02}%
  \BibitemOpen
  \bibfield  {author} {\bibinfo {author} {\bibfnamefont {Dvira}\ \bibnamefont {Segal}}\ and\ \bibinfo {author} {\bibfnamefont {Abraham}\ \bibnamefont {Nitzan}},\ }\bibfield  {title} {\enquote {\bibinfo {title} {Conduction in molecular junctions: inelastic effects},}\ }\href {\doibase 10.1016/S0301-0104(02)00687-7} {\bibfield  {journal} {\bibinfo  {journal} {Chemical Physics}\ }\textbf {\bibinfo {volume} {281}},\ \bibinfo {pages} {235--256} (\bibinfo {year} {2002})}\BibitemShut {NoStop}%
\bibitem [{\citenamefont {Jang}\ \emph {et~al.}(2008)\citenamefont {Jang}, \citenamefont {Cheng}, \citenamefont {Reichman},\ and\ \citenamefont {Eaves}}]{Jang08}%
  \BibitemOpen
  \bibfield  {author} {\bibinfo {author} {\bibfnamefont {Seogjoo}\ \bibnamefont {Jang}}, \bibinfo {author} {\bibfnamefont {Yuan-Chung}\ \bibnamefont {Cheng}}, \bibinfo {author} {\bibfnamefont {David~R.}\ \bibnamefont {Reichman}}, \ and\ \bibinfo {author} {\bibfnamefont {Joel~D.}\ \bibnamefont {Eaves}},\ }\bibfield  {title} {\enquote {\bibinfo {title} {Theory of coherent resonance energy transfer},}\ }\href {\doibase 10.1063/1.2977974} {\bibfield  {journal} {\bibinfo  {journal} {The Journal of Chemical Physics}\ }\textbf {\bibinfo {volume} {129}},\ \bibinfo {pages} {101104} (\bibinfo {year} {2008})}\BibitemShut {NoStop}%
\bibitem [{\citenamefont {Pollock}\ \emph {et~al.}(2013)\citenamefont {Pollock}, \citenamefont {McCutcheon}, \citenamefont {Lovett}, \citenamefont {Gauger},\ and\ \citenamefont {Nazir}}]{Ahsan13}%
  \BibitemOpen
  \bibfield  {author} {\bibinfo {author} {\bibfnamefont {Felix~A.}\ \bibnamefont {Pollock}}, \bibinfo {author} {\bibfnamefont {Dara P.~S.}\ \bibnamefont {McCutcheon}}, \bibinfo {author} {\bibfnamefont {Brendon~W.}\ \bibnamefont {Lovett}}, \bibinfo {author} {\bibfnamefont {Erik~M.}\ \bibnamefont {Gauger}}, \ and\ \bibinfo {author} {\bibfnamefont {Ahsan}\ \bibnamefont {Nazir}},\ }\bibfield  {title} {\enquote {\bibinfo {title} {A multi-site variational master equation approach to dissipative energy transfer},}\ }\href {\doibase 10.1088/1367-2630/15/7/075018} {\bibfield  {journal} {\bibinfo  {journal} {New Journal of Physics}\ }\textbf {\bibinfo {volume} {15}},\ \bibinfo {pages} {075018} (\bibinfo {year} {2013})}\BibitemShut {NoStop}%
\bibitem [{\citenamefont {Lee}\ \emph {et~al.}(2015)\citenamefont {Lee}, \citenamefont {Moix},\ and\ \citenamefont {Cao}}]{Cao15}%
  \BibitemOpen
  \bibfield  {author} {\bibinfo {author} {\bibfnamefont {Chee~Kong}\ \bibnamefont {Lee}}, \bibinfo {author} {\bibfnamefont {Jeremy}\ \bibnamefont {Moix}}, \ and\ \bibinfo {author} {\bibfnamefont {Jianshu}\ \bibnamefont {Cao}},\ }\bibfield  {title} {\enquote {\bibinfo {title} {Coherent quantum transport in disordered systems: A unified polaron treatment of hopping and band-like transport},}\ }\href {\doibase 10.1063/1.4918736} {\bibfield  {journal} {\bibinfo  {journal} {The Journal of Chemical Physics}\ }\textbf {\bibinfo {volume} {142}},\ \bibinfo {pages} {164103} (\bibinfo {year} {2015})}\BibitemShut {NoStop}%
\bibitem [{\citenamefont {Iles-Smith}\ \emph {et~al.}(2024)\citenamefont {Iles-Smith}, \citenamefont {Diba},\ and\ \citenamefont {Nazir}}]{Ahsan24}%
  \BibitemOpen
  \bibfield  {author} {\bibinfo {author} {\bibfnamefont {Jake}\ \bibnamefont {Iles-Smith}}, \bibinfo {author} {\bibfnamefont {Owen}\ \bibnamefont {Diba}}, \ and\ \bibinfo {author} {\bibfnamefont {Ahsan}\ \bibnamefont {Nazir}},\ }\bibfield  {title} {\enquote {\bibinfo {title} {Capturing non-markovian polaron dressing with the master equation formalism},}\ }\href {\doibase 10.1063/5.0181839} {\bibfield  {journal} {\bibinfo  {journal} {The Journal of Chemical Physics}\ }\textbf {\bibinfo {volume} {161}},\ \bibinfo {pages} {134111} (\bibinfo {year} {2024})}\BibitemShut {NoStop}%
\bibitem [{\citenamefont {Anto-Sztrikacs}\ and\ \citenamefont {Segal}(2021)}]{NickNJP}%
  \BibitemOpen
  \bibfield  {author} {\bibinfo {author} {\bibfnamefont {Nicholas}\ \bibnamefont {Anto-Sztrikacs}}\ and\ \bibinfo {author} {\bibfnamefont {Dvira}\ \bibnamefont {Segal}},\ }\bibfield  {title} {\enquote {\bibinfo {title} {Strong coupling effects in quantum thermal transport with the reaction coordinate method},}\ }\href {\doibase 10.1088/1367-2630/ac02df} {\bibfield  {journal} {\bibinfo  {journal} {New Journal of Physics}\ }\textbf {\bibinfo {volume} {23}},\ \bibinfo {pages} {063036} (\bibinfo {year} {2021})}\BibitemShut {NoStop}%
\bibitem [{\citenamefont {Iles-Smith}\ \emph {et~al.}(2016)\citenamefont {Iles-Smith}, \citenamefont {Dijkstra}, \citenamefont {Lambert},\ and\ \citenamefont {Nazir}}]{Ahsan16}%
  \BibitemOpen
  \bibfield  {author} {\bibinfo {author} {\bibfnamefont {Jake}\ \bibnamefont {Iles-Smith}}, \bibinfo {author} {\bibfnamefont {Arend~G.}\ \bibnamefont {Dijkstra}}, \bibinfo {author} {\bibfnamefont {Neill}\ \bibnamefont {Lambert}}, \ and\ \bibinfo {author} {\bibfnamefont {Ahsan}\ \bibnamefont {Nazir}},\ }\bibfield  {title} {\enquote {\bibinfo {title} {Energy transfer in structured and unstructured environments: Master equations beyond the born--markov approximations},}\ }\href {\doibase 10.1063/1.4940218} {\bibfield  {journal} {\bibinfo  {journal} {The Journal of Chemical Physics}\ }\textbf {\bibinfo {volume} {144}},\ \bibinfo {pages} {044110} (\bibinfo {year} {2016})}\BibitemShut {NoStop}%
\bibitem [{\citenamefont {Woods}\ \emph {et~al.}(2015)\citenamefont {Woods}, \citenamefont {Cramer},\ and\ \citenamefont {Plenio}}]{Chain15}%
  \BibitemOpen
  \bibfield  {author} {\bibinfo {author} {\bibfnamefont {Mischa~P.}\ \bibnamefont {Woods}}, \bibinfo {author} {\bibfnamefont {Marcus}\ \bibnamefont {Cramer}}, \ and\ \bibinfo {author} {\bibfnamefont {Martin~B.}\ \bibnamefont {Plenio}},\ }\bibfield  {title} {\enquote {\bibinfo {title} {Simulating bosonic baths with error bars},}\ }\href {\doibase 10.1103/PhysRevLett.115.130401} {\bibfield  {journal} {\bibinfo  {journal} {Physical Review Letters}\ }\textbf {\bibinfo {volume} {115}},\ \bibinfo {pages} {130401} (\bibinfo {year} {2015})}\BibitemShut {NoStop}%
\bibitem [{\citenamefont {Colla}\ and\ \citenamefont {Breuer}(2022)}]{MD22}%
  \BibitemOpen
  \bibfield  {author} {\bibinfo {author} {\bibfnamefont {Alessandra}\ \bibnamefont {Colla}}\ and\ \bibinfo {author} {\bibfnamefont {Heinz-Peter}\ \bibnamefont {Breuer}},\ }\bibfield  {title} {\enquote {\bibinfo {title} {Open-system approach to nonequilibrium quantum thermodynamics at arbitrary coupling},}\ }\href {\doibase 10.1103/PhysRevA.105.052216} {\bibfield  {journal} {\bibinfo  {journal} {Phys. Rev. A}\ }\textbf {\bibinfo {volume} {105}},\ \bibinfo {pages} {052216} (\bibinfo {year} {2022})}\BibitemShut {NoStop}%
\bibitem [{\citenamefont {Curtright}\ and\ \citenamefont {Zachos}(2015)}]{Curtright_2015}%
  \BibitemOpen
  \bibfield  {author} {\bibinfo {author} {\bibfnamefont {Thomas~L.}\ \bibnamefont {Curtright}}\ and\ \bibinfo {author} {\bibfnamefont {Cosmas~K.}\ \bibnamefont {Zachos}},\ }\bibfield  {title} {\enquote {\bibinfo {title} {Elementary results for the fundamental representation of su(3)},}\ }\href {\doibase https://doi.org/10.1016/S0034-4877(15)30040-9} {\bibfield  {journal} {\bibinfo  {journal} {Reports on Mathematical Physics}\ }\textbf {\bibinfo {volume} {76}},\ \bibinfo {pages} {401--404} (\bibinfo {year} {2015})}\BibitemShut {NoStop}%
\end{thebibliography}%
\end{document}